\documentstyle[sprocl,epsfig]{article}

\def\Journal#1#2#3#4{{#1} {\bf #2}, #3 (#4)}

\def\APPB{{\em Acta Phys. Polon.} B}
\def\AP{{\em Ann. Phys.} }
\def\APAS{{\em Acta. Phys. Austr. Supp.} }
\def\ARNPS{{\em Ann. Rev. Nucl. Part. Sci} }
\def\CPC{{\em Comp. Phys. Comm.} }
\def\EPJC{{\em Eur. Phys. J.} C}
\def\IJMPA{{\em Int. J. Mod. Phys.} A}
\def\JPG{{\em J. Phys.} G}
\def\JETPL{{\em  JETP Lett. }}
\def\JHEP{{\em  JHEP} }
\def\MPLA{{\em Mod. Phys. Lett.} A}

\def\NPBPS{{\em Nucl. Phys.} B ({\em Proc. Suppl.}) }
\def\NPB{{\em Nucl. Phys.} B}
\def\PLB{{\em Phys. Lett.} B}
\def\PS{{Physica Scripta} }
\def\PRL{{\em Phys. Rev. Lett. }}
\def\PRD{{\em Phys. Rev.} D}
\def\PR{{\em Phys. Rev.} }
\def\PRpC{{\em Phys. Rep.} C}
\def\PRp{{\em Phys. Rep.} } %***********Khoze defined it as \PR

\def\RMP{{\em Rev. Mod. Phys.} }
\def\SJNP{{\em Sov. J. Nucl. Phys.} }
\def\SPJETP{{\em Sov. Phys. JETP} }
\def\SPLIR{{Sov. Phys. Lebedev Inst. Rep.} }
\def\SSSR{{\em  Izv. Akad. Nauk SSSR, Ser. Fiz.} }
\def\ZPC{{\em Z. Phys.} C}

\def\al{\alpha}

\def\be{\begin{equation}}
\def\ee{\end{equation}}
\def\bea{\begin{eqnarray}}
\def\eea{\end{eqnarray}}
\def\lapproxeq{\lower .7ex\hbox{$\;\stackrel{\textstyle
<}{\sim}\;$}}
\def\gapproxeq{\lower .7ex\hbox{$\;\stackrel{\textstyle
>}{\sim}\;$}}
\def\epem{$e^+e^-$\ }
\def\als{$\alpha_s$\ }

\newcommand{\labl}[1]{\label{#1}}
\newcommand{\N}{{\mathcal N}}
\newcommand{\Nq}{{\mathcal N}_q}
\newcommand{\Ng}{{\mathcal N}_g}
\newcommand{\Na}{{\mathcal N}_a}
\newcommand{\Nb}{{\mathcal N}_b}
\newcommand{\Nc}{{\mathcal N}_c}
\newcommand{\Ord}{{\cal O}}

\def\lapproxeq{\lower .7ex\hbox{$\;\stackrel{\textstyle
<}{\sim}\;$}}
\def\gapproxeq{\lower .7ex\hbox{$\;\stackrel{\textstyle
>}{\sim}\;$}}

\def\nostrocostrutto#1\over#2{\mathrel{\mathop{\kern 0pt \rlap
  {\raise.2ex\hbox{$#1$}}}
  \lower.9ex\hbox{\kern-.190em $#2$}}}
\def\lsim{\nostrocostrutto < \over \sim}   %less or around ...
\def\gsim{\nostrocostrutto > \over \sim}   %greater or around...

\newcommand{\as}{\alpha_s}
\newcommand{\tij}{\vartheta_{ij}}
\newcommand{\vth}{\vartheta}

\def\eq{\begin{equation}}
\def\eqx{\end{equation}}
\def\aq{\begin{eqnarray}}
\def\aqx{\end{eqnarray}}
\def\ano{\gamma_0}

\def\al{$\alpha_s\;$ }
\def\t12{\vartheta_{12}}
\def\phi{\varphi}

\def\be{\begin{equation}}
\def\ee{\end{equation}}
\def\theta{\vartheta}
\def\th{\vartheta}
\def\th12{\vartheta_{12}}
\def\thc{\vartheta_c}

\def\k1n{(k_1,..,k_n)}

\def\upt{^{(2)}}
\def\r2{r^{(2)}}
\def\d2{d^{(2)}}

\def\a{\gamma_0}
\def\a2{\gamma_0^2}

\catcode`@=11
\newcount\@tempcntc
\def\@citex[#1]#2{\if@filesw\immediate\write\@auxout{\string\citation{#2}}\fi
  \@tempcnta\z@\@tempcntb\m@ne\def\@citea{}\@cite{\@for\@citeb:=#2\do
    {\@ifundefined
       {b@\@citeb}{\@citeo\@tempcntb\m@ne\@citea\def\@citea{,}{\bf ?}\@warning
       {Citation `\@citeb' on page \thepage \space undefined}}%
    {\setbox\z@\hbox{\global\@tempcntc0\csname b@\@citeb\endcsname\relax}%
     \ifnum\@tempcntc=\z@ \@citeo\@tempcntb\m@ne
       \@citea\def\@citea{,}\hbox{\csname b@\@citeb\endcsname}%
     \else
      \advance\@tempcntb\@ne
      \ifnum\@tempcntb=\@tempcntc
      \else\advance\@tempcntb\m@ne\@citeo
      \@tempcnta\@tempcntc\@tempcntb\@tempcntc\fi\fi}}\@citeo}{#1}}
\def\@citeo{\ifnum\@tempcnta>\@tempcntb\else\@citea\def\@citea{,}%
  \ifnum\@tempcnta=\@tempcntb\the\@tempcnta\else
   {\advance\@tempcnta\@ne\ifnum\@tempcnta=\@tempcntb \else \def\@citea{-}\fi
    \advance\@tempcnta\m@ne\the\@tempcnta\@citea\the\@tempcntb}\fi\fi}
\catcode`@=12
\begin{document}

\noindent\hspace*{9cm} DTP/00/54 \newline
\hspace*{9cm} MPI-PhT/2000 - 35 \newline
\hspace*{9cm} TPJU - 9/2000 \newline
\hspace*{9cm} August 10, 2000 \newline %\vspace{0.5cm}
\vfill
\title{ANALYTICAL QCD AND MULTIPARTICLE PRODUCTION \footnote{To be published
in "Handbook of QCD" (Ioffe Festshrift), ed. M. A. Shifman (World Scientific).}}
%\vfill
\author{ VALERY A. KHOZE}

\address{Department of Physics, University of
Durham,\\ Durham DH1 3LE, UK}

\author{ WOLFGANG OCHS }

\address{Max Planck Institut f\"ur Physik, Werner-Heisenberg-Institut,
F\"ohringer Ring 6, D-80805 Munich, Germany}

\author{JACEK WOSIEK}
\address{Marian Smoluchowski Institute of Physics, Jagellonian University,
Reymonta 4, 30-059 Cracow, Poland}

%%%%%%%%%%%%%%%%%%%%%%%%%%%%%%%%%%%%%%%%%%%%%%%%%%%%%%%%%%%%%%
% You may repeat \author \address as often as necessary      %
%%%%%%%%%%%%%%%%%%%%%%%%%%%%%%%%%%%%%%%%%%%%%%%%%%%%%%%%%%%%%%
%\vfill
\maketitle\abstracts{
\vspace*{1.5cm}
We review the perturbative approach to multiparticle production in hard
collision processes. It is investigated to what extent parton level 
analytical 
calculations at low momentum cut-off can reproduce experimental data on
the hadronic final state.  
Systematic results are available for various observables
with the next-to-leading logarithmic accuracy (the so-called modified leading
logarithmic approximation - MLLA). We introduce the analytical formalism
and then discuss recent applications 
concerning multiplicities, inclusive spectra, correlations and angular flows
in multi-jet events. 
In various cases the perturbative picture is surprisingly successful,
even for very soft particle production.
\vspace*{2.5cm}}

\newpage
\tableofcontents
  \section{Introduction}

A characteristic feature of high energy collisions is the
production of many hadrons. At existing
high energy  colliders with incoming electrons or protons the mean number of
produced hadrons ranges from 30 to 50, at colliders with incoming nuclei
it is even above 1000. For a long time multiparticle dynamics has
been studied within the framework of phenomenological theories and models
basing on the quark structure of the hadrons involved in the considered process
and the analytical structure of the scattering amplitudes, see for instance
Ref. 1. %$\,$\cite{ilk}

Within the  parton model$\,$\cite{feyn,bpas} 
each constituent of the hadron carries part of the total hadron momentum 
and could scatter on other partons or interact with leptons and  
fragment finally at larger distances into the final state hadrons 
under the action of the confinement forces. In
the multiparticle final states in the ``hard'' collisions 
which involve large momentum transfers typically 
several jets of collimated hadrons appear. 
They can be related to the partons emerging from the primary hard
interaction. The existence of spectacular jets of hadrons -- the footprints
of partons -- is among the most striking phenomena in high energy physics.
Jets from primary quarks were discovered 
at the \epem collider SPEAR at SLAC in 1975$\,$\cite{rfs}
with the angular distribution as expected from the production of
spin $\frac{1}{2}$ quarks.

With the advent of Quantum Chromodynamics$\,$\cite{QCD} (QCD)  
a quantitative treatment of many phenomena involving hadrons became
accessible. The force between the quarks is mediated by gluons
which interact also among themselves.
They may be produced in a hard collision process
and evolve into a jet of final state hadrons. The gluon jets have been
observed in \epem  collisions
 by the four PETRA experiments at DESY$\,$\cite{bbbb} in association
with the two quark jets. 

These discoveries have opened up a large field of
jet physics with detailed investigations at the \epem   colliders LEP and SLC,
the p$\overline {\rm p}$ colliders at CERN and Fermilab 
and at the ep collider HERA. It will certainly remain one of the main topics
for studies at the colliders of the future. 
In the quantitative treatment of jet production one starts from a precise
definition of a jet which refers to a  resolution parameter 
in a jet-finding and jet-³defining algorithm. In this way
the multiparticle final state is reduced to a final state with only a few 
jets characterized by their 4-momenta. In the theoretical analysis of jet
production one assumes that the jet of hadrons can be represented by a jet
of very few partons (one or two, typically) and their 
production properties are analyzed in
perturbative QCD. Because of the asymptotic freedom of QCD$\,$\cite{asy} 
the coupling
constant becomes sufficiently small in hard processes
so that reasonable results can be obtained from perturbation theory.

Today many phenomena in jet physics are quantitatively described
by perturbation theory in terms of only one basic free parameter, the coupling
constant \als  at a given energy scale; 
in addition  some structural characteristics of
hadrons at a given scale are needed as well. 
For many quantities results in the next-to-leading order have been obtained.
The  accuracy of such calculations can be illustrated by the error in the
determination of \als in jet physics which is 
less than 3\%.$\,$\cite{bethke}
The  accuracy in these studies is limited typically not by the
experimental statistics but by theoretical errors from the scale
uncertainties and, in particular, from the uncertainty in the transition
from the QCD partons to the  observed hadrons. 
This latter difficulty makes it 
especially desirable to study the transition from partons to hadrons in
greater detail.

In the popular QCD-based Monte Carlo 
models for multiparticle production (HERWIG$\,$\cite{HERWIG},
JET\-SET$\,$\cite{JETSET} or ARIADNE$\,$\cite{ARIADNE})
based on the Lund string model$\,$\cite{agis}
or cluster fragmentation$\,$\cite{HERWIG} the perturbative evolution is
terminated at some low scale $Q_0$ (typically $Q_0\sim 1$ GeV)
for a dynamical variable (such as relative
transverse momentum or parton virtuality). 
Then the non-perturbative processes take over and the transition into
hadrons is described by phenomenological models.
In these models various
parameters have to be fitted by the data which makes it sometimes difficult
to directly trace the connection of a fitted observable to the underlying QCD
dynamics. Let us emphasize that all existing phenomenological models are of
a probabilistic and iterative nature when the fragmentation process is
described in terms of simple underlying branchings. Their success in
representing the data relies on the fact that in certain approximations it
is possible to absorb the quantum-mechanical interferences into the
probabilistic schemes.

Alternatively, one can try to compare the perturbative results directly to
the experimental data without a complete hadronization model. In this way
the number of non-perturbative parameters is minimized 
(in addition to the QCD coupling $\alpha_s$ one introduces a non-perturbative
cut-off $Q_0$ in the partonic cascade).
Most importantly, an  analytical treatment  becomes
feasible which reveals the details of the QCD dynamics. Also
non-probabilistic phenomena become accessible.
In recent years two kinds of applications of this general idea have been
developed.

An application of analytical methods is 
suggested in particular for observables which
are ``infrared safe'', i.e. do not change if a parton splits into two
collinear partons or emits a soft parton. Then the dependence  on the 
non-perturbative cut-off $Q_0$ is suppressed. In the last years various
applications to shape variables such as thrust and others have been 
carried out. The perturbative results for the mean values obtain
characteristic power corrections ${\cal O}(\Lambda^k/Q^k)$ with predicted
power $k$. Together with the 2-loop calculation for the lowest orders in
perturbation theory this approach is used to obtain accurate results for the
coupling $\alpha_s$.$\,$\cite{power}

Another analytical approach concerns particle densities and
correlations which are infrared sensitive and depend explicitly on the
non-perturbative cut-off $Q_0$. The concept of ``Local Parton
Hadron Duality'' (LPHD$\,$\cite{adkt1}) has been formulated originally for
inclusive spectra and states that for small $Q_0$ of order of a few hundred 
MeV the parton
distribution calculated in perturbative QCD gives already a good description
of the observables. Indeed, the general features of the inclusive particle
distributions in jets produced in 
$e^+e^-$ annihilation, deep inelastic scattering and hadron-%
hadron collisions have been described surprisingly well within this
approach. 
Meanwhile calculations of such observables have been carried
out to many more quantities (see
presentations in Refs. 15-18.  %$\,$\cite{dkmt2,webbook,ko,kl}).
The success of these calculations implies that color confinement is 
governed by rather soft processes which allow the close similarity of
parton and hadron momenta.

This paper aims at a review of such infrared sensitive 
quantities for which there are
analytical calculations. 
Many of the topics mentioned here are discussed in more details
elsewhere.$\,$\cite{dkmt2,ko} Our main goal here is to survey the basic ideas and
to illustrate the latest phenomenological advances. So we have tried
wherever possible to refer to the very recent experimental data.

The approach discussed here is very restrictive and besides the 
QCD scale $\Lambda$ there is only
one essential non-perturbative parameter, the effective cut-off $Q_0$
which is a characteristic of the onset of non-perturbative effects.
The hope is that the study of infrared sensitive quantities can provide us
with the important information on the confinement mechanism, besides the
test of the perturbative QCD dynamics. Furthermore, 
let us emphasize that a detailed understanding of the physics of
QCD jets is also important for the design of experiments and could provide
useful additional tools to study other physics. For instance, it could play
a valuable role in digging out the signals for new physics from 
the conventional QCD backgrounds.

\section{Perturbative Parton Cascades and Jets}

\subsection{Jet Definitions and Multi-Jet Structure}

In high energy hard collisions the hadronic final state consists of many
particles. Its characteristic feature is the jet structure with bundles of
hadrons collimated along certain (jet) directions.
In order to quantitatively describe this phenomenon one has to introduce the
notion of resolution: the higher the resolution of the observation 
the more jets can be
distinguished in the final state.

Resolution dependent exclusive cross sections are already familiar from QED.
The cross section for all final states which are indistinguishable
according to a resolution criterion is finite in perturbation theory. This
result is expressed by the Bloch-Nordsiek theorem$\,$\cite{bn} but  
it has its correspondence in QCD.
Finite jet cross sections in QCD were  studied first 
 by Sterman and Weinberg.$\,$\cite{sw} They counted 
two partons (jets) as distinguishable objects if their 
fractional energies $x_i$ and relative angles 
$\Theta_{kl}$ satisfied the conditions:
\begin{equation}
x_k>\varepsilon,\quad x_l>\varepsilon \qquad {\rm and} \qquad
\Theta_{kl}>\delta
\label{epsdel}
\end{equation}
for given resolution parameters $\varepsilon$ and $\delta$.
In the case of $e^+e^-\to q\bar qg$ the configurations
satisfying (\ref{epsdel})
correspond to 3-jet events; since the soft $dk/k$ and collinear 
$d\Theta/\Theta$ singularities from the gluon bremsstrahlung
are avoided the cross section is finite. In the complementary case
the singularities in the $q\bar qg$ final states cancel against the
singularities from the virtual corrections to the $q\bar q$ 
and the rate for these indistinguishable configurations is finite
as well.          

In the last years the resolution criterion based on relative transverse
momentum  has been widely used
($k_T$ or \lq\lq Durham-algorithm''\footnote{This
algorithm was discussed within the framework of analytical calculations
at the  Durham Workshop$\,$\cite{rep}.
The very idea of $k_T$ clustering has been
already applied since the 
early eighties$\,$\cite{sjoclu}.}).$\,$\cite{bs,cdotw,bkss}
In this scheme two particles (jets) of energies $E_k, E_l$ and relative angle
$\Theta_{kl}$ are considered distinguishable if
\begin{equation}  
y_{kl}=2\; (1-\cos \Theta_{kl})\;  {\rm min} (E_k^2,E_l^2)/s\; > \;y_c.
\label{durham}      
\end{equation}
for a given cut-off parameter  $y_{c}$ where 
$s$ is the $cms$ energy squared of the full event. In this case
the collinear and soft singularities are regularized by a single 
parameter $y_{c}=Q_c^2/ s$
which corresponds to the resolution $1/y_c$. For small angles $ \Theta_{kl}$ the separation 
variable corresponds to the
transverse momentum of the lower energy  particle
with respect to the higher
energy jet particle
\begin{equation}
\Theta_{kl}\ \ll \ 1:
y_{kl}\sim k_{T,kl}^2/s \; > \; y_c\quad \textrm{or}\quad k_{T,kl}>Q_c.  
\label{durhamsh}
\end{equation}
It is an important advantage of this scheme that 
some observables can be calculated in all orders of the perturbation
series in certain logarithmic approxima\-tions,$\,$\cite{bs,cdotw} a success
not achieved within the earlier JADE algorithm.$\,$\cite{JADE} Furthermore,
the ``hadronization corrections'' obtained in popular parton shower Monte
Carlo's have been found to be quite small in the $k_T$ algorithm.$\,$\cite{bkss} 

In the applications considered here we only deal with this $k_T$ algorithm.
We emphazise
the recent development (``Cambridge algorithm''$\,$\cite{cam}) with a different
treatment of the soft particles and the  class of algorithms
for $ep$ and $pp/p\overline p$ collisions which take into account the
spectator jets, discussed in a recent survey.$\,$\cite{sey}

These resolution criteria form the basis to obtain finite results 
for jet cross sections in perturbation theory. At the same time they can be
applied to define the jet in the experimental analysis of the multihadron
final state by an iterative procedure. 
For every pair of particles one computes the corresponding distance
$y_{kl}$ 
 (restricting here to the case of a single resolution parameter).
If the smallest distance in the event is smaller than
the resolution parameter $y_c$ the two particles are combined into
a single jet according to a recombination scheme.
In the simplest prescription (E-scheme) the 4-momenta of
jets are added; alternatively, one may require that either the momenta
or the energies are rescaled in such a way that massless jets are obtained.
This procedure is repeated
until all pairs satisfy $y_{kl} > y_c$. The remaining objects are the jets
at resolution $1/y_c$.

In this way one can study the multiparticle final state at variable
resolution between the extreme limits: very narrow jets -- ultimately 
the final state hadrons -- at high resolution and \lq\lq fat jets''
at low resolution. For example, the multiplicity $N$ of jets in
$e^+e^-$ annihilation approaches
\begin{eqnarray}
\textrm{at high resolution}& \ (y_c\to 0):&\quad N\to N_{\rm hadrons};
   \qquad\qquad\qquad  \nonumber\\
\textrm{at low resolution } & \ (y_c\to 1):&\quad N\to 2
\label{resolve}
\end{eqnarray}
The study of events at variable resolution leads naturally to two different
classes of observations:

\begin{itemize}
\item[(i)] {\it Multi-jet topologies}
\end{itemize}
At low resolution ($y_c$ large) there are only a few jets and 
the cross sections can be obtained from
low order perturbation theory. In many cases higher loop calculations are
available. For example, the processes $e^+e^-\to  2$ or  3 jets 
can be calculated from the relevant matrix elements and they are fully known in
$O(\alpha_s^2)$ which corresponds to inclusion of
final states with at most four
partons; the result will then depend on the 
coupling constant $\alpha_s$, the only parameter of the theory, and the
resolution parameter $y_c$
which can be preset by the experimenter, but should be large enough in order
to justify the approximation, typically $y_c>0.02$.
The accuracy of the result and the
 range of applicability in $y_c$ can be increased if the corrections from
higher orders in $\alpha_s$ are included.

A key role in particle and jet production in QCD is played by the gluon
bremsstrahlung. 
Let us recall first the basic process, the gluon radiation
off a quark: the differential spectrum of the gluon emitted
from the quark of momentum $p$
with energy $k$ in an approximation of small transverse
momenta $k_\perp$ is given by the well known formula
\begin{eqnarray}
d w^{q \rightarrow q + g} & = & \frac{\alpha_s(Q)} %(k_\perp)}
{4 \pi} \; 2C_F \; \left [ 1 + (1-z)^2 \right ] \; 
   \frac{dz}{z} \:
\frac{dk_\perp^2}{k_\perp^2}, \nonumber\\
\alpha_s (Q%k_\perp
)& =& \frac{2 \pi}{b \ln (Q/\Lambda)},
\quad b = \frac{11}{3} C_A - \frac{2}{3} n_f
\labl{brems}
\end{eqnarray}
where $C_A=N_C$, $C_F  =  (N_C^2 \: - \: 1)/2 N_C \; = \; 4/3$,
$N_C=3$ is the number of colors and
 $k_{\mu}$ is the gluon 4-momentum and $z=k/p$.
The strong coupling constant $\alpha_s$ runs with the scale $Q$ 
which has to be chosen according to the particular problem;
$\Lambda$ is the QCD-scale and $n_f$ is the number of flavors.
In addition, the processes $g\to g$ and $g\to q \overline q$ contribute to
the evolution of the parton cascade.

A characteristic feature of the bremsstrahlung  probability (\ref{brems})
is the broad logarithmic distribution over the gluon energy $k$ and 
transverse momentum $k_{\perp}$ and the soft and collinear singularities in
these variables.
 Considering multi-jet events 
the relative transverse momenta are large $(Q_c\sim Q)$, 
so the integral over the
bremsstrahlung spectrum gives a factor of $O(1)$, furthermore,
the coupling is small and the perturbation series converges rapidly. 
\begin{equation}
\textrm{Topology with extra jet:}
\qquad k_\perp\sim k\sim p \quad \to \quad
{\rm prob}\sim \frac{\alpha_s}{\pi}\ O(1)\ \ll \ 1.
\label{multi}
\end{equation}
The rate  for   
multi-jet events will decrease with jet multiplicity like $\alpha_s^n$.

\begin{itemize}
\item[(ii)]  {\it Inside jet activity}
\end{itemize}
%
%\noindent {\it b) Inside jet activity}\\
If we increase the resolution (lower $y_c$ parameter) more and more jets
will become resolved with decreasing relative transverse momentum.
The integral
over $k_\perp$ down to a low cut-off $Q_c$ will generate large double
logarithms from the collinear and soft singularities in (\ref{brems})
\begin{equation}
\textrm{Inside jet activity:} \qquad k_\perp\ \ll k\ \ll \ p \quad \to \quad
{\rm prob}\sim \frac{\alpha_s}{\pi}\ \ln^2( p/Q_0) \ \gsim 1.
\label{intra}
\end{equation}
Therefore, the probability for additional gluon emission is not small
because of the typically large logarithms.
In this case, the higher order terms
have to be resummed to get reliable results. 
Besides the $q\to qg$ process, there is another ``double logarithmic''
splitting $g\to gg$ and the single-logarithmic $g\to q\overline q$.
These processes build up the parton cascade. 

By lowering the cut-off $Q_c$
the number of sub-jets inside a primary ``fat'' jet increases
and one may ask how far down in transverse momentum 
can perturbation theory be applied; at some point all hadrons will be
resolved and the sub-jets coincide with the final hadrons. 
We will see later on, 
that indeed various features of the hadronic final state can
be described by perturbation theory for partons with low cut-off $Q_0$ of 
a few hundred MeV.
In this review our main interest is 
not so much in the precise description of multi-jet events which are
important for the accurate determination of $\alpha_s$ but
in the low scale phenomena at high resolution inside jets
and in between the jets.

\subsection{QCD Cascade and Evolution Equations} 
The concept of the evolution equation plays a central role in perturbative
QCD. Therefore we begin with a short account of how it appeared  
in the context of QCD cascade. 
After the discovery of Bjorken scaling$\,$\cite{bj,exp} in
Deep Inelastic Scattering (DIS), and the birth of the parton model,$\,$\cite{feyn,bpas} 
the idea of a parton carrying a finite fraction $x$ of the parent hadron
 momentum was well established. In the Bjorken limit the structure functions 
 became independent of the large momentum transfer $Q^2$ and were directly 
 related to the parton distributions $q(x)$, for example
 \eq
 \nu W_2(x,Q^2)\rightarrow F_2(x)=x\sum_q e_q^2 q(x).
 \eqx
Quantum Chromodynamics has introduced a subtle modification
to the above relations. With partons as quarks and gluons possessing nonabelian,
asymptotically free  interactions, the densities acquired a very gentle, logarithmic
dependence on the large scale $Q$ involved in a process,
 $q(x)\rightarrow q(x,Q^2)$.$\,$\cite{OPERGDIS}
Subsequent, experimental confirmation of these scaling violations is 
considered as a great thriumph of perturbative QCD. %$\,$\cite{viol} 
  Correspondingly
it also marked the beginning of further active developments in the theory and its
applications. 

Predictions of the scale dependence of the partonic distributions are customarily 
given in the form of the  DGLAP evolution equation$\,$\cite{gl,ap,yld1}
\eq
{\partial\over\partial t} q(x,t) ={\alpha_s(t)\over  2\pi}\int_x^1
 {dz\over z} P(x/z,\alpha_s(t) )
 q(z,t),  \label{geveq}
\eqx
where $t=\ln{Q/\mu}$, with some arbitrary {\em factorization scale} $\mu$
 (see below). The {\em splitting function} $P(x,\alpha_s)$ is calculable
as a power series in $\alpha_s$. Eq. (\ref{geveq}) is generic:  the complete 
theory  of
gluons coupled to $n_f$ flavors of quarks leads to a system of $2n_f+1$ coupled 
equations with $P$ elevated to the $(2n_f+1)\times(2n_f+1)$ transition matrix. 
 The above equation was derived originally with the aid of the 
operator product expansion and the renormalization group methods in the 
space-like region.$\,$\cite{asy} Subsequently it was established 
with the diagrammatic techniques, also in the time-like domain and 
for higher (two loop) order,$\,$\cite{lls,ddt2,fur2}
(for a comprehensive account, see for example, Ref.  38. %$\,$\cite{pok})
 The $Q$ dependence of partonic  distributions,  in a local theory like QCD, 
 is caused by weakly damped
transverse momenta of partons, {\em c.f.} (\ref{brems}). Consequently, they are
 effectively bounded only by the phase space for a given process which
  introduces the logarithmic dependence on the scale $Q$.  Fortunately, all
  the logarithms generated in this way obey the powerful factorization 
  theorem.$\,$\cite{cs,fac,fac2}
   Namely, they are independent of the particular hard process under 
   consideration, hence they naturally belong to the external partons involved 
  in the reaction. This is how the quark densities become $Q$ dependent. 
The original cross section can be split into a universal,  scale dependent part, 
and  a non-universal hard part which however does not contain large logarithms.
The universal part will then renormalize the external densities
so that the final, renormalized densities  become dependent on  $Q$.  
In the process 
of splitting  a supplementary factorization scale $\mu$ is introduced.  
The final physical result is independent of $\mu$ and the Eq. (\ref{geveq})
 can be considered as a renormalization group equation expressing 
 this independence.
 
 On the other hand, Eq. (\ref{geveq}) has yet another,  physically appealing
interpre\-tation.$\,$\cite{ap,yld1} Namely it can be regarded as a master equation 
for the Markov
process where, at any moment of a fictitious time $t=\ln{Q/\mu}$, any parton
 can emit another parton with a probability, per unit of time, related to 
 the splitting function $P$.  In this way a QCD cascade develops and the 
 dependence of the partonic densities on $t$ is easily understood. 
Even more importantly, this interpretation follows directly from the
 diagrammatic derivation of Eq. (\ref{geveq}). In a particular class of gauges,
called physical gauges,  large logarithms which generate the evolution
are contained only in diagrams which reduce to absolute squares of the
amplitudes hence allowing for the above probabilistic interpretation. 
Moreover, it was found
that even in the case where the interference effects are important, giving rise to
the {\em angular ordering} the net contribution can be presented in  
 a probabilistic form (see later).

 The statistical interpretation of Eq. (\ref{geveq}) is even more evident upon closer 
inspection of the splitting function $P$. In the simplest case of the 
$q\rightarrow q$ transition, which is relevant for the evolution of the
{\em Non-Singlet} densities, $P$ reads in the lowest order in $\alpha_s$
\eq
P(x)=C_F \left( {1+x^2 \over 1-x} \right)_+ = \hat{P}(x) - 
\left( \int_0^1 \hat{P}(z) dz\right) \delta(1-x),   \label{split}
\eqx
where $\hat{P}(x)= C_F(1+x^2) /(1-x)$ and the last part of the equation is defined
 only in a distribution sense. Hence the Eq. (\ref{geveq}) can be rewritten as
\eq
{\partial\over\partial t} q(x,t) ={\alpha_s(t)\over  2\pi}\left[ \int_x^1
 {dz\over z} \hat{P}(x/z) q(z,t) - q(x,t) \int_0^x {dy\over x} \hat{P}(y/x) \right] .
  \label{gain}
\eqx
This is the standard form of the gains -- losses (or master) equation in stochastic
processes. The {\em change} in the number of partons (with given $x$) 
is the result of two opposite effects: increase due to the emissions from 
partons with momenta higher
than $x$, and decrease caused by emissions 
towards smaller than $x$
momenta.$\,$\cite{wz}  In the original diagrammatic derivation the two terms
 are given  by the real and virtual emission diagrams respectively. In fact, each of 
 these terms separately is logarithmically divergent corresponding to the infrared
 divergence of exclusive (here elastic) cross sections. 
 The final sum in Eq. (\ref{split})
  of the elastic and inelastic contributions is infrared finite$\,$\cite{bn} 
  and is conveniently represented by the $(\dots)_+$ prescription.

Another important ingredient, which emphasizes the probabilistic nature 
of the evolution equation, is the {\em Sudakov form factor} 
$S(Q^2,\mu^2)$ first calculated in QED$\,$\cite{suda},   and intensively  
 used  in perturbative  QCD calculations, see for instance.$\,$\cite{dkmt2,ddt2}
Let us construct a probability that
a parton with momentum fraction $x$ {\em does not emit} any other parton  
during the evolution betwen $t_1$ and $t_2$, say. 
Since $\hat{P}$ is the probability of emission per unit
time, the probability that there is no emission in a small interval $\Delta t$
reads
\eq
r(t,t+\Delta t)=1-\Delta t {\as(t)\over 2\pi} \int_0^x  {dy\over x} \hat{P}(y/x).
\eqx
The probability $R(t_1,t_2)$ that there is no emission in a finite interval 
$(t_1,t_2)$ is given by the product
\begin{equation}
R(t_{1},t_{2})=\prod_{i} r(t_i,t_i+1)=\exp {\left( - \int_{t_1}^{t_2} 
\int_0^1 {\alpha_s(t) \over 2\pi} \hat{P}(z) dz dt \right) }, % \equiv S(t_1,t_2),
\end{equation}    
which is just the Sudakov form factor 
\begin{equation}
R(t_{1},t_{2})=S(t_{1},t_{2}).
\end{equation}
Hence the Sudakov form factor, which originally was derived by  
diagrammatic calculations, has also the simple probabilistic interpretation.
 The Sudakov form factor 
is infrared divergent since it 
 describes the elastic process. Therefore one usually
introduces a small infrared cut off $\varepsilon$ which defines so called
 {\em resolved} emissions. As an example consider a quark with
a momentum fraction $z>1-\varepsilon$ emitted from a parent quark. 
This must be accompanied by an emission of a soft gluon
with $x_g < \varepsilon$. Such a process is considered as unresolved. Hence
\eq
S_{\varepsilon}(t_1,t_2)=\exp{\left( - \int_{t_1}^{t_2} 
\int_0^{1-\varepsilon} {\alpha_s(t)\over 2\pi} \hat{P}(z) dz dt \right)} ,
\eqx      
gives the probability for no resolved emissions in $(t_1,t_2)$ and consequently
 contains any number of very soft, unresolved gluons.\footnote{In some cases
  also the very
low momentum fractions should be exluded "$z>\varepsilon$" if they correspond to 
the unresolved emissions, e.g. in the $g \rightarrow gg$ splitting.}

   Evolution equations can be written in many equivalent forms depending 
on the foreseen application. For example, one can eliminate the virtual emission
term from Eq. (\ref{gain}) with the aid of the Sudakov form factor
\eq
{\partial\over\partial t} q(x,t) ={\alpha_s(t)\over  2\pi}\int_x^1
 {dz\over z} \hat{P}(x/z)q(z,t) +
  {q(x,t)\over S(t_0,t)}{\partial\over\partial t} S(t_0,t),  \label{evsud}
\eqx   
or
\eq
{\partial\over\partial t}\left({q(x,t)\over S(t_0,t)}\right)
 ={\alpha_s(t)\over  2\pi}\int_x^1
 {dz\over z} \hat{P}(x/z){q(z,t)\over S(t_0,t)} . \label{diffev}
\eqx
This can be readily integrated to give 
\eq
q(x,t)=S(t_0,t) q(x,t_0)+\int_{t_0}^t d t' S(t',t)
\int_x^1
 {dz\over z}{\alpha_s(t')\over  2\pi} \hat{P}(x/z)q(z,t'). \label{intev}
\eqx
This integral form of the evolution equation has again a straightforward 
probabilistic interpretation with $t'$ being the instant of the last
emission. Note that only the real processes, described by $\hat{P}$ 
and by the Sudakov form factor, appear in this formulation. For that reason
 Eqs. (\ref{diffev},{\ref{intev})
are the suitable basis of very successful Monte Carlo simulations of the QCD cascade.
$\,$\cite{HERWIG,JETSET}

\subsection{Evolution Equations for Jet Observables}

The success of the parton description of  Deep Inelastic Scattering 
prompted an avalanche of detailed studies of the 
 QCD cascade.
 In particular, more and more 
properties of the final states produced in various high energy collisions are being
confronted with QCD predictions. This required the development of the evolution 
equations for the time-like region and their generalizations 
to other observables.  Complete information about any
 multiparticle process is contained in the 
{\em generating functional}$\,\,$\cite{ahm1}
 \begin{equation}
  Z(\{u\})=\sum_n \int d^3k_1 \ldots d^3k_n
  u(k_1)\ldots u(k_n) P_n(k_1,\ldots,k_n)
  \labl{gendef}
\end{equation}
where $P^{(n)}(k_1,\ldots, k_n)$ is the probability density for
exclusive production of particles with 3-momenta $k_1\ldots k_n$ and $u(k)$
is an auxiliary profile function. Global characteristics like total multiplicity
and its various moments are obtained by differentiating (\ref{gendef})
with respect to the {\em constant} profile parameter $u$.  On the other hand
the $k$ dependence contains the information about  the differential densities
\begin{equation}
 D^{(n)} (k_1,...,k_n)= \delta^n Z(\{u\})/\delta u(k_1)...
\delta u(k_n)\mid_{u=1} ,    \labl{dndef}
\end{equation}
and correlation functions (or cumulants)
\begin{equation}
\Gamma^{(n)} (k_1,\ldots,k_n)= \delta^n \ln Z(\{u\})/\delta u(k_1)\ldots
\delta u(k_n)\mid_{u=1}      \labl{gamdef}
\end{equation}
of arbitrary order. 

Originally partonic distributions were calculated in the so called
Leading Logarithmic Approximation (LLA). It applies for both DIS 
and $e^+e^-$ annihilaton in the  region of finite momentum fractions,
 $ 0.1 \lsim x < 1. $, say. Since  historically the first calculations were 
 done for the space-like Deep Inelastic Scattering we shall explain the basic
  principle of this approach on this example.
Since partonic densities measured there represent
the total cross section, the infrared divergences cancel. The remaining terms 
are of the type $\ln{x}\ln{Q^2}$. In the finite $x$ region they yield single
logarithms  of the form $\alpha_s^n \ln(Q^2/\mu^2)^n$. These terms are summed 
in the Leading Logarithmic Approximmation giving for example Eq.(\ref{geveq}).
% The above approach is valid as long as
%the logarithms $\ln{x}$  remain small, hence the limitation of finite $x$.
 The $\ln{x}$ terms are the remainders of the infrared cancellations
mentioned above. They are small  for larger x. Therefore one can also say 
that the LLA is valid 
when the infrared cancellation between gains and losses terms in Eq. (\ref{gain}) 
is large, {\em i.e.} when the effect of the Sudakov form factor is important. 
 As mentioned above  the Leading Logarithmic Approximation was also extended
to the $e^+e^-$ annihilation and the evolution equation
(\ref{geveq}) can be used for fragmentation functions for finite 
momentum fractions $z$.
The higher corrections containing powers of $\alpha_s$ unbalanced
by the large logarithms may then be systematically included.

%The Leading Logarithmic Approximation applies for larger
%$x$, where the effect of the Sudakov term is important and 
%the cancellation between
% gains and losses in Eq. (\ref{split}) is large. These $x$ values span
%a fixed range $ .1 \lsim x < 1. $ say,  independent of the large scale $Q^2$.
%This is reminiscent of the original Bjorken scaling limit. The same approach
%works also in the $e^+e^-$  annihilation for finite momentum fractions $z$.

On the other hand, the more detailed characteristics of the final states,
 like for example the global and differential multiplicities,
 are dominated by the region of small $z \sim 1/\sqrt{Q}$.
 This is the region
 we are mostly concerned with in the present review. 
 These soft emissions reveal a new, very characteristic feature of the 
 QCD cascade
 -- the angular ordering.$\,$\cite{ef,ahm2} It is basically a nonabelian generalization
  of the well
known Chudakov$\,$\cite{orig} effect in QED -- the soft radiation from a relativistic
 $e^{+}e^{-}$ pair is confined to a cone bounded by the electron and positron
momenta. Analogously,  a soft  gluon in a fully 
developed cascade is emitted only inside a cone bounded by the momenta of
its two immediate predecessors. Mathematically, this is caused
by a negative interference outside the above cone. Interestingly, these quantum 
interference effects can be again (at least in the large $N_C$ limit) 
cast into a probabilistic scheme described 
by the evolution equations. 

In the soft region,  $z \sim 1/\sqrt{Q}$, the Sudakov term in Eq. (\ref{evsud})
 is not important in the first approximation. 
  In such a case the infrared logarithms do not cancel and each power of \al
is accompanied by the two (soft and collinear) logarithms, {\em i.e.} multiplicities
are not infrared safe.
These contributions are summed by the Double Logarithmic 
Approximation.$\,$\cite{fur1,bcm1,dfk1}
Together with the angular ordering it reproduces
qualitatively the most important properties of the QCD cascade and retains
conceptual and technical simplicity. Hence DLA remains an important
tool of perturbative QCD. 

The Double Logarithmic Approximation is valid quantitatively only at
asymptotically high,  energies. Fortunately an important class of corrections
which bring the applicability range down to the presently available energies,
does not spoil the probabilistic interpretation. The new scheme, known as
the Modified Leading Logarithmic Approximation (MLLA), contains all 
next-to-leading logarithmic corrections,$\,$\cite{dt,ahm1} and
 is now considered 
as a standard in quantitative tests of perturbative predictions. 
 Formally, it includes 
consistently all $\Ord{(\alpha_s)}$ terms in addition to the DLA
 $\Ord{(\sqrt{\alpha_s})}$
contributions  in the exponential factors. 

 The complete MLLA description of the QCD cascade has a form of the system
of  two coupled integral evolution equations for the generating functions 
$Z_a$, each describing the cascade originating from the highly virtual 
time-like
parton $a=q,g$ with momentum $P$$\,$\cite{dt,dkt4}\footnote{Simplified versions of
this equation had been obtained already before.$\,$\cite{dfk1,bcm} } 
\begin{eqnarray}
\lefteqn{Z_a (P, \Theta; \{u (k)\}) =  e^{- w_a (P \Theta)}  u_a (p) }\nonumber \\
 &&+ \frac{1}{2!}
  \; \sum_{b,c} \; \int_0^{\Theta} \; \frac{d
\Theta^\prime}{\Theta^\prime} \; \int_0^1 \; dz \: e^{-w_a (P
\Theta)  +  w_a (P \Theta^\prime)} \labl{2.15} \nonumber \\
&& \times \frac{\alpha_s (k_\perp^2)}{\pi} P_a^{bc}(z) 
 Z_b (zP, \Theta^\prime;\{ u\}) \: Z_c ((1 - z) P,
\Theta^\prime;\{ u\})  \Theta (k_\perp^2 - Q_0^2). \label{zmlla}
\end{eqnarray}
The  $P_a^{bc}(z)$ are the splitting functions$\,$\cite{ap,yld1}
\begin{eqnarray}
P_q^q(z)&=& P_q^g(1-z)=C_F\frac{1+z^2}{1-z}, \labl{ap1}\\
P_g^q(z)&=& P_g^q(1-z)=T_R[z^2+(1-z)^2],       \labl{ap2}\\
P_g^g(z)&=& P_g^g(1-z)=2
C_A\left[z(1-z)+\frac{1-z}{z}+\frac{z}{1-z}\right],  \labl{ap3}
\end{eqnarray} 
with $C_F$ as in (\ref{brems}) and $T_R=1/2$.

The first term in
the $r.h.s.$ of (\ref{zmlla}) corresponds to the case when the $a$-jet consists
of the parent parton only.  The integral term describes the first
splitting $a \rightarrow b + c$ with angle $\Theta^\prime$
between the products.  The Sudakov form factor  guarantees this 
decay to be the first one:  it is the probability to emit
{\it nothing} in the angular interval between $\Theta^\prime$ and
$\Theta$.  The two last factors account for the further
evolution of the produced subjets $b$ and $c$ having smaller energies
and smaller $\Theta^\prime$ than the opening angle as required by angular
ordering.

Using the
normalization property of the GF
\begin{equation}
Z_a (P, \Theta; \{ u \})|_{u (k) \: \equiv \: 1} \; = \; 1
\labl{2.18}
\end{equation}
the MLLA  Sudakov formfactors can be found from 
\begin{eqnarray}
w_q & = & \int^\Theta_{Q_0/P} \; \frac{d \Theta^\prime}{\Theta^\prime}
\;
\int_0^1 \; dz \; \frac{\alpha_s (k_\perp^2)}{ \pi} \;
P_q^q
(z), \;\;\;
\labl{2.19a}\\
& & \nonumber \\
w_g & = & \int^\Theta_{Q_0/P} \; \frac{d \Theta^\prime}{\Theta^\prime}
\;
\int_0^1 \; dz \; \frac{\alpha_s (k_\perp^2)}{ \pi} \; \left
[
\frac{1}{2} \; P_g^g (z) \: + \: n_f  P_g^q (z) \right ].
\labl{2.19b}
\end{eqnarray}
 Collinear and soft singularities in Eqs.\
(\ref{2.15}),%
(\ref{2.19a}) and (\ref{2.19b}) %(2.19)
 are regularized  by the transverse
momentum restriction
\begin{equation}
k_\perp \;  > \; Q_0 
%p_{min} & = & min \{z, (1 - z)\} \; p. \nonumber
\labl{2.20}
\end{equation}
where $Q_0$ is a cut-off parameter in the cascades and
$k_\perp\approx z(1-z)P \Theta^\prime$ for small angles $\Theta^\prime$.

Differentiating the product $Z_a \: \exp \: \left[ w_a (P
\Theta)
\right] $ with respect to $\Theta$ and using Eq.\ (\ref{2.15})
one arrives at
the
Master Equation 
\begin{eqnarray}
\lefteqn{\frac{d}{d  \ln  \Theta}  Z_a (P, \Theta)  = 
\frac{1}{2}
\; \sum_{b,c} \; \int_0^1 \; dz } \nonumber \\ 
 & &\times \frac{\alpha_s (k_\perp^2)}{ \pi} \: P_a^{bc} (z)  
\left [Z_b (zP, \Theta) \: Z_c ((1 - z)P, \Theta) \: - \: Z_a
(P,\Theta) \right ] \labl{2.23}
\end{eqnarray}
which gives, as one of the applications, the MLLA counterpart of the
 DIS evolution equation (\ref{geveq}) 
for single parton densities.  % $\,$\cite{gl,ap,yld1}

Contrary to the differential evolution equations the integral evolution 
equations determine also the initial conditions of the system. 
In this case they follow from Eq. (\ref{2.18}) 
\begin{equation}
Z_a (P, \Theta; \{ u \})|_{P \Theta  =  Q_0}  =  u_a( P)
\labl{2.16}
\end{equation}
with the simple interpretation that the jet originating from a parton $a$
contains only this parton at the lowest virtuality $Q_0$.

Generating functions $Z_q$ and $Z_g$ form the building blocks
sufficient to describe the complete final states realized in 
physical reactions. 
For instance, the $e^+ e^-$ annihilation into hadrons at the
total cms energy $W = 2P$ is described by
\begin{equation}
Z_{e^+ e^-} \: (W; \{ u \}) \; = \; \left [ Z_q (P, \Theta \sim
\pi; \{ u \}) \right ]^2.
\labl{2.21}
\end{equation}
%while the final particles produced in the $\Upsilon$ decay 
%are represented by
%\begin{equation}
%Z_{ggg}^{\Upsilon} \: (W; \{ u \}) \; = \; \left [ Z_g (E_g, \Theta =
%\pi;
%\{ u \}) \right ]^3
%\labl{2.22}
%\end{equation}
The equations (\ref{2.23}) are now actively exploited and various applications 
will be discussed below.

In the Double Logarithmic Approximation the parton $b$,  say,  emitted at the 
elementary vertex $a\rightarrow b + c$ is considered so soft that it
 does not influence the original parton  $a$. Consequently $1-z \rightarrow 1$ 
 and $c\rightarrow a$ in Eq.(\ref{2.23}). This soft energy assumption also
implies that the splitting functions can be replaced by their most singular parts at
$z\sim 0$.
 This yields a
simpler equation which can be integrated to give the DLA Master Evolution
 Equation$\,$\cite{dfk1,dfk2,vsf}
(with $d^3k=d\omega d^2k_\perp$)
\begin{eqnarray}
  Z_p(P,\Theta;\{u\}) &=& u(P)\exp
     \biggl(\int\limits_{\Gamma(P,\Theta)}
     \frac{d\omega}{\omega}\frac{d^2k_\perp}{2\pi k^2_\perp} %\cdot 
     \nonumber\\
&\times &  c_p \frac{2\alpha_s (k^2_\perp)}{\pi}\;
 [Z_g(k,\Theta_k;\{u\})-1]\biggr),  
\labl{dlaz}
\end{eqnarray}
 $c_p$ refers to the respective color factors $C_A$ and $C_F$.
 The secondary gluon $g$ is emitted into the interval $\Gamma(P,\Theta)$:
$\omega<E=|\vec P|$ and $\Theta_k<\Theta$. Due to the angular ordering
constraint the emission of this gluon is  
bounded by its angle
$\Theta_k$ to the primary parton $p$. 
As mentioned earlier the DLA, although formally correct 
only at the asymptotically high energies,  reproduces satisfactorily the general 
structure of the QCD casacade and allows for an important analytical insight.

At this point we would like to emphasize a deep and beautiful universality 
of the above methods. Even though historically the LLA was first used in the
Deep Inelastic Scattering and DLA and MLLA in $e^+e^-$ annihilation, 
the LLA also applies to the latter at finite $z$. Similarly  
the double logarithmic asymptotics also successfully describes the DIS 
structure functions at large  $\ln{(1/x)} \sim \ln{Q^2} $, 
and MLLA master equation can be used to derive the DGLAP evolution 
equation which is also applicable in DIS. 

At the same time the region lying yet "beyond" the DLA asymptotics
in DIS, {\em i.e.}
$\ln{(1/x)} \gg \ln Q^2 $, is being intensively studied. It is described by the
BFKL evolution equation and is important for our understanding of the
emergence Pomeron trajectory in QCD. 

%     formulas for evolutio eqs. for GFs
%   4. Ev.eqs. for cut moments

%\bibitem{ahm1} A. H. Mueller [24] in our jets.

%*************************************

%\input{shifdual4}

\subsection{Parton Hadron Duality Approaches}
At present the application of QCD to multiparticle production
is not possible without additional assumptions about the hadronization
process at large distances which is governed by the color-confinement
forces. The simplest idea is to treat hadronization as long-distance
process, involving only small momentum tranfers, and to compare
directly the perturbative predictions at the partonic level with the
corresponding measurements at the hadronic level. This can be
applied at first to the total cross sections, and then to jet
production for a given resolution; here the partons are compared
to hadronic jets at the same resolution and kinematics. This approach has
led to spectacular successes and has built up our present confidence in
the correctness of QCD as the theory of strong interactions. 
In such applications the resolution or cut-off scale is normally a fixed 
fraction of the primary energy itself.

It is then natural to ask whether such a dual correspondence can be
carried out further to the level of partons and hadrons themselves.
The answer is, in general, affirmative for ``infrared and
collinear safe'' observables which do not change if a soft
particle is added or one particle splits into the collinear particles.
Such observables become insensitive to the cut-off $Q_0$ for small $Q_0$.
%and can become independent of the final state of the jet
%evolving assuming that the laws of energy-momentum and flavour
%quantum numbers at hadron level follow those at parton level. 
Quantities of this type are energy flows and correlations and
global event shapes like thrust etc.

In the next step of comparison between partons and hadrons we consider
observables which count individual particles, for example, particle
multiplicities, inclusive spectra and multiparton correlations.
Such observables depend explicitly on the cut-off $Q_0$ (the smaller
the cut-off, the larger the particle multiplicity).

The very assumption of the hypothesis of 
Local Parton Hadron Duality$\,$\cite{adkt1} 
is that the particle yield is described by a parton
cascade where the conversion of partons into hadrons occurs at a 
low virtuality scale, of the order of hadronic masses $(Q_0\sim $ few
hundred MeV),
independent of the scale of the primary hard process, and involves only
low-momentum transfers; it is assumed that the results obtained for
partons apply to hadrons as well.

%Originally it was supposed that the correspondence of the partonic 
%properties to the hadronic ones should only be considered in an inclusive
%and average sense.

Within the LPHD approach, PQCD calculations have been carried out in the
simplest case (at asymptotically high energies) in the Double Logarithmic
Approximation %(DLA) $\,$\cite{dfk1,dfk2,vsf,bcm} 
or in the
Modified Leading Logarithmic Approximation %(MLLA)$\,$\cite{dt,ahm1,adkt1} 
which includes higher order terms
of relative order $\sqrt{\alpha_s}$ (e.g.\ finite energy corrections);
they are essential for quantitative agreement with data at realistic
energies.
According to LPHD, the shape of the so-called ``limiting'' spectrum which
is obtained by formally setting $Q_0=\Lambda$ in the parton evolution
equations, should be mathematically similar to that of the
inclusive hadron distribution.

In this review we examine, in particular, applications of the LPHD
scenario concerning ``infrared sensitive quantities''. To deal with the
cut-off $Q_0$ one can proceed in different ways. If the cut-off
dependence factorizes (for example, multiplicity) one can again get
``infrared safe'' predictions after a proper normalization. In other
cases (for example, inclusive momentum spectra) the observables
become insensitive to the cut-off at very high energies if
appropriately rescaled quantities are used.

More generally, one can test parton-hadron duality 
relations between partonic and hadronic
characteristics of the type of
\begin{equation}
 O(x_1,x_2,\ldots)|_{hadrons}= K~O(x_1,x_2,\ldots, Q_0, \Lambda)|_{partons}
\label{lphdeq}
\end{equation}
where the non-perturbative cut-off $Q_0$ and the ``conversion coefficient''
$K$ should be determined by experiment (for review, see Ref. 17). %$\,$\cite{ko}). 
An essential point is
that this conversion coefficient should be a true
constant independent of the hardness of the underlying process.

The hypothesis of LPHD lies in the very heart of the analytical
perturbative scenario, but a the same time this key hypothesis
could be considered as its Achilles heel as it remains outside
of what can be derived within the established framework of QCD
today. One motivation of LPHD is the ``pre-confinement'' property
of QCD$\,$\cite{av} which ensures that color charges are compensated locally
and color neutral clusters of limited masses are formed within the
perturbative cascade. On the other hand, LPHD fits naturally into
the space-time picture of the hadroformation in QCD jets to be discussed
below.

When comparing differential parton and hadron distributions there
can be a mismatch near the soft limit caused by the
mass effects. This mismatch can be avoided by a proper choice
of energy and momentum variables. In a simple model$\,$\cite{lo,klo}
partons and hadrons are compared at the same energy (or transverse mass)
using an effective mass $Q_0$ for the hadrons, i.e.
\begin{equation}
E_{T,parton}=k_{T,parton}\Leftrightarrow E_{T,hadron}=
\sqrt{k^2_{T,hadron} + Q^2_0}
\label{partonhadron}
\end{equation}
then, the corresponding lower limits are $k_{T,parton}\to Q_0$
and $k_{T,hadron}\to 0$.

Finally, let us recall that within LPHD approach there is no convincing way
to introduce the different hadron species. For this one must
resort to models. These are also vital for the practical 
purposes, for instance, for unfolding parton distributions from hadron
spectra. At the moment all the so-called WIG'ged = (With Interfering
Gluons) Monte Carlo models (HERWIG$\,$\cite{HERWIG}, JETSET$\,$\cite{JETSET}, 
ARIADNE$\,$\cite{ARIADNE})
are very successful in the representation of the existing data and
they are intensively used for the predictions of the results of present and
future measurements. It is worthwhile to mention that for many 
observables the LPHD concept is quantitatively realized
within these algorithmic schemes.

\subsection{Space-Time Picture of Jet Evolution}

To exemplify the space-time structure of the development of the
QCD jets$\,$\cite{dkmt2,dkt4,yia2}
 let us consider the process $e^+e^-\to q\bar q$. This
may be viewed as the decay of a highly virtual photon with mass
$Q$, or as a real $Z^0$, the decay time scale is short, $t_{annih}\simeq
1/Q\sim 10^{-3}-10^{-2} fm$. The $q\bar q$ pair is kicked out of the
vacuum as bare (at scale $1/Q$) color charges until
the gluon field has had time to regenerate out to a typical
hadron size $R\sim 1 fm$. Allowing for the Lorentz boost this takes a time
$t_{had}\simeq Q/m\times R\approx QR^2\sim 10^2$ fm,
where the second approximation
is appropriate to light hadrons.

Since $t_{had} \gg t_{annih}$ the question arises of how the color
charges are conserved over the space-like separated distances
involved. The primary quarks will radiate gluons and
here two new scales are relevant.

First, from the virtuality prior to emission, the formation time of
a gluon of energy $k$ is $t_{form}\simeq k/k^2_\perp$. Secondly, for the
gluon to reach a transverse separation of $R$ and become independent
of the emitter takes a time $t_{sep}=(k_\perp R)\cdot t_{form}$, whilst
the hadronization time may be written as $t_{had}(\simeq kR^2)=(k_\perp)^2
\cdot t_{form}$. For such quark-gluon picture to make sense we require
$k_\perp > R^{-1}$ so that $t_{form}< t_{sep} < t_{had}$. 
Within this scenario the first
hadrons are formed at the time $t\sim t_{crit}\sim R$. It is the
moment when the distance between the outgoing $q$ and $\bar q$ approaches
$R$. At $t>t_{crit}$ the two jets are separated as globally blanched, and the
parton cascades develop inside each of them. The gluon
bremsstrahlung becomes intensive only when the transverse distance
between any two color partons exceeds $R$.

With increasing time the partons with larger and larger
energies $k\sim \frac{t}{R^2}$ hadronize (inside-outside chain).
%$\,$\cite{}.
%
If $k_\perp<R^{-1}$ then within perturbative scenario we can say 
nothing. On the borderline are quanta with $k_\perp R\sim 1$ (though with 
arbitrary large energies). These do not have enough time to
behave as free perturbative partons because their hadronization
time is comparable with the formation time, $t_{form}\sim t_{hadr}\sim kR^2$.

We distinguish these quanta from the essentially perturbative gluons by the
name gluers.$\,$\cite{dkmt2} Contrary to conventional QCD partons gluers
do not participate in perturbative cascading and their formation is
a signal of switching on the real strong interactions
$(\alpha_s\sim O(1)$).

In this picture soft particles with $k\sim R^{-1}$ produced at the 
lower edge of the perturbative phase space play a very special role.
Their production rate is unaffected by the QCD cascading, and in some
sense they can be considered as the eye-witnesses of the beginning
of the ``hadronization wave''.

It is an interesting question, whether the contribution from
non-perturbative emission (a \lq\lq pedestal'' in the rapidity distribution)
not included in the perturbative calculation (quanta with $k_\perp>Q_0$)
manifests itself in the present data. The studies on rapidity gaps
 do not require such an addition to the perturbative result
$\,$\cite{os} within the measurement accuracy and similar conclusions could be
drawn from the soft particle production in jets.$\,$\cite{klo}

\section{Multiplicities}

The simplest global characteristic of the hadronic final state is the particle
multiplicity. The mean multiplicity of partons in a jet is derived
from the generating functional
according to the general rules (\ref{dndef}) 
by single differentiation. Then one obtains 
from the master equation (\ref{2.23}) the following coupled system of 
evolution equations for
the multiplicities $\Na$ in quark and gluon jets ($a=q,g$) 
\begin{eqnarray}
\frac{d \Na(Y)}{d Y}  & = &
\frac{1}{2}   
 \sum_{B,C} \int_0^1  dz 
\frac{\alpha_s (k_\perp)}{\pi}  P_a^{bc} (z) \nonumber \\
%& & \labl{eveqmult}\\
 \; &\times &
\left [\Nb (Y+\ln z) + \Nc (Y+\ln (1 - z))  -  
\Na (Y) \right ].  \labl{eveqmult}
\end{eqnarray} 
The multiplicities $\Na$ depend on the jet virtuality $\kappa\approx
P\Theta$ and the cut-off $Q_0$ or on
\begin{equation}
Y=\ln\frac{P\Theta}{Q_0} \qquad \lambda=\ln\frac{Q_0}{\Lambda}
\labl{logvar}           
\end{equation}            
where $\Theta$ denotes the maximum  angle between the outgoing partons $b$
and $c$ -- for large angles $\Theta$ the variable 
$Y \to \eta=\ln (\kappa/Q_0),
\ \kappa=2P\sin (\Theta/2)$ have also been suggested.$\,$\cite{do}
The integral boundaries in (\ref{eveqmult}) 
are further reduced by the condition $k_\perp>Q_0$. The transverse momentum
is usually taken as 
 $k_\perp=z(1-z)\kappa$, in some applications also
$k_\perp=\min(z,1-z)\kappa$ (``Durham $k_\perp$'').
The initial condition for solving this system of equations reads
\begin{equation}
\Na (Y)|_{Y=0}  = 1
\labl{initmult}
\end{equation} 
which means there is only one particle in a jet at threshold.

This set of equations determines the multiplicities of partons 
in quark and gluon jets
in absolute terms for a given cut-off parameter $Q_0$. In principle, one can
solve the equations by iteration starting with (\ref{initmult}) which yields
the perturbative expansion of this quantity. In some approximate schemes
this series can actually be resummed analytically.

\subsection{Asymptotic Behavior}
At high energies the solutions of (\ref{eveqmult}) 
can be written
\begin{equation}
\N_g(Y) \sim  \exp\left(\int^Y\gamma(y)dy\right) \labl{nasy}
\end{equation} 
where the anomalous dimension $\gamma$ has the expansion in
$\gamma_0\sim\sqrt{\alpha_s}$
\begin{equation}
\gamma=\gamma_0(1-a_1 \gamma_0 - a_2 \gamma_0^2 - a_3 \gamma_0^3 \ldots),
\labl{gamma}
\end{equation}
likewise the ratio of gluon and quark multiplicity 
\begin{equation}
r\equiv \frac{\Ng}{\Nq}=\frac{C_A}{C_F}(1-r_1\gamma_0 -
r_2\gamma_0^2-r_3 \gamma_0^3 \ldots).
\labl{rgq}
\end{equation}
The coefficients can be found from (\ref{eveqmult}) by expanding the $\N$'s
at large $Y$. The leading$\,$\cite{ahm2,dfk1,bcmm} (DLA) and next to
leading$\,$\cite{bw1,adkt1} %\cite{ahm2,bw1,adkt2}
(MLLA) coefficients $a_i$ 
lead to the multiplicity growth 
expressed in terms of the running coupling %$\,$\cite{bw1}
\begin{equation}
%\ln  \N  \sim  \sqrt{ \frac{32 \pi  N_C}{\alpha_s (Y)}}
%%
%\frac{1}{B}  +  \left ( \frac{B}{2}  -  \frac{1}{4}
%\right ) \; \ln \: \alpha_s (Y) \: + \: \Ord (\sqrt{\alpha_s}).
\ln  \N(Y)  \sim  c_1/\sqrt{\alpha_s (Y)} +  c_2\ln \alpha_s (Y) +c_3
\label{lnn2}
\end{equation}
where $c_3$ is an arbitrary constant and
\begin{equation}
c_1=\sqrt{96\pi}/b,  \qquad   c_2=\frac{1}{4}+\frac{10}{27}n_f/b.  
\label{lnncoef}
\end{equation}
In this next-to-leading high energy approximation (MLLA) 
$\Ng\sim \Nq$ since the
difference would be a $\sqrt{\alpha_s}$ correction as given by the terms
$r_1$ and $a_2$. 
The asymptotic limit $ r=C_A/C_F=9/4$ in (\ref{rgq}) 
has appeared first in the discussion of
radiation from color octet and triplet sources$\,$\cite{bg} and in the jet
calculus in Leading Log Approximation$\,$\cite{kuv}. 
Terms of higher order have been obtained in
next-to-leading order$\,$\cite{ahm1,mw}, next-to-next-to-leading order$\,$\cite{gm,dn1} 
and in one yet higher order (3NLO)$\,$\cite{cdgnt}.  

The Eq. (\ref{eveqmult}) completely determines the leading term (DLA) 
and the second term (MLLA) in (\ref{gamma}) which are important for the high energy behavior.
The  terms of yet  higher order    (see also the
recent review Ref. 71) %              $\,$\cite{dg}) %(\ref{eveqmult}) 
are not completely determined by the single jet evolution equation 
 (\ref{eveqmult}) because they are affected by large angle emission processes.
However, these terms are of some relevance nevertheless as they 
include energy conservation in improved accuracy. 
Furthermore, the summation of the full perturbation series 
allows one to take into account the initial
conditions (\ref{initmult}) at threshold. These results will be discussed
next.

\subsection{Full Solution in DLA}
In this approximation only the most singular terms in the splitting
functions $\sim 1/z$ are kept, i.e. $P_g^g\simeq 2C_A/z,\
P_q^g\simeq C_F/C_A P_g^g$, c.f. Sect 2.3. The recoil is neglected, i.e. the
incoming parton retains its energy and angle in the final state. The 
maximal angle $\Theta$ in the parton splitting is then the half
opening angle of the jet.
Using the logarithmic variables (\ref{logvar})
and the anomalous dimension
\begin{equation}
\gamma_0^2(Q)=\frac{2N_C\alpha_s(Q)}{\pi} =
\frac{\beta^2}{\ln(Q/\Lambda)},\quad 
\beta^2=\frac{16 N_C}{b},\quad b=\frac{11}{3} N_C-\frac{2}{3}n_f,
\labl{gamma0}
\end{equation}
we obtain the evolution equation for the gluon multiplicity
\begin{equation}                   
\frac{d \Ng(Y)}{d Y}   = 
 \int_0^Y  dy \gamma_0^2(y)
\Ng (y),
 \labl{dlaeveq}% \nonumber
\end{equation}
where the integral is over the intermediate parton momenta $k=zP$, or
$ y=\ln(k\Theta/Q_0)$.
After second differentiation 
\begin{equation}
\Ng'' (Y)-\gamma_0^2(Y) \Ng(Y)=0, 
\labl{dlad2eq}
\end{equation}
\begin{equation}
\Ng(0)=1,\quad \Ng'(0)=0.
\labl{dlainit}
\end{equation}
In case of fixed coupling this leads simply to
\begin{equation}
\Ng (Y)=\cosh(\gamma_0 Y)\ \simeq \ 
\frac{1}{2}\left(\frac{P\Theta}{Q_0}\right)^{\gamma_0},
\labl{dlafix}
\end{equation}
with a power behavior at high energies. This result was
found already prior to QCD in fixed coupling field theories$\,$\cite{pol}. 
For running coupling the solution of (\ref{dlad2eq}) is found
in terms of modified Bessel functions
\begin{equation}    
\Ng (Y)=\beta \sqrt{Y+\lambda} 
   \{ K_0(\beta \sqrt{\lambda}) I_1(\beta \sqrt{Y+\lambda}) + 
      I_0(\beta \sqrt{\lambda}) K_1(\beta \sqrt{\lambda}) \}.
\labl{dlarun}       
\end{equation}      
At high energies (large $Y$) the asymptotic expressions
$
I_\nu(z)\approx e^z/\sqrt{ 2\pi z}$ and 
$K_\nu(z)\approx e^{-z}\sqrt{ \pi/2 z}$ apply. Then
the second term in (\ref{dlarun}) can be neglected whereas the first term
yields the exponential growth of multiplicity
\begin{equation}
\Ng (y)\sim \exp\sqrt{\frac{16 N_C}{b} (Y+\lambda)},
\labl{dlaasy}                                                           
\end{equation}
corresponding to the first term in (\ref{lnn2}).
Because the coupling is decreasing  with increasing energy the multiplicity
growth is slower than the power (\ref{dlafix}) of the fixed coupling theory
but still larger than a
logarithm as in a ``flat plateau'' model. Using this formula for hadrons
$(Q_0\sim {\cal O}(\Lambda),\  \lambda\ll Y$) the dependence of the cut-off
parameter $\lambda$ factorizes and determines the absolute normalization.
 
We may also apply Eq.  (\ref{dlarun}) to jet multiplicities.
This means we consider $Q_0\to Q_c$ ($\lambda\to \lambda_c$) 
as variable cut-off for the relative 
transverse momenta between jets 
within the Durham jet algorithm (see Sect. 2.1). At high energies with the
above approximations we find the behavior in the two limits 
(\ref{resolve}) for $y_c=(Q_c/P\Theta)^2$
\begin{eqnarray}
\textrm{at high resolution}&  (Q_c\to Q_0):  &    \N\sim
(\beta^2Y)^{1/4} \ln\left(\frac{2}{\beta\sqrt{\lambda_c}}\right)
     \exp{\sqrt{\beta Y}} \phantom{abc}  \labl{hres}\\
\textrm{at low resolution } &  (y_c\to 1):  & 
    \N\to 1,
\label{resolvedla}
\end{eqnarray}
where we used $K_0(z)\simeq \ln(2/z)$ for small $z$. At high resolution
for $Q_c~\to~\Lambda$ ($\lambda_c\to 0$) the parton 
multiplicity diverges logarithmically because
of the Landau pole appearing in the running coupling. The pole is shielded
by the cut-off $Q_c=Q_0$ and at this value the particle multiplicity
reaches the hadron multiplicity $\N/K$ according to the LPHD prescription 
(\ref{lphdeq}). It should be noted that for small  $\lambda$
the coupling becomes large $\sim {\cal O}(1)$. The higher order terms are
sufficiently suppressed by phase space
so that the perturbative
series can be resummed as demonstrated for DLA by Eq. (\ref{dlarun}).
This overall features of the DLA are similar to the 
more precise calculations and the experimental findings to be discussed
below.

\subsection{Modified Leading Logarithmic Approximation (MLLA)} 

The next order terms are generated by the non-singular terms in the splitting
functions, the $z$ dependence of the coupling and the inclusion of energy
conservation in the parton splitting.
One can modify the
evolution equation and keep only terms of $\Ord (\sqrt{\alpha_s})$ and
$\Ord(\alpha_s)$$\,$\cite{dkmt2,do} neglecting those of higher order assuming 
$\alpha_s$ to be small. Then
one can derive the
multiplicity again from a differential equation and express the 
result$\,$\cite{dt,adkt1} in a
compact form 
%in terms of modified Bessel (MacDonald) functions
%$I_\nu (x)$ and $K_\nu (x)$
\begin{equation}
\Ng (Y, \lambda)  =  %C_A^g  
   z_1  \left (
\frac{z_2}{z_1}
\right )^B \; \{I_{B + 1} (z_1)  K_B (z_2) \: + \: K_{B + 1}
(z_1)  I_B (z_2)\},
\labl{mllamult}
\end{equation}
\begin{equation}
z_1  =  \sqrt{\frac{16 N_C}{b} (Y +
\lambda)},  \quad z_2  =  \sqrt{\frac{16 N_C}{b}\lambda},
 \quad B=\frac{a}{b},
\quad a=\frac{11}{3} N_C+\frac{2n_f}{3N_C^2}.
 \labl{mllanot}
\end{equation}
This expression preserves the initial condition $\Ng(0,\lambda)=1$.
A simplification occurs in the case $\lambda\to 0$ (``limiting
spectrum'') where one finds from (\ref{mllamult}) using 
$K_B(z)\to \Gamma(B)(z/2)^{-B}/2$ and $I_B(z)\to 0$ for $z\to 0$ 
the finite limit
\begin{equation}
\Ng^{{\rm lim}} (Y) = \Gamma(B)
  \left ( \frac{z_1}{2} \right )^{(-B+1)} I_{B + 1} (z_1). 
\labl{lsmult}
\end{equation}
At high energies this result and in the same way 
the first term of Eq. (\ref{mllamult})
yield the asymptotic form 
\begin{equation}
\N (Y, \lambda) \sim Y^{-B/2+1/4} \exp\sqrt{\frac{16N_C}{b}Y}
\labl{mllaasy} 
\end{equation}
which is equivalent to (\ref{lnn2}).
The DLA result is recovered from  (\ref{mllamult}) for $B\to0$.
On the other hand, within the high energy MLLA approximations
assuming small $\alpha_s$, the logarithmic singularity for $\lambda\to 0$
present in the general equation (\ref{eveqmult}) 
and in (\ref{dlarun}) 
of the DLA has disappeared. Generally, the MLLA results are expected to
differ from the exact solution in the kinematic region of large coupling
$\alpha_s$.
A common solution to the MLLA simplified coupled 
evolution equations for quark and
gluon jets has been derived as well and can be represented by  expressions
in terms of modified Bessel functions.$\,$\cite{cdfw}.

At threshold the second condition $\N'(0,\lambda)=0$ should hold which follows
directly from the evolution equation (\ref{eveqmult}).
A shortcoming at first sight of the full analytical MLLA solutions is 
a violation of this threshold condition and actually $\N(0,\lambda)'<0$. 
The reason is again 
that  $\alpha_s$ becomes large near threshold and taking only the first
two terms of the $\sqrt{\alpha_s}$ expansion is not justified. This is one
of the motivations to solve the evolution equation more precisely using
numerical methods. 

\begin{figure*}[bt]%[hbt]
\begin{center}
\mbox{\epsfig{file=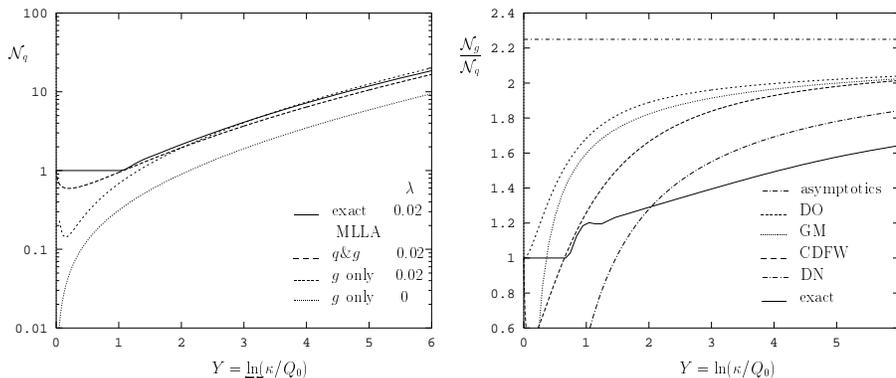,bbllx=3.5cm,bblly=13.cm,bburx=20.5cm,%
bbury=19.cm,width=12.cm}}  %16.cm}}
          \end{center}
\caption{Comparison$\,$\protect\cite{lo3} 
of different approximations to the master equation for
jet evolution$\,$\protect\cite{dkmt2} 
as a function of the jet energy variable $Y=\ln(\kappa/Q_0)$
($\kappa=2E\sin(\Theta/2),\; \Theta=\pi/2$):
(a) Multiplicity in quark jets; the exact numerical solution%
$\,$\protect\cite{lo2}, 
the MLLA results from the coupled Eqns. for $q\&g$ jets$\,$\protect\cite{cdotw},
the solution for $g$ jets only, multiplied by 4/9 and the limiting spectrum
with $\lambda=0$, multiplied by 4/9$\,$\protect\cite{dkmt2};
(b) Multiplicity ratio for exact
solution$\,$\protect\cite{lo2}
compared with asymptotic value 9/4$\,$\protect\cite{bg}, the
MLLA results using the normalization at threshold
(DO$\,$\protect\cite{do} and CDFW$\,$\protect\cite{cdotw}) and asymptotic
expansions  
(GM$\,$\protect\cite{gm} and DN$\,$\protect\cite{dn1}) for $\lambda=0.02$.}
\label{plotmult}
\end{figure*}

\subsection{Numerical Solutions}

The MLLA approximate solutions have been compared with the numerical
solution of the coupled system of equations in (\ref{eveqmult}).$\,$\cite{lo2}
For small $\lambda\sim 0.02$  good agreement is found already shortly
above threshold (Fig. \ref{plotmult}a$\,$\cite{lo3}): 
for the numerical solution with the exact integration
limits the inelastic threshold with
$\N>1$ starts at $Y>0$, for the analytic solutions
with approximate integration limits this occurs already at
$Y=0$ with $\N'<0$. A larger discrepancy by a factor 2 
is found for the ``limiting spectrum'' with $\lambda=0$.

On the other hand,
the ratio of gluon and quark multiplicities is rather sensitive to the type
of approximation as this difference is a sub-leading effect 
(see Fig. \ref{plotmult}b). At LEP energies ($Y\sim 5$) the inclusion
of higher order terms decreases the ratio $r$ from the asymptotic value
$r=9/4$. The asymptotic solutions,
``GM''$\,$\cite{gm} and ``DN''$\,$\cite{dn1}, which have no 
normalization condition at any finite energy 
eventually will reach unphysical values $r<1$ at low energies.
The fully resummed results are normalized
to $r=1$ at threshold. The numerical solution takes the lowest value of the
ratio $r$ at high energies. 

\subsection{Experimental Results on Quark Jets}

Tests of the QCD predictions for multiplicities are available from the
final states in $e^+e^-$ annihilations and in the current fragmentation
region in deep inelastic $ep$ scattering. In MLLA accuracy 
the asymptotic predictions
(\ref{lnn2}) can be taken over from the single gluon jet to the  
$e^+e^-$ final state. In most fits of (\ref{lnn2}) to the data the 2-loop
formula for  $\alpha_s$ is used although the leading log calculations are
only accurate to one loop.
\begin{figure*}[bt]%[hbt]
\begin{center}
\mbox{\epsfig{file=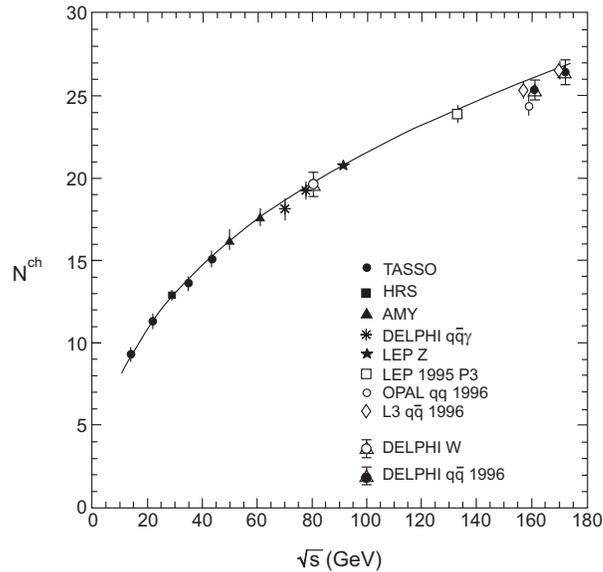,bbllx=1.0cm,bblly=6.cm,bburx=19.2cm,%
bbury=24.2cm,width=8.cm}}  %16.cm}}
          \end{center}
\vspace{-0.3cm}
\caption{Mean charge particle multiplicity in $e^+e^-$ annihilation as
function of total energy compared with the MLLA prediction. The data at
energies $\geq M_Z$ have been corrected for increased multiplicity from $b$
and $c$ decays.$\,$\protect\cite{delphimult}}  
\label{figmult}
\end{figure*}

The agreement with the data is generally good. 
As an example we show data in the range from 20 to 180 GeV 
in Fig. \ref{figmult}.$\,$\cite{delphimult}
The curve represents Eq. (\ref{lnn2}) where $\N$ is multiplied 
with $(1+d \sqrt{\alpha_s})$ to allow for the possibility of a 
next-to-next-to-leading-order correction.
Fitted parameters are the overall normalization, $\alpha_s(m_Z)=0.119\pm0.003$ 
and  $d=1.11\pm 0.39$. Recent measurements at LEP-2$\,$\cite{opalmult}
follow the extrapolation from lower energies although they are 
systematically a bit at the
low side.
For example, the fit to data 
between 12 and 161 GeV predicts $\N_{ch}=27.6$ at 189 GeV which is to be
compared with the measurement$\,$\cite{opalmult}
$ \N_{ch}=26.95$ $\pm$ 0.16(stat.) $\pm$ 0.51(syst.).
Recently a comparison of data with the prediction including
the higher order terms up to 3NLO has been performed.$\,$\cite{dg} Fitting the
normalization and the coupling leads  to a good result with reasonable
$\Lambda$ in the range 0.12-0.26 GeV.

%Writing$\,$\cite{lo}
%\begin{equation}
%\N_{ch}^{e^+e^-}\sim K_{ch} 2\frac{4}{9}\Ng+c
%\labl{nch} 
%\end{equation}
%with the factor $K$ from (\ref{lphdeq}), the factors 2 and 4/9 for the two
%hemispheres and the quark jet and using 
%the limiting spectrum form (\ref{lsmult}) for the multiplicity in a single
%jet one finds good agreement with data from low energies up to the $Z$
%with $K\sim 2$ (assuming $ K_{ch}\sim 2/3 K$). The other fit parameter is
%then $Q_0=\Lambda\sim 270$ MeV. This means that one parton in
%the cascade (with cut off Q0) corresponds to about two final hadrons.
%
%\begin{figure*}[htb]
\begin{figure*}[t]
\begin{center}
%lower left. upper right positions
%\mbox{\epsfig{file=ochs1.ps,width=8.2cm,bbllx=2.cm,bblly=11.8cm,bburx=20.cm,bbury=23.2cm}}
\mbox{\epsfig{file=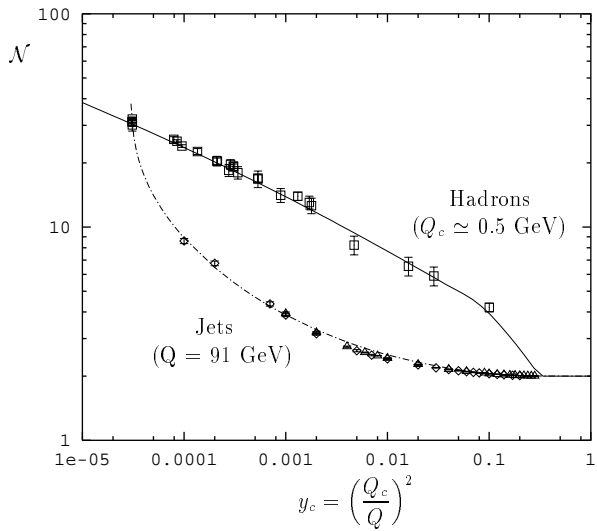,width=8.0cm,bbllx=3.2cm,bblly=9.2cm,bburx=18.cm,bbury=22.8cm}}
\end{center}
\vspace{-0.7cm}
\caption[]{
Data on the average jet multiplicity $\protect\N$ at $Q$ = 91 GeV 
for different resolution parameters $y_c$ (lower set) and
the average hadron multiplicity (assuming $\protect\N = \frac{3}{2}
\protect\N_{ch}$)
at different $cms$ energies between $Q=3$ and $Q=91$ GeV using
$Q_c=Q_0$ = 0.508 GeV in the parameter $y_c$ (upper set).
The  curves follow from the evolution equation (\protect\ref{eveqmult})
with $\Lambda$ = 0.5 GeV; the upper curve for hadrons is based on 
the duality picture (\ref{lphdeq}) with $K=1$ and parameter
$\lambda=\ln(Q_0/\Lambda)=0.015$
(Figure from Ref. 74).% $\,$\cite{lo2}) %p.25 
}
\label{hjetmult}
\end{figure*}
\begin{figure*}[hbt]
\begin{center}
\noindent
\begin{minipage}{5.8cm}%.4\linewidth}
\mbox{\epsfig{bbllx=0bp,bblly=45bp,bburx=285bp,bbury=280bp,%
file=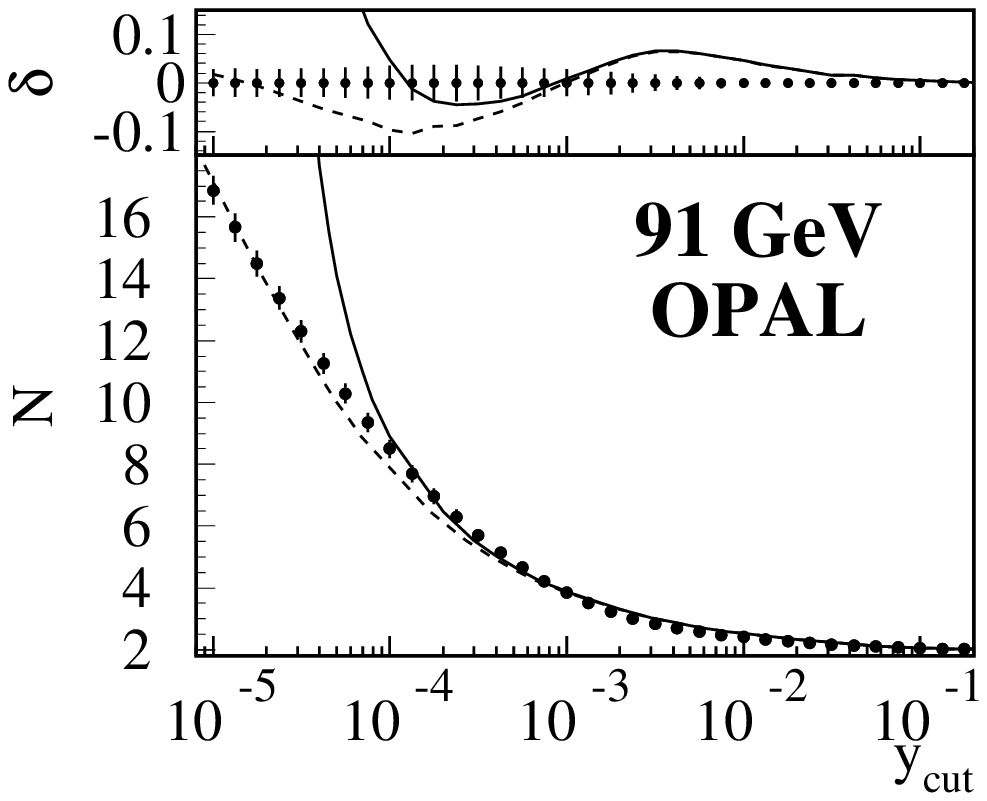,
width=5.8cm} }
\end{minipage}
\hfill 
\begin{minipage}{5.8cm} %.4\linewidth}
\mbox{\epsfig{bbllx=0bp,bblly=45bp,bburx=285bp,bbury=280bp,%
file=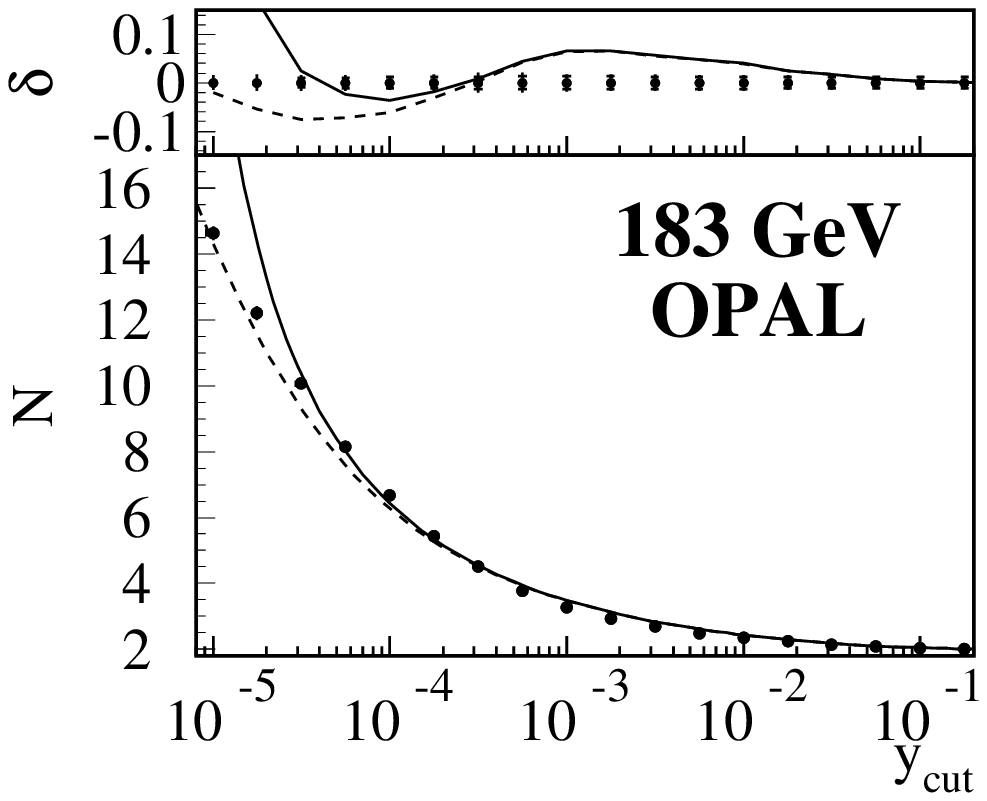,
width=5.8cm}}
\end{minipage}
\end{center} 
\vspace{-0.3cm}
\caption[]{
Jet multiplicities 
extending towards lower $y_{cut}$ parameters; full lines as in
Fig.~\protect\ref{hjetmult} for jets, 
dashed lines the same predictions but shifted $y_{cut}\to
y_{cut}-Q_0^2/Q^2$ according to the different kinematical boundaries  
as in (\protect\ref{partonhadron}), with parameters as in
Fig.~\protect\ref{hjetmult} 
(data from OPAL$\,$\protect\cite{opaljet,Pfeifenschneider}).
}
\label{multlowy}
\end{figure*}

Alternatively, one can compare the data with predictions from the full
parton cascade  using the normalization at threshold. 
The numerical solution of the pair of evolution equations (\ref{eveqmult})
 with (\ref{initmult}) has been compared with
data on \epem annihilation; in this calculation the term of order 
\als had been
replaced by the result from the full matrix element of the same
order.$\,$\cite{lo2}
The predictions from this analysis for both, jets and hadrons,
are shown in Fig. \ref{hjetmult}
together with the data.

The lower set of data refers to the jet multiplicity at
variable resolution $y_{cut}=(Q_c/Q)^2$ calculated at fixed total energy $Q$.
For $y_{cut}\to 1$ all particles are
combined into two jets and therefore  $\N\to 2$ as discussed in
(\ref{resolve}). 
On the other hand, for
$y_{cut}\to 0$ all hadrons are resolved and $\N\to \N_{had}$.
The theoretical prediction describes the data reasonably well down to 
 $y_{cut}\sim 10^{-4}$ and is determined fully by the parameter
$\Lambda$. 

An improvement of the description for large $y_{cut}$ is possible
if 2-loop results are used. 
For $y_{cut} \gsim 0.01$ the data are well described by  %\gtrsim
the complete matrix element calculations  to $O(\alpha_s^2)$. 
%(first results of this kind in $\,$\cite{kl}) 
 In the region  $y_{cut}\gsim 10^{-3}$
the resummation of the higher orders in $\alpha_s$ is important.$\,$\cite{cdotw}
The calculation at large  $y_{cut}$
allows the precise determination of the
coupling or, equivalently, of the QCD scale parameter
$\Lambda_{\overline{MS}}$.
$\,$\cite{L3jmul,opaljmul}

The multiplicity diverges
 for small cut-off  $Q_{cut}\to \Lambda$
as in this case the coupling $\alpha_s(k_T)$ diverges. 
The divergence is shielded by the cut-off $Q_0$ and 
according to the duality picture in (\ref{lphdeq})
the parton multiplicity represents the hadron multiplicity at this scale.
It is found from the data that this happens for  $Q_c=Q_0\simeq 0.5$ GeV
if the total hadron multiplicity is taken as 3/2 of the charged multiplicity.
The corresponding
calculation at lower $cms$ energies is in 
agreement with the hadron multiplicity 
data down to $Q=3$~GeV with  the same parameter $Q_0$
as shown  
by the upper set of data and the theoretical curve in Fig. \ref{hjetmult}.
Interestingly, the normalization constant in (\ref{lphdeq}) can be chosen as
\begin{equation}
K\approx 1,
\labl{kunity}
\end{equation}
whereas in previous approximate calculations using the limiting spectrum
(\ref{lsmult}) the value 
$K\approx 2$
has been obtained.$\,$\cite{lo} The parameter $K$ is correlated
with  $Q_0$ and can be varied within about 30\%.
The result $K=1$ 
implies that the hadrons, in the duality picture, can be viewed as
very narrow jets with low resolution parameter $Q_0\sim$
a few 100 MeV.\footnote{The precise value of $Q_0$ depends on the expression
used for the $k_\perp$ scale in the argument of $\alpha_s$ and varies typically
between 250 and 500 MeV.}

The behavior of multiplicities in Fig. \ref{hjetmult} is close to the
qualitative expectations from the DLA discussed in Sect. 3.2. The running of the
coupling \als is crucial for the results. Namely, for 
constant $\alpha_s$ both
curves for hadrons and jets would coincide
and follow a power law in the ratio of available scales 
$Q_{cut}/Q$ as in~(\ref{dlafix}). With running $\alpha_s(k_T/\Lambda)$ the
absolute
scale of $Q_{cut}$ matters: $\alpha_s$ varies most strongly for
$Q_{cut}\to\Lambda$ for jets at small $y_{cut}$ 
and for hadrons near the threshold of the process at large  $y_{cut}$
(small $Q$) where again $\alpha_s \gsim 1$.

Recently, data became available at smaller $y_{cut}$ in a wide energy range
from 35 to 183 GeV$\,$\cite{opaljet,Pfeifenschneider} and examples are shown in
Fig. \ref{multlowy}.  In the theoretical calculation all hadrons are resolved
for $Q_{cut}\to Q_0$ whereas in the experimental quantities this happens for 
$Q_{cut}\to 0$. This kinematical mismatch can be avoided$\,$\cite{lo2}
by a shift in $y_{cut}$
according to (\ref{partonhadron}). The shifted (dashed) curves in Fig. 
\ref{multlowy}
describe the data rather well
 whereby the $Q_0$ parameter has been taken from
the fit to the hadron multiplicity before; the predictions fall a bit
below the data at the lower energies like 35 GeV. The
non-perturbative $Q_0$ 
correction becomes negligible for $Q_c>1.5$ %\gtrsim 1.5$ GeV.

It appears that the final stage of hadronization in the jet evolution can be
well represented by the parton cascade with small cut-off $Q_0$ and 
with the standard 1-loop running coupling.
This description clearly goes beyond
standard perturbation theory which is determined entirely by the QCD scale
 $\Lambda$. The calculation in the soft region rather corresponds to a 
non-perturbative model which involves a hadronization scale $Q_0$.
In some kinematic regions (small $k_\perp$) the coupling becomes large
$\alpha_s/\pi \sim \Ord (1)$ but -- as experienced with the analytical DLA
results -- there is good convergence of the leading log summation also in
this region as the higher order terms are suppressed by soft gluon
coherence effects. 

%
%\begin{figure*}[htb]
\begin{figure*}[t]
\begin{center}
%lower left. upper right positions
\mbox{\epsfig{file=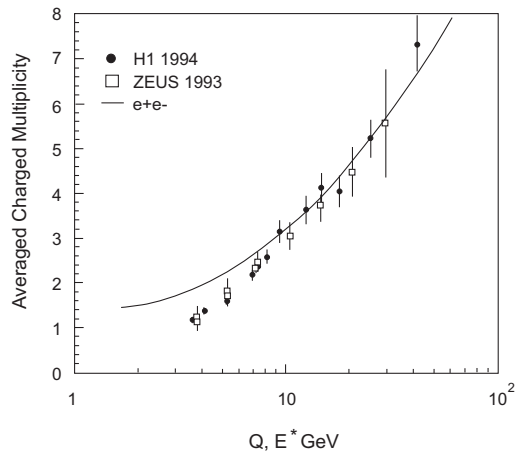,width=7.0cm,bbllx=1.7cm,bblly=8.5cm,bburx=19.cm,bbury=24.2cm}}
\end{center}
\vspace{-0.7cm}
\caption[]{
Average charged particle  multiplicity in the current fragmentation region 
of the Breit frame as measured by H1 and ZEUS Collaborations at momentum
transfer $Q$ in comparison with \epem results at $cms$ energy 
$E^*=Q$.$\,$\protect\cite{H1breit}
}
\label{dismult}
\end{figure*}

\subsection{Test of Jet Universality}
Results on quark jet fragmentation are expected to be universal in the parton
model.$\,$\cite{feyn} 
If we consider the particle multiplicity the soft particles are
included and in this case the reference frame becomes important. 
In deep inelastic lepton proton scattering at momentum transfer $Q^2$
the particles in the current fragmentation region in the Breit frame
should be compared with the multiplicity of one hemisphere in \epem
collisions at $cms$ energy $E^*=Q$ (see, for example,
Refs. 1,82-84)               %$\,$\cite{ilk,swz,abas,gdkt}).  %Ref.$\,$\cite{Breit}). 
Results of such a comparison are shown in Fig. \ref{dismult}. The DIS
results approach those from \epem for energies $Q\gsim 10$ GeV.
At lower energies processes not available in \epem annihilation like photon
gluon fusion are important. The agreement at higher energies confirms the
universality of the jet fragmentation and the relevance of the Breit frame. 

\subsection{Comparison of Quark and Gluon Jets}

The experimental results on multiplicities in quark jets are derived
as half of the total multiplicity of \epem annihilation or the multiplicity in
the current region of DIS. These experimental results are well met by the
perturbative calculations (and their non-perturbative extensions).

The situation is more difficult for gluon jets and the following results
have been presented: 

1. A fully inclusive
measurement is possible at lower energies $Q$
from radiative decays of 
$\Upsilon(1S)$ which are assumed to proceed in lowest perturbative 
order through $\Upsilon\to \gamma gg$. This yields measurements at 
$Q\sim 5$ GeV of the $gg$ system.$\,$\cite{cleo1} 
Similarly, results have been obtained
near 10 GeV from the decay $\Upsilon(3S)\to \gamma \chi_{b2}$,$\,$\cite{cleo2}
assuming $ \chi_{b2}\to 
gg$. At $Q\sim 5$ GeV the ratio $r$ is still compatible with  unity.

2. At higher energies the multiplicities have to be extracted from gluon jets
in a more complex multi-jet environment, either 3-jet events in \epem or
high $p_T$ jets in $pp$ colliders. A simple situation is met again in 
$e^+e^-\to$ 3 jets  with two quark jets (taken as identified $b$ quark
jets) recoiling together against the gluon. This possibility has been
pointed out some time ago$\,$\cite{dkt11} and worked out in detail
for \epem annihilation.$\,$\cite{jwg} Results on the multiplicity in the gluon
hemisphere have been obtained in this way by OPAL$\,$\cite{opalglu} and the
ratio $r$ is found to be
$r=1.514\pm0.019\pm0.034$ at  $Q\sim 80$ GeV.

3. Furthermore, results have been obtained from 
symmetric 3 jet events in Y-configuration with a quark in one hemisphere and
quark and gluon in the other one.$\,$\cite{delphiglu} In this case the gluon jet
multiplicity is obtained as difference of the 3 jet and the known
$q\overline q$ multiplicity as suggested by a perturbative leading order
analysis.$\,$\cite{dkt11,ko}
The results obtained in the intermediate
 energy region  interpolate between the $Q=10$ and $Q=80$ 
results above.

4. The ratio $r$ has also been determined from high $p_T$ jet production
with dijet masses in the range 100-600 GeV  at the
TEVATRON by CDF.$\,$\cite{cdfmult} The energy dependence of the multiplicity
has been assumed to follow the ``limiting spectrum'' formula (\ref{lsmult}).
Using the known composition of quark and gluon jets the energy dependence
of the multiplicity is then predicted for a given ratio $r$. The curves in Fig.
\ref{cdffig}, calculated for fixed ratios, are then compared to the data
which yields the estimate $r=1.7\pm0.3$ over this energy range.   

The experimental results 1. and 2. do not use further theoretical input
and have been compared with the 
asymptotic predictions and with the numerical solutions of the evolution
equations;  
 both are found not fully satisfactory.

1. In the first approach$\,$\cite{dg} the 3NLO asymptotic expression 
is fit to the gluon
jet data at $Q\sim5,10$ and $80$ GeV by adjusting the normalization and
$\Lambda$.
Using the corresponding expression
for $r=\Ng/\Nq$ and taking the same parameters 
one can predict the multiplicity in quark jets. 
This is now found lower over the full energy region by 20-30\%.
Correspondingly the ratio $r$ is predicted larger than the measurements.

2. In the second approach$\,$\cite{lo2} the numerical solution of the evolution
equations (\ref{eveqmult}) is fit to the \epem data and the  
parameters $Q_0$ and $\Lambda$ are determined. The fits are found
satisfactory (see Sect. 3.5). The gluon jet multiplicity from the same
solution are now compared with data.
There is a good agreement with the OPAL result at $Q\sim 80$ GeV
but the prediction for CLEO at $Q\sim 5$ GeV is too large by 20\%.
It has been argued that at low energy the exact treatment of the $\Ord
(\alpha_s)$ term is important as was found explicitly for quark jets.
For the gluon jets at the $\Upsilon$ such calculations have not been
performed yet. Another explanation could be a large non-perturbative
effect for gluons which ``freezes'' the gluon degrees of freedom 
at low energies. 
It is interesting to note that the result from the HERWIG Monte Carlo
shows a similar behavior with too large ratio $r$ at low energies.

At  higher energies the numerical solutions yield a rise from $r=1.54$ 
at $Q=100$ GeV to $r=1.68$ at $Q=600$ GeV which is consistent with the CDF
result within the large error.  

%
%\begin{figure*}[htb]
\begin{figure*}[t]
\begin{center}
%lower left. upper right positions
\mbox{\epsfig{file=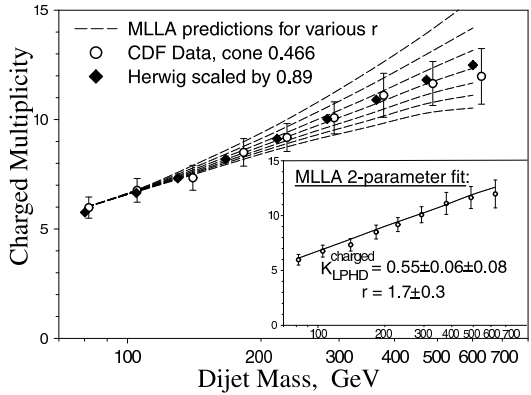,width=7.0cm,bbllx=0cm,bblly=7.5cm,bburx=18.5cm,bbury=22.cm}}
\end{center}
\vspace{-0.7cm}
\caption[]{
Charged multiplicity in a high $p_T$ jet with cone opening 0.466 rad 
for different dijet masses in
comparison with MLLA predictions for different values of the ratio
$r=\Ng/\Nq$. The curves corresponds 
to $r=1.0,1.2,1.4,1.6,1.8,2.0,2.25$.$\,$\cite{cdfmult}
}
\label{cdffig}
\end{figure*}

Taking these results together there is a clear evidence for the difference
of quark and gluon jet multiplicities. The ratio $r$ at presently accessible
energies is much smaller than the asymptotic value $r=9/4$.
It is a considerable success that at the higher LEP energies
there is agreement
with perturbative QCD calculations if all non-leading logarithmic 
terms from the MLLA evolution equation are included.  

\section{Momentum Spectra of Particles Inside Jets}

We start from the MLLA evolution equation for particle spectra and 
consider solutions relevant for the small $x$ region.  The early DLA results
on the approximately Gaussian shape of the distribution 
(``hump-backed plateau'')$\,$\cite{dfk1,bcmm} in the variable
\begin{equation}
\xi=\ln\frac{1}{x_p}, \qquad x_p=\frac{p}{E_{jet}},
\labl{xidef}
\end{equation}
for momentum $p$ of the particle
has been an important milestone in the application of perturbative QCD to 
the multiparticle phenomena. They demonstrated the relevance of
angular ordering and color coherence to the particle spectra at low momentum.
The success of the improved calculations 
within MLLA accuracy gave strong support to the concept of LPHD.
 %LPHD.%$\,$\cite{adkt1}. 
This Gaussian shape is found quite universally
in jets from various collision processes and also for all particle species;
however, the differences observed for various species is not yet
quantitatively understood.
We will discuss here some MLLA based 
computations and compare them with
typical recent results from $e^+e^-$, $ep$ and $\overline pp$ colliders
whereby we  include 
 some striking phenomena with the very soft
particles ($\lsim$ 1 GeV), as well as the energy evolution 
of spectra from threshold to
asymptotic energies. 

Whereas this analysis is focused on the small $x$ domain 
of the particle spectra there is a
complementary approach which treats the evolution equation 
 in the larger $x_p$ region (say, $x\gsim 0.1$) 
(``fragmentation function'') for which results in 2-loop
accuracy for the running coupling are available. 
 Such studies are an important tool for the
determination of the strong coupling.$\,$\cite{nw,bkk,kkp} This approach is
expected to be more precise at large $x_p$ as compared to the 1-loop MLLA
calculations, on the other hand  
it does not take into full account
 the contributions important for the soft region. Furthermore, 
in this approach an ansatz
involving several parameters  is required
for the fragmentation function(s) at a particular finite energy
whereas the MLLA/LPHD approach has only one essential non-perturbative parameter 
$Q_0$.

\subsection{Evolution Equation for $\xi$-Spectra and Approximate Solutions}

The inclusive distribution of partons $b$
in a jet from parton $a$ is obtained from the generating functional through
\begin{equation}
x {D}_a^b (x)
= E_b  \frac{\delta}{\delta
u(k_b)}  Z_a (E_a, \Theta; \{u\})|_{u=1} 
\labl{dbar}
\end{equation}
and its evolution equation  from Eq.(\ref{2.23}). In the approximation
$1-z\approx 1$ it takes the form
\begin{equation}
\frac{d}{dY}  x {D}_A^B (x, Y) =
\sum_{C = q, \overline{q}, g}  \int_0^1 
dz  \frac{\alpha_s (k_\perp)}{2 \pi}  \Phi_A^C (z)  \left
[ \frac{x}{z}  {D}_C^B  \left ( \frac{x}{z}, Y+\ln z  \right ) \right ],
\labl{2.24}
\end{equation}
 where $Y=\ln(P\Theta/Q_0)$ and
$\Phi_A^C$ stands for the DGLAP splitting functions.
The boundary condition for (\ref{2.24}) reads
\begin{equation}
x {D}_A^B (x) |_{Y=0} = \delta (1 - x)  \delta_A^B.
\labl{2.25}
\end{equation}
  The scale of the coupling is given by the
transverse
momentum taken as $k_\perp \simeq z (1 - z) E \Theta$.  The shower
evolution
is cut off by the parameter $Q_0$, such that $k_\perp \geq Q_0$
and this restriction is understood in (\ref{2.24}).

The integral equation can be solved by Mellin transform 
\begin{equation}
D_\omega (Y, \lambda)  =  \int_0^1  \frac{dx}{x}   
x^\omega  \left [x {D} (x, Y, \lambda) \right ] =
\int_0^Y  d \xi e^{- \xi \omega} \: D (\xi, Y, \lambda).
\labl{2.26}
\end{equation}
% with
%\renewcommand{\theequation}{2.27}
%\begin{equation}
%Y \: = \: \ln \frac{E \Theta}{Q_0}, \;\;\; \lambda \: = \: \ln
%\frac{Q_0}{\Lambda}, \;\;\; \xi \: = \: \ln \frac{E}{k}
%\labl{2.27}
%\end{equation}
%\noindent and parton momentum $k$.

In flavor space the valence quark and ($\pm$) mixtures of sea
quarks and gluons evolve independently with different \lq\lq
eigenfrequencies" $\nu_{\pm}(\omega)$.  At high energies and $x \ll 1$, the
dominant
contribution to the inclusive spectrum comes from the \lq\lq
plus"-term, which we denote by $D_\omega (Y, \lambda) \equiv
D_\omega^+ (Y, \lambda)$.  In an approximation where only the
leading singularity plus a constant term is kept 
$\nu_+(\omega)\approx 4N_C/\omega -a$, one obtains in a sequence of steps
from (\ref{2.24}) 
the following evolution equation$\,$\cite{dkmt2}
\begin{equation}
\left ( \omega + \frac{d}{d Y} \right )   \frac{d}{d Y}  
D_\omega (Y, \lambda)  -  4 N_C   \frac{\alpha_s}{2 \pi}  
D_\omega (Y, \lambda)   % \nonumber \\
 =  - a   \left ( \omega + \frac{d}{dY} \right )  
\frac{\alpha_s}{2 \pi}   D_\omega (Y, \lambda),
\labl{2.28}
\end{equation}
where
\begin{equation}
a =  \frac{11}{3}N_C + \frac{2n_f}{3 N_C^2}.
\labl{amldef}
\end{equation}
and the boundary condition corresponding to (\ref{2.25}) is
$D_\omega(0,\lambda)=1$.
Restricting to the leading term by setting
 $a=0$, i.e. dropping the $r.h.s.$ in (\ref{2.28}), yields the DLA evolution
equation. One can now 
derive asymptotic solutions at high energies in leading-order and 
next-to-leading order
or a full solution of the evolution equation  (\ref{2.28}) 
including the boundary
condition at threshold. 

Introducing the anomalous dimension $\gamma_\omega$
according to
\begin{equation}
D_\omega (Y, \lambda)  =  D_\omega (Y_0, \lambda) 
\exp  \left ( \int_{Y_0}^Y  dy  \gamma_\omega  \left [
\alpha_s (y) \right ] \right ),
\labl{2.29}
\end{equation}
the evolution equation (\ref{2.28}) can
be expressed in terms of a differential equation for $\gamma_\omega$
\begin{equation}
(\omega + \gamma_\omega) \gamma_\omega  -  \frac{4 N_C
\alpha_s}{2 \pi}  =  - \beta (\alpha_s)  \frac{d}{d
\alpha_s} \gamma_\omega  -  a (\omega + \gamma_\omega)
\frac{\alpha_s}{2 \pi},
\labl{2.30}
\end{equation}
where $\beta (\alpha_s) = \frac{d}{dY} \: \alpha_s (Y)
\simeq - b \: \frac{\alpha_s^2}{2 \pi}$.  
For the DLA one finds
\begin{equation}
(\omega + \gamma_\omega) \gamma_\omega  -  \gamma_0^2  =
0, \quad \gamma_\omega^{DLA} (\alpha_s)  = \frac{1}{2} 
\left ( - \omega + \sqrt{\omega^2 + 4 \gamma_0^2} \right );
\labl{2.31}
\end{equation}
 with
\begin{equation}
\gamma_0^2  =  \gamma_0^2 (\alpha_s)  \equiv 2 N_C 
\frac{\alpha_s}{\pi}.
\labl{2.32}
\end{equation}
Choosing the (+) sign for the square root yields the solution which
dominates at high energies whereas the solution with the (--) sign would die
out. 

 The next-to-leading order  result
in MLLA follows from Eq. (\ref{2.30}) including the $r.h.s.$ where the first
term proportional to the $\beta$-%
function keeps trace of the running coupling effects while the
second accounts for the hard corrections to the soft singularities in the
splitting function. In comparison to the leading term they are 
of relative order
$\sqrt{\alpha_s}$ and the MLLA correction to DLA reads
\begin{equation}
\gamma_{\omega} \; = \; \gamma^{DLA}_{\omega} \: + \: \frac{\alpha_s}{2 \pi}
\:
\left [ - \frac{a}{2} \: \left ( 1 +
\frac{\omega}{\sqrt{\omega^2
+ 4 \gamma_0^2}} \right ) \: + \: b \frac{\gamma_0^2}{\omega^2
+
4 \gamma_0^2} \right ] \: + \: O (\alpha_s^{3/2}).
\labl{2.33}
\end{equation}

Instead of these asymptotic solutions one can also find directly
an analytical expression for the exact solution
of the differential equation
(\ref{2.28}) in terms of confluent
hypergeometric functions
$\,$\cite{dkmt2,dkt5}
\begin{eqnarray}
D_{\omega} (Y, \lambda) & = & \frac{\Gamma (A + 1)}{\Gamma (B +
2)}  z_1 z_2^B \{ \Phi (- A + B + 1, B + 2, -z_1)  \Psi
(A, B
+ 1, z_2) \nonumber \\
%                              & & \labl{2.34}\\
& + & e^{z_2 - z_1} (B + 1)  \Psi (A + 1, B + 2, z_1) 
\Phi
(- A + B + 1, B + 1, - z_2) \}, \nonumber \\  \labl{2.34}
\end{eqnarray}
where we have used the notation
\begin{equation}
A  =  \frac{4 N_C}{b \omega}, \quad B  = \frac{a}{b}, \qquad
z_1  =  \omega (Y + \lambda), \quad z_2  =  \omega \lambda.
\labl{2.35}
\end{equation}
From these representations of $D_\omega$ one can reconstruct the $x$ (or
$\xi$) distributions by inverse Mellin transformation
\begin{equation}
\left [ x {D} (x, Y,\lambda) \right ]  \equiv  {D}
(\xi, Y, \lambda)  =  \int_{\epsilon - i \infty}^{\epsilon + i \infty} 
 \frac{d \omega}{2 \pi i}  x^{- \omega}  D (\omega, Y, \lambda)
\labl{2.37}
\end{equation}
where the integral runs parallel to the imaginary
axis to the right of all singularities of the integrand in the
complex $\omega$-plane.

\subsection{Moments}
It is convenient to analyze the properties of the 
$\xi$ spectrum in greater detail
in terms of the normalized moments$\,$\cite{fw,dkt5}
\begin{equation}
  \xi_q \equiv <\xi^q>
    = \frac{1}{\N}\int d\xi \xi^q  D(\xi).
  \labl{ximomdf}
\end{equation}
Also one defines the cumulant moments
$K_q $ %$\,$\cite{so}, 
or the reduced cumulants  $k_q \equiv
K_q/\sigma^q$,
which  are given for $q\leq 4$ by
\begin{eqnarray}           
K_1 & \equiv & \overline{\xi} \; \equiv \; \xi_1 \nonumber\\
K_2 & \equiv & \sigma^2 \; = \; < (\xi - \overline{\xi})^2
>, \nonumber \\            
K_3 & \equiv & s\sigma^3 \; \equiv \;
< (\xi - \overline{\xi})^3 >, \nonumber \\
K_4 & \equiv & k \sigma^4 \; \equiv \;
< (\xi - \overline{\xi})^4 > \: - \: 3 \sigma^4,
\labl{2.66}                
\end{eqnarray}             
where the third and forth reduced cumulant moments are
the skewness $s$ and the kurtosis $k$ of
the                        
distribution.

The cumulant moments can be found
%using the explicit solution
%of MLLA Evolution Equation
using the expansion of the Mellin-transformed spectrum
$D_\omega (Y, \lambda)$ in (\ref{2.26}):
\begin{equation}           
\ln \: D_\omega (Y, \lambda) \; = \; \sum_{q = 0}^{\infty} \;
   K_q (Y, \lambda) \; \frac{(- \omega)^q}{q !}.
\labl{2.68}
\end{equation}
The high energy behavior of moments in next-to-leading order can be
obtained with $K_q (Y, \lambda)= \left . (-\partial/\partial \omega)^q \ln 
D_\omega (Y, \lambda) \right |_{\omega = 0}$.
%
%\begin{equation}
%K_q (Y, \lambda) \; = \; \left . \left ( -
%\frac{\partial}{\partial \omega} \right )^q \; \ln \: D_\omega
%(Y, \lambda) \right |_{\omega = 0}.
%\labl{2.69}
%\end{equation}
From the high energy approximation of the anomalous dimension
$\gamma_\omega$ in (\ref{2.33}) using (\ref{2.29})
%This yields for the cumulant moments
\begin{equation}
K_q  =
 \int_0^{Y} dy  \left . \left ( - \frac{\partial}{\partial \omega} \right
)^q  \gamma_\omega (\alpha_s (y)) \right |_{\omega = 0},
\labl{2.71}
\end{equation}
which shows the direct dependence of the
moments on $\alpha_s (Y)$.  For fixed $\alpha_s$, for example,
one obtains simply $K_q (Y)  \propto  Y$ for high
energies. Alternatively, one can derive evolution equation for the
moments$\,$\cite{fw} and derive the high energy behavior in 
next-to-leading order.
Approximate forms for the yet higher order terms can be obtained from Eq.
(\ref{2.34}) for arbitrary $\lambda$ 
and in particular for the limiting spectrum with $\lambda=0$.
   
The mean of the $\xi$ distribution in MLLA 
proves to have an energy dependence of the form$\,$\cite{fw,dkt5,dkfb}
\begin{equation}
\overline\xi  =  Y \left [ \frac{1}{2} +
\sqrt{\frac{C}{Y}}  +  \frac{\overline a}{Y}  +  \overline b
Y^{- 3/2} \right ]
\labl{2.50}
\end{equation}
with
\begin{equation}
C  =  \frac{a^2}{16  N_C b}  =  0.2915  (0.3513)
\; {\rm for} \; n_f  = 3 (5).
\end{equation}
The moments evolve in different energy regions according to the relevant
number of flavors $n_f$. It turns out that even at the energies of
LEP-2 the approximation $n_f=3$ is rather good, in any case better than
$n_f=5$.$\,$\cite{lo,klo}

A quantity closely related to $\overline\xi$ 
is the position of the maximum $\xi^*$ which
differs from $\overline\xi$ only in higher order terms $\overline a,
\overline b$ in
(\ref{2.50}). For the limiting spectrum one finds$\,$\cite{fw,dkfb,dkt7} 
%The subleading coefficients $a_i$ have been computed
%$\,$\cite{dkt5,dkfb,fw,dkt7} but the next term can be obtained
%only for certain combinations of the mean, median and peak.  For peak
%position $a_{\rm max} = - C$ and it appears that in the
%available
%energy range the expression
%\renewcommand{\theequation}{2.51}
\begin{equation}
\xi^*  =  Y \left [ \frac{1}{2} \: + \: \sqrt{
\frac{C}{Y}} \: - \: \frac{C}{Y} \right ].
\labl{2.51}
\end{equation}
This form leads to a nearly linear dependence of $\xi^*$ on
$Y$. 
It is worthwhile to mention that in the large $N_C$ limit, when 
$11 N_C\gg 2n_f$ (cf. Eqs.(\ref{brems}) and (\ref{amldef}))
the parameter $C$ becomes independent on both $n_f$ and $N_C$ and approaches
its asymptotic value of $C=\frac{11}{3}\frac{1}{2^4} \simeq 0.23$.
Therefore in this limit the effective gradient of the straight line is
determined by such a fundamental parameter of QCD as the celebrated 
$\frac{11}{3}$ factor (characterizing the gluon self interaction)
in the coefficient $b$. 

The shape parameters $\sigma, k, s$ for the limiting spectrum have been
derived in  
next-to-leading order at  $N_C = 3$ as %$\,$\cite{fw,dkt5}
\begin{eqnarray}
\sigma & = & \frac{Y}{\sqrt{3 z}} \;  \left ( 1 \: - \:
\frac{3}{4 z} \right ) \; + \; O (Y^{- 1/4})   \labl{2.56a} \\
s & = & - \: \frac{a}{16} \; \frac{1}{\sigma} \; + \; O (Y^{-
3/4}) \labl{2.56b} \\
k & = & - \: \frac{27}{5 Y} \; \left ( \frac{Y}{\sqrt{z}}
\; - \; \frac{b}{24} \right ) \; + \; O (Y^{- 3/2}) \labl{2.56c}
\end{eqnarray}
where $z=\sqrt{16N_CY/b}$.
It is worthwhile to notice that the next-to-leading effects
are
very substantial at present energies.  In particular, the
spectrum significantly softens because of energy conservation effects.
This influences the rate of
particle multiplication which is strongly overestimated by the DL
approximation.

The asymptotic behavior $(n \geq 1)$ can be obtained from (\ref{2.71}) with
(\ref{2.33})
\begin{eqnarray}
\sigma^2 & \sim & Y^{3/2} \labl{2.72a}\\
k_{2 n + 2} & \sim & \left ( \sqrt{Y} \right )^{- n} \labl{2.72b}\\
k_{2 n + 1} & \sim & \left ( \sqrt{Y} \right )^{- n - 1/2}.
\labl{2.72c}
\end{eqnarray}
One concludes that the higher cumulants $(n > 4)$
appear to be less significant for the shape of the spectrum in
the hump region $\delta \lapproxeq 1$.

The higher order terms in the series expansion (\ref{2.56a}-\ref{2.56c}) left
out are still numerically sizable at LEP-1 energies
$\,$\cite{lo} ($\sim$ 10\% contribution from 
next-to-MLLA corrections to $\bar \xi$ and $\sigma^2$)
and increase towards lower energies. Therefore, it is appropriate in a
comparison with the data over a larger energy interval to use the full result
%(\ref{ximomk}) 
from the MLLA solution (\ref{2.34}) including the boundary
condition (\ref{2.25}). A further discussion of the higher moments 
is found in the review.$\,$\cite{ko}

\subsection{Predictions for the $\xi$-Spectra}
Next we present analytical results for some interesting limits.

\begin{itemize}
\item[(i)]  {\it Asymptotic Gaussian}
\end{itemize}
%
%\noindent{\it a) Asymptotic Gaussian}\\
%\noindent 
The simplest example is the DLA prediction  %using (\ref{2.29}) with (\ref{2.31})
which corresponds to the spectrum at very high energies. Near the maximum
one finds an approximately Gaussian shape$\,$\cite{dfk1,bcmm}
\begin{equation}
D(\xi,Y) \simeq \frac{\N(Y)}
{\sqrt{2\pi\sigma^2}} \exp \left(-\frac{1}{2}\delta^2\right)
\labl{gauss}
\end{equation}
where
\begin{equation}
\delta=\frac{(\xi-\overline \xi)}{\sigma},\qquad 
\overline \xi=\frac{Y}{2}, \qquad 
\sigma^2=\frac{(Y+\lambda)^{3/2}-\lambda^{3/2}}{12(N_C/b)^{1/2}}
\labl{gausspar}
\end{equation}
with  multiplicity $\N(Y) \sim \exp ( \sqrt{16 N_C (Y+\lambda)/b})$
at high energies as in (\ref{dlaasy}). This approximately Gaussian shape 
(\lq\lq hump-backed plateau'') is an important
prediction of QCD.  The drop of the spectrum towards small momenta 
(large $\xi$) is a consequence of the coherent emission of soft gluons from
the faster ones in the jet and we will come back to this phenomenon in more
detail below. 

\begin{itemize}
\item[(ii)] {\it Limiting Spectrum}  
\end{itemize}
%
%\noindent {\it b) Limiting Spectrum}\\
%\noindent
In perturbation theory one is usually restricted to regions
where  $\alpha_s$ is small, this would require $\lambda\gg 1$. However, 
one can see that in Eq. ({\ref{gauss}) the shape has a smooth limit for
$\lambda\to 0$, i.e. $\alpha_s\to \infty$. Therefore, the shape is
in this sense infrared safe. 
%There is though a logarithmic singularity in
%this limit $\lambda\to 0$ in the total multiplicity $\N$ as discussed before.

In this limit $\lambda\to 0$ one can derive an analytic expression
for the $\xi$ spectrum from the full MLLA equation (\ref{2.34})
using an integral representation for the hypergeometric function $\Phi$,%
$\,$\cite{dt,dkmt2,dkt9}
\begin{eqnarray}
{D}^{\lim}(\xi,Y) & = & \frac{4 N_C}{b} \: \Gamma (B) \;
\int_{-
\frac{\pi}{2}}^{\frac{\pi}{2}} \; \frac{d \ell}{\pi} \: e^{- B
\alpha} \; \left [ \frac{\cosh \: \alpha + (1 - 2 \zeta) \sin
h
\: \alpha}{\frac{4 N_C}{b} \; Y \; \frac{\alpha}{\sinh \:
\alpha}}
\right ]^{B/2} \nonumber \\
& & \nonumber\\
& & \times \; I_B \; \left ( \sqrt{ \frac{16 N_C}{b} \; Y \;
\frac{\alpha}{\sinh \: \alpha} \; [\cosh \: \alpha + (1 - 2
\zeta) \: \sinh \: \alpha ]} \right ).
\labl{2.44}
\end{eqnarray}
Here $\alpha = \alpha_0 + i \ell$ and $\alpha_0$ is
determined by $\tanh \: \alpha_0 = 2 \zeta - 1$ with $\zeta =
1
- \frac{\xi}{Y}$.  $I_B$ is the modified Bessel function of
order $B$.
In the present approximation this distribution describes the gluon
spectrum in a gluon jet. For the quark jet this distribution 
is to be multiplied by $C_F/N_C=4/9$. The spectrum (\ref{2.44})
reproduces the Gaussian behavior near the maximum.

There is one caveat on the distribution (\ref{2.44}). As pointed out
in Sect. 3.2 the multiplicity $\N\sim \int D(\xi)d\xi$ diverges
logarithmically for $\lambda\to 0$. In the MLLA evolution equation the
terms beyond next-to-leading order 
in the $\sqrt{\alpha_s}$ expansions are dropped
which is justified for small $\alpha_s$. The gluon emission near threshold
at $\xi\sim Y$ (small momentum $p\sim Q_0$) involves large $\alpha_s$
and is therefore not expected to be well approximated in (\ref{2.44}).

\begin{itemize}
\item[(iii)]  {\it Distorted Gaussian}
\end{itemize}
%
%\noindent {\it c)Distorted Gaussian}\\ 
%\noindent
The spectrum near the maximum can be represented by a distorted Gaussian
distribution and one finds for small
$\delta\lsim 1$ in terms of the cumulant moments$\,$\cite{fw}
\begin{equation}
{D}(\xi)  \simeq  \frac{\N}{\sigma   
\sqrt{2 \pi}}   \exp  \left [ \frac{1}{8}  k  - 
\frac{1}{2}  s \delta  -  \frac{1}{4}  (2 + k)
\delta^2
 +  \frac{1}{6}  s \delta^3  +  \frac{1}{24}  k
\delta^4 \right ].
\labl{2.54}
\end{equation}
 Relative to the leading-order predictions, the moments in MLLA accuracy
(\ref{2.56a})-(\ref{2.56c}) 
imply that the peak in the $\xi$-distribution is
shifted up (i.e.\ to lower $x$), narrowed, skewed towards lower
$\xi$, and flattened, with tails that fall off more rapidly than a
Gaussian.

\begin{figure*}[tb]%[hbt]
\begin{center}
\mbox{\epsfig{file=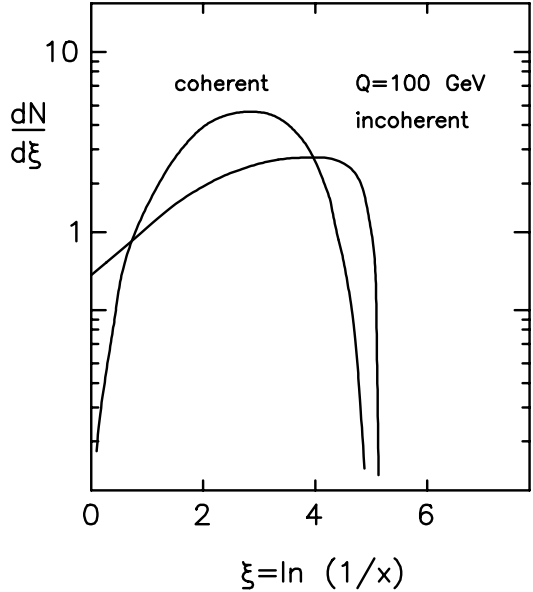,bbllx=3.5cm,bblly=9.7cm,bburx=14.3cm,%
bbury=22.2cm,width=7.cm}} 
          \end{center}
\vspace{-0.3cm}
%\vspace{5cm}
\caption{
Spectrum in $\xi=\ln(1/x)$ as obtained from the MC calculation for a
parton cascade including the soft gluon coherence which shows
 the approximately Gaussian shape. Also shown is
the spectrum for a cascade without coherence.$\,$\protect\cite{marw}}
\label{figmarw}
\end{figure*}

\begin{itemize}
\item[(iv)]  {\it Numerical results}
\end{itemize}
%
%\noindent{\it d) Numerical Results}\\
The evolution of the parton jet in its probabilistic approximation
but taking into account soft gluon interference
can also be treated numerically with Monte Carlo methods. 
In Fig. \ref{figmarw}
we show the $\xi$ spectrum for $e^+e^-$ annihilation at energy $Q=100$ GeV
as obtained from the Marchesini-Webber$\,$\cite{marw} MC which includes coherence
in the soft gluon radiation. One can clearly see the typically Gaussian
shape. On the other hand, if the coherence is \lq\lq switched off''
in the MC
the distribution looses its Gaussian shape and instead peaks near the edge
of phase space for the soft particles ($\xi\lsim Y$, $p\gsim Q_0$).

\begin{itemize}
\item[(v)]  {\it The Soft Limit of the Particle Spectrum}
\end{itemize}
%
%\noindent{\it e) The Soft Limit of the Particle Spectrum}\\
In this limit the coherence of the soft gluon emission plays an important
role. If a soft gluon is emitted from a $q\overline q$ two-jet
system then it cannot resolve with its large wave length
all individual partons but only \lq\lq sees''
the total charge of the primary partons $q\overline q$. 
In the analytical treatment, this property follows from the dominance of the
Born term of $\Ord(\alpha_s)$ and one expects a nearly energy independent
spectrum.$\,$\cite{adkt1} 

This property has been
studied recently in greater detail.$\,$\cite{lo,klo} 
An analytical formula applicable for the low momenta $p$ and $p_T$
has been given and its general behaviour reads
\begin{equation}         
\frac{dn}{dydp_T^2}\ \sim \ C_{A,F} \frac{\alpha_s(p_T)}{p_T^2}
       \left( 1+\Ord\left(  
\ln\frac{\ln (p_T/\Lambda)}{\ln(Q_0/\Lambda)}\
\ln\frac{\ln(p_T/(x\Lambda))}{\ln(p_T/\Lambda)}\right)\right)
\label{Born}             
\end{equation}           
for rapidity $y$ and momentum fraction $x$
where the second term is known within MLLA and vanishes for $p_T\to Q_0$.
 In this way the energy independence of the spectrum follows in the soft
limit.

A crucial prediction$\,$\cite{klo} from this approach is the
dependence of the soft particle density on the color of the primary partons 
in  (\ref{Born}): In the soft limit the particle density
in gluon and quark jets should approach the ratio
\begin{equation}
R(g/q)\to \frac{C_A}{C_F}=\frac{9}{4}\qquad {\rm for}\qquad p_T\to Q_0
\labl{rfin}
\end{equation} 
i.e. for the minimal $p_T$ or $p$. As discussed above, the prediction
(\ref{rfin}) has been obtained in the DLA for the total
multiplicities at asymptotic energies. For the soft particles it is
expected already at finite energies! The consequences have been worked out
also for soft particle production in DIS and hadron hadron
collisions.$\,$\cite{klo} Further results on the soft particle production
are discussed in a recent review.$\,$\cite{wor99} 

\subsection{Comparison with Experimental Data}

We begin with a discussion of the $\xi$ spectra. The observation of the
Gaussian shape of these spectra and a good agreement with MLLA predictions
in the PETRA energy range was a first success of the LPHD
concept.$\,$\cite{adkt1} The approximately Gaussian shape is confirmed in the
meantime for the spectra in jets in the full variety of 
the hard collisions studied and also for different particle species. 
In general, the limiting spectrum provides a fairly good description
 of the data. We
emphasize a few recent results:
\begin{figure*}[tb]%[hbt]
\begin{center}
\mbox{\epsfig{file=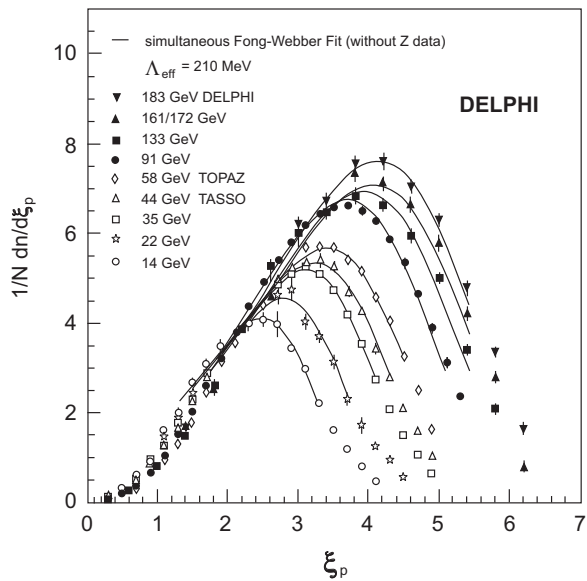,bbllx=1.3cm,bblly=7.cm,bburx=19.0cm,%
bbury=24.5cm,width=8.cm}}  %16.cm}}
          \end{center}
\vspace{-0.3cm}
\caption{
Distribution in $\xi_p=\ln(1/x_p)$ for charged particles
in $e^+e^-\to$ hadrons. The full lines
represent the result of the simultaneous distorted Gaussian fit 
 to all energies using moments within MLLA accuracy$\,$\protect\cite{fw} 
 as performed by the
DELPHI Collaboration.$\,$\protect\cite{delphixi}}
\label{delphihbp}
\end{figure*}

In $e^+e^-$ annihilations the study of spectra has been continued towards the
higher energies of LEP-2. The DELPHI Collaboration$\,$\cite{delphixi} has fitted
the distorted Gaussian parametrization (\ref{2.54}) to their data up to 183 GeV
and those at lower energies$\,$\cite{topaz,tassox} down to 14 GeV
using the moments in next-to-leading order
(\ref{2.50},\ref{2.56a}-\ref{2.56c}). The fit has been performed to the
central region where the distribution was larger than 60\% of  maximum height.
The fitted parameters are $\Lambda=210\pm 8$ MeV, the constant
$\overline a$ in (\ref{2.50}) and the normalization using $n_f=3$.
As can be seen from Fig. \ref{delphihbp} the shape of the hump 
at the different  energies is rather well described.

The limiting spectrum typically fits in a broader $\xi$ range 
and gives a good description of the spectra up to the LEP-1$\,$\cite{opal1},
also to LEP-1.5$\,$\cite{klo}; the recent result from OPAL$\,$\cite{opalmult}
at 189 GeV  
  indicates a slightly broader distribution (about 10\% larger width) 
than expected from the extrapolation of lower energies.

\begin{figure*}[bt]%[hbt]
\begin{center}
\mbox{\epsfig{file=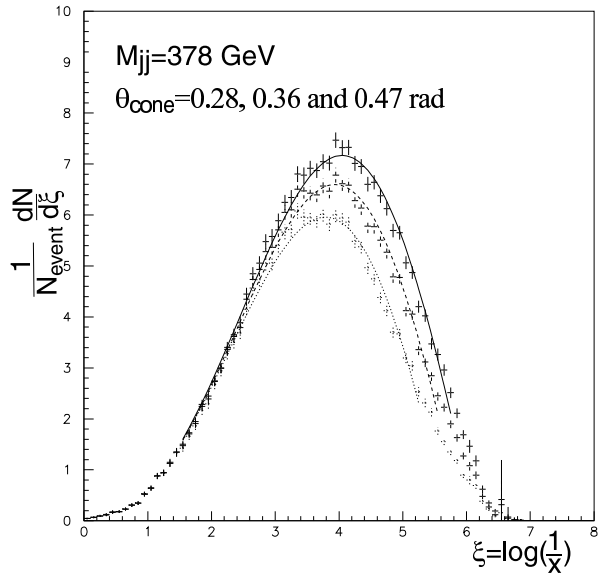,bbllx=1.5cm,bblly=5.cm,bburx=20.0cm,%
bbury=23.2cm,width=8.cm}}  %16.cm}}
          \end{center}
\vspace{-0.3cm}
\caption{
Inclusive  $\xi$ spectrum for charged particles in different 
jet opening cones fitted to the limiting spectrum in MLLA at the dijet mass
of 378 GeV as measured by the CDF collaboration.$\,$\protect\cite{cdfmult}}
\label{figcdfxi}
\end{figure*}

Jets with high transverse energies $E_T$ from 40 to 300 GeV 
in $p\overline p$ collisions  
have been investigated by the CDF Collaboration$\,$\cite{cdfmult} at the
TEVATRON
whereby also the jet opening angle $\Theta$ has been varied between
$\Theta=0.168$ and  $\Theta=0.466$. The fitted value for $\Lambda=Q_0$
in the limiting spectrum was found rather stable $Q_0=240\pm 40$ MeV for the
45 data sets (5 cones $\times$ 9 dijet mass cuts). 
As an example, 
the evolution of the spectrum with the jet opening angle is shown in
Fig. \ref{figcdfxi}. 
The curves are seen to represent the data well within a
few \%. Similar results are obtained at other energies whereby the 
$\xi$ range of the successful fit increases with jet energy and opening
angle.    

\begin{figure*}[bt]%[hbt]
\begin{center}
\mbox{\epsfig{file=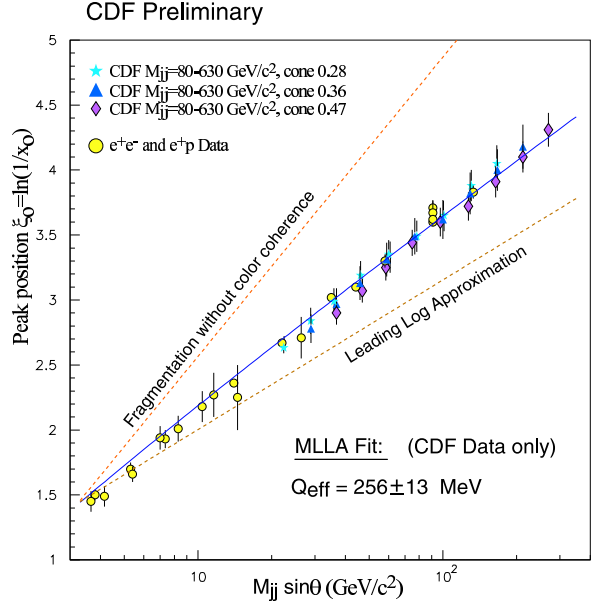,bbllx=0.5cm,bblly=4.2cm,bburx=19.0cm,%
bbury=23.4cm,width=8.cm}}  %16.cm}}
          \end{center}
\vspace{-0.3cm}
\caption{
Peak position $\xi^*$ of the inclusive $\xi$ distribution plotted against
di-jet mass $\times$ $\sin(\Theta)$ in comparison with the MLLA prediction
(central curve); also shown are the double logarithmic approximation (lower
curve with asymptotic slope $\xi^*\sim Y/2$) and expectation from cascade
without coherence. Result by CDF Collaboration.$\,$\protect\cite{cdfmult}}
\label{cdfxist}
\end{figure*}

A comparison of the peak position $\xi^*$ 
for the various jet energies and opening angles is shown in Fig.
\ref{cdfxist}. The spectra depend on the evolution variable
$Y=\ln(P\Theta/Q_0) $ for small angle; a comparison 
is also performed with data at full angle $\Theta=\pi/2$ 
from $e^+e^-$ and $ep$ collisions whereby 
the variable $Y=\ln(P\sin\Theta/Q_0)$ has been used. The data scatter around the
expected curve (\ref{2.51}) for $n_f=3$. Taking instead the scaling variable
$Y=\ln(2P\sin(\Theta/2)/Q_0) $ the full angle data would be shifted to the
right by a factor $2\sin(\pi/4)\sim 1.4$. This would 
correspond essentially to a change of the 
next-to-next-to-leading order term in (\ref{2.51}) but
would not change the slope. 

The slope is nicely confirmed and the leading DLA
contribution ($\xi^*\sim Y/2$) is shown for comparison as the lower curve in
Fig. \ref{cdfxist}, adjusted in height; the upper curve represents 
the spectrum for the incoherent cascade which peaks near the maximum
($\xi^*\sim Y)$. Apparently the data support  the prediction
from the parton cascade with suppression of soft particles due to coherent
gluon emission in a large energy range $2E_{jet}\sin\Theta\sim~4-300$~GeV.

%In the application of the analytical results for  the particle spectrum near
%the soft limit (\ref{Born}) the model    
%(\ref{partonhadron}) for  mass effects has been used. 
The analytical results for the particle spectrum near the soft limit
(\ref{Born}) are nicely confirmed by the data. In these calculations
the model    
(\ref{partonhadron}) for  mass effects has been used. 
The experimental data from the
available range of energies in $e^+e^-$ annihilation ($Q=3-183$ GeV) and 
Deep Inelastic Scattering ($Q<20$ GeV)$\,$\cite{lo,klo,H1plateau,delphixi}
are well described by the calculations. The data are consistent 
with approaching the energy independent limit 
for small momenta $p\to 0$ and this conclusion does not depend on choosing
a particular model for the mass effects. 

It is more difficult to test prediction (\ref{rfin}) as a $gg$ system is not
easily available. From the study of 3-jet events in $e^+e^-$ annihilation
with the gluon recoiling against a $q\overline q$ jet pair an estimate of
this ratio of the soft particle yield in gluon and quark jets 
can be obtained.$\,$\cite{opalca} 
It approaches 
$R(g/q)\sim 1.8$ for $p\lsim 1$ GeV 
which is above the global multiplicity ratio $\sim $ 1.5
in the quark and gluon jets but still
below the ratio $C_A/C_F=9/4$. It is plausible that this 
difference is due to the finite angle between the quark jets,
i.e. a deviation from exact collinearity.$\,$\cite{wor99} 
 It will be interesting to study the radiation pattern in more detail.
A first result from such studies will be discussed in Section 7.

\section{Multiparton Correlations Inside Jets}
\subsection{The General Structure of n-Parton Angular Correlations} 

The angular ordering is one of the most characteristic features 
of the parton cascade, {\em c.f.} Sub-section 2.3. 
  This basic property of the elementary 
emissions has a direct counterpart in the angular dependence of multiparton correlations. 
They were derived in the analytical form by solving corresponding evolution equations
 with the aid of the resolvent (Greens function) method. This led to the 
 integral representations used also in the jet calculus approach.$\,$\cite{kuv} 
 The remaining integrals can be solved in the limit of the strong angular ordering  
 providing a closed form of many multiparton distributions.    
For example, the fully differential angular correlation function of two gluons 
in a jet $(\vec{P},\Theta)$ reads$\,$\cite{ow1,jw} 
\begin{eqnarray}
\Gamma^{(2)}(\Omega_1,\Omega_2)&=&D^{(2)}(\Omega_1,\Omega_2) 
-D^{(1)}(\Omega_1)D(\Omega_2)    \label{Gas}\nonumber\\   
&=& \frac{\gamma_0^2}{2(4\pi)^2} \frac{1}{\vartheta_{12}^2}
\left( \frac{E\vartheta_{12}}{Q_0} \right)^{2\gamma_0}
\left ( \frac{1}{\vartheta_{1}^2}
 \left (\frac{\vartheta_1}{\vartheta_{12}}\right )^{\gamma_0/2}
 \Theta(\vartheta_{1}-\vartheta_{12}) \right.
 \nonumber \\
&+& \left.
\frac{1}{\vartheta_{2}^2}
\left (\frac{\vartheta_2}{\vartheta_{12}}\right )^{\gamma_0/2}
 \Theta(\vartheta_{2}-\vartheta_{12})
 \right), \label{ancorr}
\end{eqnarray}
where $\vartheta_{i}$ denotes the angle between the momentum of i-th final parton
and the momentum of the initial parton $\vec{P}$ and $E=|\vec{P}|$.  
The two terms in (\ref{ancorr}) describe two possible ways to produce the final
configuration, namely
$P\rightarrow 1\rightarrow 2$ and $P\rightarrow 2\rightarrow 1$. 
The relative angle $\vartheta_{12}$ is always restricted by the previous emission
 angle, 
i.e.  $\vartheta_{ij}<\vartheta_i, i=1,2$,  {\em differently} in both terms. 
 The above result was derived as the high energy limit of the DLA with the 
 constant $\alpha_s$. 
 However  the generic  structure  of Eq. (\ref{Gas}) remains the same also in more 
 quantitative approximations and for arbitrary number of partons. 
  This is seen in the general DLA result for n-parton connected correlation function
  also in the running \al\  case

\begin{equation}
\Gamma^{(n)}(\{\Omega\})\simeq
\left({f\over 4\pi}\right)^n {1\over n}
\sum_{i=1}^n {\gamma_0^{n\over 2}(E\vartheta_i)\over \vartheta_i^2}
\exp{\biggl(2\beta \omega(\epsilon_i,n)\sqrt{\ln E\vth_i/\Lambda}\biggr)}
   F_i^n (\{\chi\})\/.
\label{gamom}
\end{equation}
Again, all the relative angles are limited by the corresponding polar angles
of previous emissions.
Here $\epsilon_i=\ln(\vth_i/(\tij)_{{\rm min}})/\ln (E\vth_i/\Lambda)$
and $F_i^n(\{\chi\})$ 
is homogeneous of degree $p=-2(n-1)$ in
the relative angles built from factors $1/\tij^2$. 
Similarly to the two parton case, Eq. (\ref{Gas}), the correlation is a sum of $n$ terms,
 the  $i$-th term describing a configuration with
the parton $i$  emitted  from (connected to) the initial parton at the polar angle $\vth_i$.
  All other partons in such a configuration are  either
connected to $i$ (factor $\tij^{-2}$) or among themselves
(factor $\vth_{jl}^{-2}$). The function $\omega(\epsilon,n)$ is a particular feature 
of the running \al and will be discussed later. For small $\epsilon$ (large angles)
the simple power behavior analogous to Eq. (\ref{Gas}) emerges. 

  The above power dependence of the connected correlations is another characteristic
property of a QCD cascade. From a simple picture of a branching process one may expect 
that the resulting parton distributions have a self similar, fractal structure. 
Indeed, Eq. (\ref{Gas}) proves 
that such expectations are correct --- QCD cascade
is strictly self similar for constant \al  case. 
The evolution of a QCD cascade can be also easily visualized
in a fractal phase space which has been constructed within the
LUND dipole cascade model$\,$\cite{lufractal}.
The {\em R\'{e}nyi dimension}$\,$\footnote{This generalization 
 of the Hausdorff dimension allows the fractal dimension to depend on $n$.
 The Hausdorff dimension reduces to the integer, geometrical dimension
 for non-fractal objects.}  
of
the inclusive n-parton configuration has been derived %$\,$\cite{ow1,ow2,bp,dd} 
in Refs.~109,111,113,114 as
\eq
D_n={n+1\over n }\gamma_0. 
\eqx 
On the other hand, the running \al introduces  a scale in the problem, 
hence the self similarity is only approximate.

Dynamics of QCD provides then a unique prediction for the structure of the 
multiparton correlation. This  distinguishes between various phenomenological
models$\,$\cite{dwk} of multiparticle production. In fact the way how the n-body correlation 
emerges from the two-body ones  is very similar to the model of
 Van Hove proposed ten years ago.$\,$\cite{vh}

\subsection{Distribution over the Relative Angle - Comparison with the Experiment}

Predictions of the type of Eq. (\ref{Gas}) allow  to test the Local Parton Hadron Duality
on a more differential, hence more detailed level. However full multibody correlations of higher order
are notoriously difficult to measure experimentally.\footnote{On the other hand,
 various higher {\em moments} of multiplicity distributions have been measured. 
 We do not review this  subject here.}
 Nevertheless the measurements of particular
sections of the two-body correlations inside a jet already exist and provide 
important checks of the theory.
The distribution over the relative angle $\t12$ is such an object, hence we will discuss
it in more detail. 
Consider two partons inside a jet $(\vec{P},\Theta), E=|\vec{P}|$ at fixed relative angle $\t12$. 
The density of $\t12$ has been 
calculated in various approximations. In the simplest case (same as in Eq.(\ref{Gas})) 
 it reads
\aq
{dN^{(2)}\over d\t12}&\equiv&\int D^{(2)}(\Omega_1,\Omega_2)
\delta(\t12-\Theta_{12})\nonumber \\
  &=&{\ano\over 2\t12}\left( \frac{E\vartheta_{12}}{Q_0} \right)^{2\gamma_0}
                                  \left (\frac{\Theta}{\vartheta_{12}}\right )^{\gamma_0/2}.  \label{rho12fx}
\aqx
The angular ordering condition takes here the form $Q_0/E\ll \t12<\Theta$.  A more complex expression
valid under a weaker  restriction $Q_0/E<\t12<\Theta$	is also available. Again the fractal nature of the
well developed QCD cascade manifests itself via a simple power dependence on the relative angle.
The absolute density, Eq. (\ref{rho12fx}), depends on the cut-off $Q_0$. Therefore it is advisable
 to define normalized, $Q_0$ independent distributions
\begin{equation}
r(\vartheta_{12})=\frac{dN^{(2)}/d \vartheta_{12}} 
                  {dN_{\rm prod}^{(2)}/d\vartheta_{12}}, \;\;\qquad 
\hat r(\vartheta_{12})=\frac{dN^{(2)}/d \vartheta_{12}}
                       {N^2}
\labl{r2}
\end{equation}
In the first ratio, $r$, the normalizing quantity is defined as in
(\ref{rho12fx}) but with $ D^{(2)}(\Omega_1,\Omega_2)$
replaced by $ D^{(1)}(\Omega_1) \;  D^{(1)}(\Omega_2)$;\footnote{
In the experiment one takes the angle $\vartheta_{12}$
between particles of different events (``event mixing").} 
in the second observable, $\hat r$, the square 
of the particle multiplicity $N^2(E,\Theta)$ in the
forward cone is used  as the normalizing factor. 
  DLA predictions with the running \al for these quantities read
  \begin{eqnarray}
 r(\vartheta_{12})&=&\exp \left(\bar{b}
\left(\omega(\epsilon,2)-2\sqrt{1-\epsilon}\right)
    \right)
     \labl{rres},\\
 \hat r(\epsilon)&=&\bar b \exp \left(\bar b
     \left(\omega(\epsilon,2)-2\right)
    \right), \quad \bar b=2\beta\sqrt{\ln(E\Theta/\Lambda)}
     \labl{rhres}
\end{eqnarray}
with $\beta$ as in Eq. (\ref{gamma0}).
Therefore the dependence on the relative angle enters only through
a new scaling variable
\eq
\epsilon={\ln{(\Theta/\t12)}\over\ln{(E\Theta/\Lambda)}}, \;\;\;\;\; 0\leq\epsilon
\leq{\ln{(E\Theta/Q_0)}\over\ln{(E\Theta/\Lambda)}},
\eqx
which uniquely combines the three experimentally controlled parameters
$E,\;\Theta$ and $\t12$. The function $\omega(\epsilon,n)$ is a solution of a simple
algebraic equation hence its behavior is known.$\,$\cite{ow2} (to good approximation
$\omega(\epsilon,n)=n\sqrt{1-\epsilon}(1-(2n^2)^{-1}\ln{(1-\epsilon)}$). 
 In particular, for the developed cascade, i.e. for  $\t12\sim\Theta$, $\epsilon\sim 0$,
 $\omega$ is linear
 \eq
 \omega(\epsilon,n)=n-{n^2-1\over 2 n}\epsilon,
 \eqx
 and we recover the power scaling for normalized correlations
 \begin{equation}
 r(\vartheta_{12}) \simeq 
 \left({\Theta\over\vartheta_{12}}\right)^{{\gamma_0(Q)\over 2}},\qquad
\hat r (\vartheta_{12}) \simeq  {2\gamma_0(Q) \over \vartheta_{12}}
\left({\Theta\over\vartheta_{12}}\right)^{-{3\gamma_0(Q)\over 2}},
  \labl{rpow}
\end{equation}
 which coincides with the constant \al result provided one replaces the fixed
  \al by the running \al at the scale 
 $Q=E\Theta/\Lambda$. 
 
 Both ratios, Eq. (\ref{r2}), have been measured now by DELPHI in $e^+e^-$ 
 and  by ZEUS in the current fragmentation region of DIS.$\,$\cite{delphi12,zeus12}
 The first quantity, $r(\t12)$, 
is more sensitive to the genuine correlations as it
measures the deviations from the distribution of uncorrelated
pairs, but it depends more critically on the choice of the jet
axis in the definition of the angles $\Omega_i$. The
second quantity, $\hat{r}(\t12)$, depends only weakly on the jet axis through the opening
angle $\Theta$, however a large (but not a whole) part of its $\t12$ dependence follows from 
 the simple kinematics of uncorrelated pairs. 
 Experimentally,   finding a jet axis is not a problem for ZEUS where
it is determined by the virtual photon. In DELPHI analysis the sphericity axis
was used.  Scaling in $\epsilon$, revealed in Eqs. (\ref{rres},\ref{rhres}),  
is a strongly constraining
 feature of QCD predictions. In particular, it can be tested by 
 changing {\em both} $\Theta$
and $E$ at fixed $\epsilon$. This is  suitable for DIS kinematics
where the momentum of a recoiled jet in the Breit frame, $P=Q/2$, can be
readily controlled. Hence the ZEUS data span a range of jet energies
$ E  \lsim 30$ GeV and DELPHI provides results for the two highest energies
$E=45.5$ and $E=91.5$ GeV. Both groups use $15^{\rm o} < \Theta < 90^{\rm o} $. 

         ZEUS results for $\hat{r}(\t12)$ in a scaling form are displayed in  
 the left columns of Fig. \ref{zeusfig}. 
Indeed the data show only a residual dependence on $E$ at fixed $\epsilon$.
The shape of the scaling curve is reproduced rather satisfactorily for 
higher jet energies. Data clearly indicate the running of $\alpha_s$. 
It is worth noting that the running observed here is in the energy range
$\sim 1$ GeV which is  substantially lower than in other applications. Yet the 
perturbative description is satisfactory. 
DELPHI results (cf. Fig. \ref{delphifig}a)   %, are consistent with ZEUS and
 confirm the QCD scaling predictions 
on $\hat{r}(\t12)$ at their highest energies as well. 
Both groups report that the $\Theta$ dependence is not strong,
 nevertheless the $\epsilon$  is favored as a scaling variable.   
 
  The situation is more complex with the genuine correlations $r(\t12)$. 
 ZEUS and DELPHI results are shown in the right columns of Fig. \ref{zeusfig} 
 and Fig. \ref{delphifig} respectively. 
 Clearly at low energies ZEUS data have no resemblance to the high energy 
 predictions.
 However a clear transition to the scaling form is seen at higher energies. 

DELPHI high energy data ( between $E=45.5$ and $91.5$ GeV ) indicate a scaling behavior
 for large angles ($\epsilon\lsim 0.25$) albeit below the DLA prediction. 
 Considerable scaling 
 violations persist throughout the full energy range of ZEUS and DELPHI at small angles
 (large $\epsilon$). Apparently, the scale breaking is related to the normalizing  
 $N^{(2)}_{prod}$  in (\ref{r2}). This factor is built from single particle densities 
 $D(\vartheta)$ and strongly depends on the jet axis. This dependence can be
 reduced by measuring the {\em energy-multiplicity-multiplicity}
  correlations$\,$\cite{dkmw,ow3} (see next section). 
 
    Also the theoretical part is based on the rather crude Double Logarithmic 
  Approximation.  Complete phase space, full matrix element and next to
   leading corrections are not included in (\ref{r2}). 
 
  With this in mind we find the present state of the confrontation 
 between theory and experiment rather satisfactory. At the same time we expect that
  it will stimulate further effort of both parties.   

%\clearpage  
\begin{figure}[p]%[hbt]
\begin{center}
\mbox{\epsfig{file=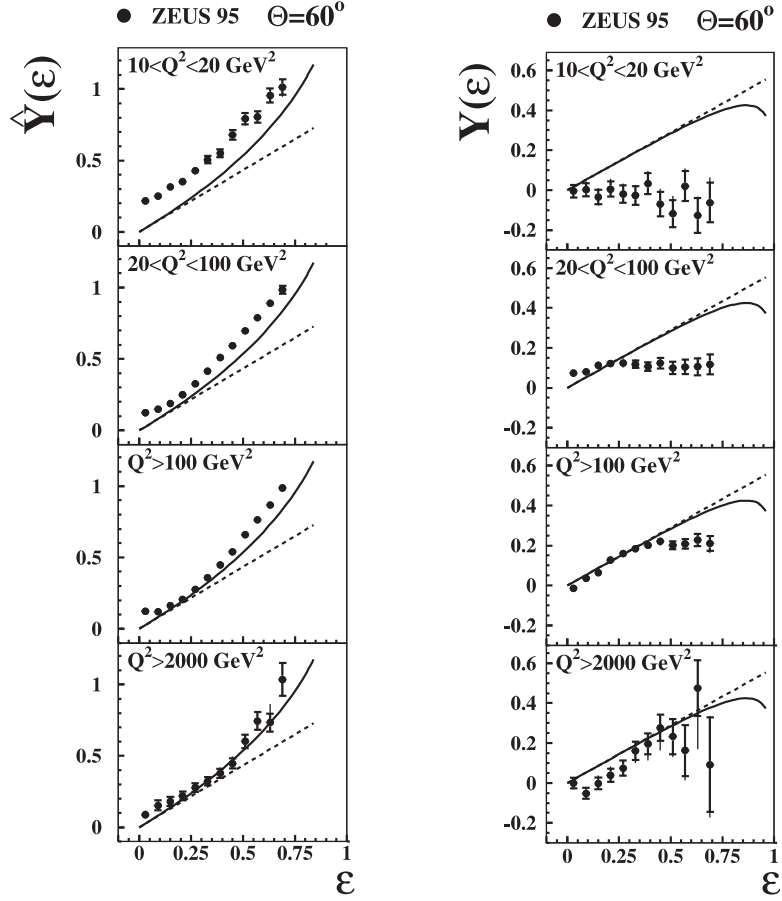,bbllx=2.5cm,bblly=10.cm,bburx=19.5cm,%
bbury=25.cm,width=11.cm}}  %16.cm}}
%\mbox{\epsfig{file=fig51.ps,bbllx=3.cm,bblly=8.5cm,bburx=18.7cm,
%bbury=27.2cm,width=12.cm}} 
          \end{center}
\caption{ZEUS results for the scaling functions
 $\hat{Y}(\epsilon)=-\beta\ln{(\hat{r}/\tilde{b})}/\tilde{b}$ 
 (left) and
$Y(\epsilon)=2\beta\ln{r}/\tilde{b}$ 
(right) for various jet energies $E=<Q>/2$;
 $\tilde{b}=2\beta\sqrt{\ln{(E\sin{\Theta/\Lambda})}}$. Solid lines give the 
scaling QCD predictions, namely  $\omega(\epsilon,2)-2$, 
Eq. (\protect\ref{rhres}) and 
 $\omega(\epsilon,2)-2\sqrt{1-\epsilon}$, Eq.\ (\protect\ref{rres}).  
Dotted line shows 
the constant \al version of the asymptotic forms, Eq. (\protect\ref{rpow}).}
\label{zeusfig}
\end{figure}
%\clearpage
% 
 \begin{figure}[bt]%[hbt]
\begin{center}
\mbox{\epsfig{file=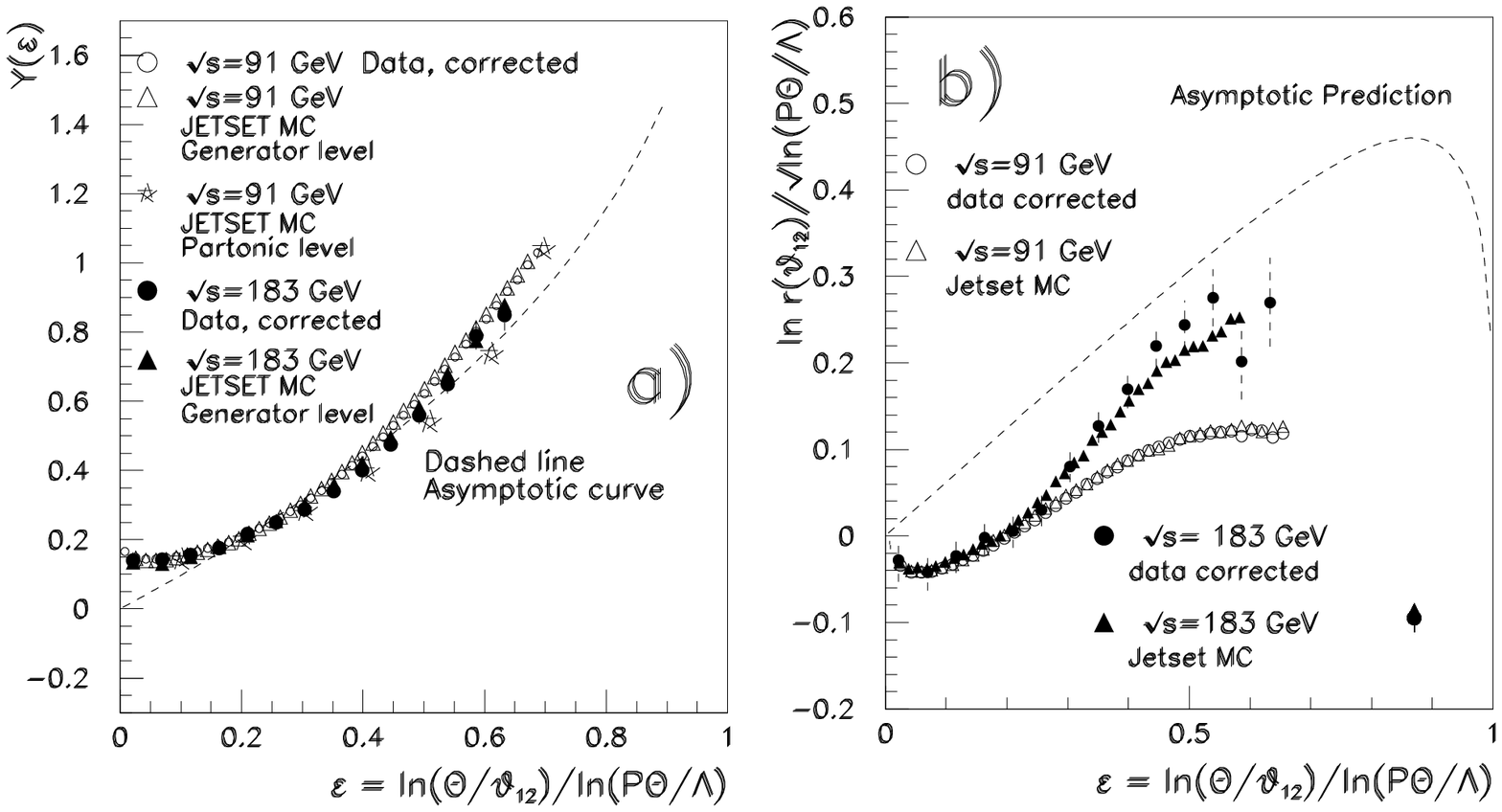,bbllx=2.5cm,bblly=19.cm,bburx=19.5cm,%
bbury=29.cm,width=10.cm}}  %16.cm}}
          \end{center}
\caption{The same scaling functions as in Fig.\ref{zeusfig}, as measured
by the DELPHI Collaboration for two highest jet energies. Results of the 
Monte Carlo simulations are also shown.}  
\label{delphifig}
\end{figure}
  
\subsection{Azimuthal Correlations to the Next-to-Leading Order}
  
         Complete next-to-leading results are available only for the azimuthal correlations 
among gluons emitted from a hard $q\bar{q}$ pair in an $e^+e^-$ 
annihilation.$\,$\cite{dkmw,dmo} This problem is closely related
to the string-drag effect and will be discussed in detail later. 
In short, consider emission of a soft gluon associated with the
hard $q\bar{q}g$ system. Particle density is reduced around the direction opposite to the hard gluon due to the negative 
interference effects. In the limit when the hard gluon becomes softer interference is still destructive and one expects
 similar suppression. 
 
 The lowest order result for the normalized correlation reads in the soft gluon limit
 \eq
 C(\eta,\phi)=1+{N_C\over 2 C_F} {\cos{\phi}\over \cosh{\eta}-\cos{\phi}},
 \eqx
 with $\eta$ and $\phi$ being the rapidity and azimuthal angle differences of two soft 
 gluons.
 
 Theoretical improvements of this formula  involve: i) corrections to the multiplicity flow, 
 ii) corrections from the hard matrix element, and iii) resuming
  $\Ord(\alpha_s \ln^2\theta_{gg})$ 
 terms similar to $\theta_{12}$ singularities considered in the previous section. 
 Moreover,  the sensitivity to the jet axis was eliminated by introducing a concept of 
 the {\em energy-multiplicity-multiplicity} correlations.$\,$\cite{dkmw} 
 Additional weighting  by the energy effectively defines the jet axis as the direction of a fast particle.
 Such a definition is simple to implement experimentally and retains all features of the QCD evolution.

  \begin{figure*}%[hbt]%[hbt]
\begin{center}
%\mbox{\epsfig{file=fig53.ps,bbllx=2.cm,bblly=12.cm,bburx=20.5cm,%
%bbury=22.cm,width=10.cm}}  %16.cm}}
\mbox{\epsfig{file=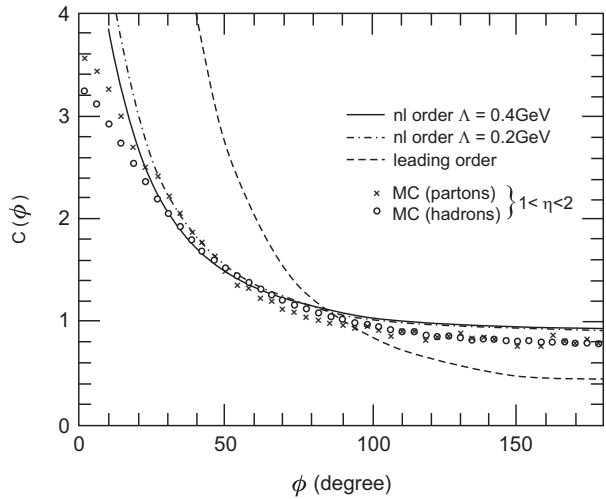,bbllx=3.4cm,bblly=13.cm,bburx=17.8cm,%
bbury=25.5cm,width=8.cm}}  %16.cm}}
          \end{center}
\caption{Azimuthal correlations of partons in a jet near central rapidity $\eta$,
 at leading and next-to-leading
orders for different $\Lambda$ and comparison with the Monte Carlo (HERWIG) 
result$\,$\protect\cite{dmo}. }

\label{fifig}
\end{figure*}
 
 In Fig.\ \ref{fifig} theoretical calculations are compared with the MC results. 
  Indeed the back-to-back configuration ($\phi\sim 180^{\rm o}$) 
 is suppressed. The effect of non leading corrections is rather dramatic especially for the aligned gluons. 
 At $\phi=\pi$ the non leading corrections almost double the leading order result 
 bringing theoretical predictions 
 to within $10\%$ of the MC simulations. Measurements by OPAL and ALEPH have  
 confirmed the above behavior.$\,$\cite{opal,aleph}
 Experiment agrees with Monte Carlo up to $5-10\%$ in the whole 
 range of $\phi$ which provides satisfactory support 
 for higher order calculations and for the LPHD. In particular,
 experimental numbers for the back-to-back (in the azimuth)
 configuration read
 \bea
 C(\pi)&=&0.787\pm 0.002\pm 0.004\;\; (\rm{OPAL}), \nonumber \\
 C(\pi)&=&0.783\pm 0.001\pm 0.016\;\; (\rm{ALEPH}),
 \eea
where the first error is statistical and the second is systematic.
This is to be compared with the MC result $C_{MC}=0.8$
and analytical improved $C_{NL}=0.93$.

% \begin{figure*}[bt]%[hbt]
%\begin{center}
%%\mbox{\epsfig{file=fig53.ps,bbllx=2.cm,bblly=12.cm,bburx=20.5cm,%
%%bbury=22.cm,width=10.cm}}  %16.cm}}
%\mbox{\epsfig{file=fig53.ps,bbllx=3.4cm,bblly=13.cm,bburx=17.8cm,%
%bbury=25.5cm,width=12.cm}}  %16.cm}}
%          \end{center}
%\caption{Azimuthal correlations of partons in a jet near central rapidity $\eta$,
% at leading and next-to-leading
%orders for different $\Lambda$ and comparison with the Monte Carlo (HERWIG) 
%result.$\,$\protect\cite{dmo} }
%
%\label{fifig}
%\end{figure*}

  \subsection{The Critical Angle and the Local Parent Child Correspondence}
    
 The success of perturbative QCD in describing hard processes only enhances
our curiosity and the need to understand soft phenomena as well. 
It is  an experimental fact that many properties of hadronic final states
are well described by equivalent partonic observables also at the scales close to, or even beyond,
the standard range of applicability of perturbative QCD.  This is the foundation of the
Local Parton Hadron Duality hypothesis, which still inspires many theoreticians. Is it possible, that some
properties of the soft, confining interactions are already hidden in the perturbative cascade? The only example
of such a mechanism is provided by the pre-confinement scenario 
in early days 
of perturbative QCD.$\,$\cite{av} In this section 
we would like to discuss another intriguing property of the perturbative
cascade. Namely there exist a small critical angle
which is still within the range of perturbative calculations,  and below which the 
distribution of next generation of partons
is {\em equal} to the distribution of their parents.$\,$\cite{ow3} This precise coincidence of 
the distributions of partons from 
{\em different generations}, which we call
 the Local Parent Child Correspondence, has some interesting consequences.
    
    Consider again the  distribution of the relative angle
$D^{(2)}(\th12,P,\Theta)$ discussed earlier.  In the double logarithmic
approximation this correlation function obeys the integral equation which can be derived
from the evolution equation (\ref{dlaz}).$\,$\cite{ow2}
\begin{equation}
D^{(2)}(\theta_{12},P,\Theta)=d\upt(\theta_{12},P,\Theta)
     + \int_{Q_0/\theta_{12}}^{P}
\frac{dK}{K} \int_{\theta_{12}}^{\Theta} \frac{d\Psi}{\Psi}
  \a2(K\Psi) D\upt(\theta_{12},K,\Psi).
\labl{rho2}
\end{equation}
 The inhomogenous
term $\d2$ is constructed from the product of 1-particle angular
distributions and found as
$d^{(2)}(x)=D^{(1)}(x)
\overline{N}(x),\;\; x=P\th12/\Lambda$ and
with the multiplicity $\overline{N}$ in the cone
of half-angle $\th12$ around the primary parton axis. In exponential
accuracy $d^{(2)} \sim \overline{N}^2$. 

The  solution of Eq. (\ref{rho2}) can be written as$\,$\cite{ow2}
\begin{equation}
D^{(2)}(\th12,\Theta,P)=
%d_2\left({P\th12\over\Lambda}\right)+
\int_{Q_0/\th12}^{P} \frac{dK}{K}
%\overline{
R
%}
\left({P\over K}, {\Theta\over\th12},
 {K\th12\over\Lambda} \right)
\d2\left({K\th12\over\Lambda}\right).
 \labl{res}
\end{equation}
The resolvent
$
%\overline{
R
%}
\left({P\over K}, {\Theta\over\th12},
 {K\th12\over\Lambda} \right)$ is nothing but the momentum distribution of the  
parent partons $K$ in the jet $(P,\Theta)$ with the condition, that 
their virtuality is bound from below by $K\th12$ and {\em not} by the
elementary cut-off $Q_0$. This is equivalent to restricting their emission
angle to be greater than $\th12$. 

It turns out that this distribution is non analytic
  at some small value of the relative angle
$\th12=\thc$. To see this  consider the dependence of the integrand
of Eq. (\ref{res}) on the parent parton momentum $K$. The interplay
between the decrease of the parent density $\overline{R}$ and the increase
of the children densities $\d2$ with $K$ produces a sharp maximum 
which is however
$\th12$ dependent. The important element 
of the running \al analysis is that the position of this maximum is {\em
independent of} $Q_0$, since both distributions do not depend on $Q_0$.
Now, for small relative angles the above maximum shifts {\em below}
the lower limit $Q_0/\th12$ of the momentum integration. At that point
the result changes non analytically (albeit very gently --- analogously to the
second order phase transitions in statistical systems). Namely, for
$\th12$ bigger than $\thc$, $\rho^{(2)}$ is given (  to the exponential
accuracy) by the $Q_0$ independent value of the integrand at the
maximum. For $\th12$ below $\thc$, however, the result is given by the
value of the integrand {\em at the lower bound} and obviously depends on
$Q_0$. Moreover, since $d_2(1)=1  $ in our exponential approximation, 
the density of pairs below
$\thc$ is {\em given entirely by the density of parent partons} at the
minimal momentum $K=Q_0/\th12$  
\eq
\rho^{(2)}(\t12)\sim R\left({P\t12\over Q_0},{\Theta\over\t12},{Q_0\over\Lambda}\right) .   \label{lpcd}
\eqx
The value of the critical angle derived from these considerations reads$\,$\cite{ow3}
\eq
\thc={Q_0\over P} \left( P\Theta\over Q_0\right)^{1/5}, \;\; 
\thc={Q_0\over P} \left({\ln{ (P\Theta/ \Lambda)}\over\ln{(Q_0/\Lambda)}}\right)^{4/9}, 
\eqx
for the constant and running \al respectively. The transverse momentum associated
with this angle grows weakly with the jet momentum $P$ hence the perturbative
description applies also in the new region  $ Q_0/P < \t12 < \thc $. 
Even though some features of this
result (e.g. the emergence of a sharp singularity) would be modified in more 
refined approximations, the 
existence of the two angular regions with different properties is genuine. 
In the large angle region,
the distribution scales in $\epsilon$ (see previous section) and is independent of $Q_0$.
 On the other hand, below $\thc$ the distribution depends on $Q_0$, which can be 
 interpreted as the sensitivity to the soft interactions where the hadronization
 and confinement start becoming important. 
 
 The numerical value of the critical angle
  is rather small at high energies, $\thc\lsim 1^{\rm o}$ at LEP energies. 
It is interesting to note that the scaling in $\epsilon$ is indeed best satisfied
at large angles  ($\vartheta_{12} \gg \vartheta_c$), see Fig. \ref{delphifig},
 which is consistent with this analysis.
 
  Sensitivity to the cut-off $Q_0$ at small angles leads to an interesting prediction 
  for the particle mass 
  dependence of the produced pair. Assuming that the particle mass acts as an 
  effective transverse momentum cutoff, we expect that the spectrum of pairs 
  with large relative angles is
 mass independent, while for angles smaller than $\thc$ the density depends 
 on the mass  and is suppressed for higher masses.

% It is therefore tempting to think of a partonic pair
%with  $\t12<\thc$ as a prototype of a hadron and consider the relation, Eq.(\ref{lpcd}) 
%as a perturbative counterpart of the LPHD. 

\section{On Inclusive Properties of Heavy Quark Jets}
Until now we have discussed mainly the properties of jets in $e^+ e^-$ collisions obtained 
by analyzing the full hadron data sample with contributions from all flavors.  In the last 
few years intensive experimental studies have been performed with the heavy-quarks $Q (c, b)$.  
Experiments at LEP and SLC have produced a wealth of new interesting results on the profiles of 
jets initiated by heavy quarks.  The accuracy of measurements started to be comparable with that 
in the $q\bar{q}$ events.  This is mainly related to the steady improvement in the heavy-quark 
tagging efficiencies.  Further progress is expected from the measurements at hadronic colliders 
and a future linear $e^+ e^-$ collider.  The principal physics issues of these studies are 
related not only to testing the fundamental aspects of QCD, but also to their large potential 
importance for measurements of heavy-particle properties:  lifetimes, spatial oscillations of 
flavor, searching for CP-violating effects in their decays etc.

A good understanding of $b$-initiated jets is of primary interest for analysis of the 
final-state structure in $t\bar{t}$ production processes.  A detailed knowledge of the $b$-jet 
profile is also essential for the Higgs search strategy.

The physics of heavy-quarks has been traditionally considered as one of the best testing 
grounds for QCD.  Note that since the mass of $b$-quark is much larger than the QCD scale 
$\Lambda$, the $b$-quark fragmentation proves to be an especially useful tool for studying
the 
perturbative scenario.  In this section we will restrict ourselves to the discussion of some specific 
features of heavy-quark-initiated events which are related to our previous considerations of 
multiparticle production and LPHD.

\subsection{QCD Bremsstrahlung Accompanying Heavy Quark Production}

When examining the perturbatively based structure of the heavy-quark jets, one has to 
understand first how the development of the parton cascades generated by $Q$ depends on 
the quark mass $M$.  It was demonstrated$\,$\cite{dfk1,dkt7,dkt3,dkt12} that the difference in 
many properties of hadronic jets initiated by heavy quarks (excluding the products of the weak 
decay of the heavy quark iteself), from that of light quarks, originates from the restriction 
of the phase space available to gluon radiation associated with the kinematic effects of the 
heavy-quark mass.

As well known, the radiation pattern of the primary soft gluon with energy $\omega$ from a 
massive relativistic quark with energy $E_Q \gg M$ and small emission angle $\Theta \ll 1$ 
is given by
\be
\label{eq:a1}
d \sigma_{Q \rightarrow Q + g} \; = \; \frac{\alpha_S}{\pi} \: C_F \frac{\Theta^2 
d \Theta^2}{(\Theta^2 + \Theta_0^2)^2} \: \frac{d \omega}{\omega},
\ee
where
\be
\label{eq:a2}
\Theta_0 \; = \; \frac{M}{E_Q}.
\ee
\ From Eq.~(\ref{eq:a1}) one concludes that the large double-logarithmic contribution comes 
only from the region of relatively large radiation angles
\be
\label{eq:a3}
\Theta \; \gg \; \Theta_0,
\ee
where emission becomes insensitive to $M$ and appears to be identical to that for the 
light $q$-jet
\be
\label{eq:a4}
d \sigma_{Q \rightarrow Q + g} \; = \; \frac{\alpha_S}{\pi} \: C_F \: 
\frac{d \Theta^2}{\Theta^2} \: \frac{d \omega}{\omega}.
\ee
For the region
\be
\label{eq:a5}
\Theta \; < \; \Theta_0
\ee
the angular integration is no longer logarithmic.

As it follows from (\ref{eq:a1}) the forward gluon radiation is
suppressed.$\,$\cite{dkt3,dkt12}  
This phenomenon is characteristic for bremsstrahlung off a massive particle.  It reflects the 
conservation of the projection of the total angular momentum on the massive fermion momentum:  
the soft radiation does not change the quark helicity and the forward emission is forbidden (in 
analogy to the well-known \lq\lq $0-0$ transition\rq\rq\ phenomenon).  As it is discussed 
below, after taking into account parton cascades generated by a primary gluon, the relative 
yield of final particles due to this part of phase space is ${\cal O} (\sqrt{\alpha_S})$.  
Since the main logarithmic contribution is absent and the differential cross section 
(\ref{eq:a1}) vanishes in the forward direction, it is natural$\,$\cite{dkt7} to call this region the 
\lq\lq dead cone\rq\rq$\,$\cite{} --- a relatively depopulated cone around the quark 
direction with an opening angle $\Theta \sim \Theta_0$.

Notice that the standard expression for the soft bremsstrahlung spectrum
\be
\label{eq:a6}
d \sigma_{Q \rightarrow Q + g} \; = \; \frac{2 C_F \alpha_S}{\pi} \: 
\frac{v^2 \sin^2 \Theta}{(1 - v^2 \cos^2 \Theta)^2} \: \frac{d \omega}{\omega} \: d \cos \Theta
\ee
with
\be
\label{eq:a7}
v \; = \; \sqrt{1 - \frac{M^2}{E_Q^2}}
\ee
describes adequately the non-logarithmic region of finite emission angles $\Theta \sim 1$.  
This part of the radiation phase space contributes as ${\cal O} (\sqrt{\alpha_S})$ too.  
Corresponding corrections will be taken into full account later when we derive formulae for 
particle multiplicities outside the scope of the double logarithmic 
approximation.

Recall that the angular pattern in (\ref{eq:a6}) is exactly the same as in the QED case of 
classical photon bremsstrahlung off the massive electric charge.$\,$\cite{ll}

As we mentioned above, the structure of the primary gluon radiation at large angles 
(\ref{eq:a3}) appears to be identical to that for the light $q$-jet.  The same holds true for 
the internal structure of secondary gluon subjects (angular-ordered cascades).

\subsection{Specific Features of Events Containing Heavy Quarks}

As was discussed in detail in the previous sections, the dominant role of the perturbative 
dynamics has been very successfully tested in studies of light-hadron distributions in 
QCD jets.  Within the LPHD approach the quark mass-induced restriction on the gluon 
radiation leads to a number of important consequences for the $Q$-quark jets.
\begin{itemize}
\item[(i)] {\it Explicit visualization of the dead cone} 
\end{itemize}
\ From the measurements of the \lq\lq companion\rq\rq\ particle distributions inside fixed 
cones around the $Q$ direction one can learn about the expected depletion of the forward-particle 
production for angles $\Theta < \frac{M}{E_Q}$.  Note that a study of the forward hadroproduction 
in $Q$-jets which is free from \lq\lq trivial\rq\rq\ perturbative bremsstrahlung effects, may 
help in the investigation of the subtle features of some non-trivial confinement scenarios.

The first comparison of the primary particle angular distribution in the $b$ and light-quark 
jets in $e^+ e^-$ annihilation has been reported by DELPHI.$\,$\cite{vpn}  In the data analysis 
special care has been taken in order to exclude the decay products of the $b$-hadrons from 
consideration.  Preliminary results look quite encouraging.  However further detailed studies 
are needed for establishing the perturbative-QCD origin of the observed depopulation of the 
forward-particle production in the $b$-quark initiated jets.

\begin{itemize}
\item[(ii)] {\it Leading particle effect in heavy-quark fragmentation}
\end{itemize}
Suppression of small angles restricts emission of hard gluons with relatively small transverse
momenta 
$k_\perp$, where radiation is normally most intensive as being proportional to $\alpha_S (k_\perp^2)$.  
This results in a decrease in the energy fraction that the heavy-quark $Q$ is sharing with the secondary 
bremsstrahlung particles.$\,$\cite{dkt1,dkt12,dkt7}  Thus, the inclusive distribution of $Q$ is 
expected to peak near $x_Q \approx 1$, the so-called leading particle effect, which first was predicted 
within the framework of the parton model$\,$\cite{afk} and quantified as the so-called Peterson 
fragmentation.$\,$\cite{cpm6}  Recall that at the hadronization stage the 
heavy-quark loses only a momentum 
fraction of order $\Lambda/M$ when forming a heavy-light hadron.$\,$\cite{afk}  The leading heavy 
particle phenomenon has been clearly observed experimentally in the $b\bar{b}$ and $c\bar{c}$ events 
in $e^+ e^-$ annihilation.$\,$\cite{sld1,delphi6}  As expected, at the $Z^0$ the $b$-quark inclusive 
spectra peak very close to one, around $x_Q \simeq 0.8-0.9$.

The record accuracy in determination of the $b$-quark spectrum has been achieved in the recent 
SLD measurement.$\,$\cite{sld1}  As a result, this allows to discriminate between current parton-shower 
plus hadronization models.

As discussed in detail,$\,$\cite{dkt3} within the LPHD scenario the effects of perturbative gluon 
radiation allow to reproduce the shape of the heavy-quark inclusive spectrum provided the 
perturbative prediction is extrapolated smoothly down to the region of small gluon transverse momenta 
within the universal low-scale $\alpha_S$ approach.  As well known such approach proves to be quite 
successful in the description of the jet event shape, for a discussion, see
for example, Ref.~13   %$\,$\cite{power} 
and the article by V. Braun and M. Beneke in this book.

  The Sudakov form-factor suppression of the 
quasi-elastic region at $x_Q \rightarrow 1$ results in the distribution which is qualitatively similar 
to the parton-model-motivated Peterson fragmentation function.$\,$\cite{cpm6}  Note that all-order 
resummation of the collinear and Sudakov logs have been considered also.$\,$\cite{mn}  It is 
worthwhile to mention, that as far as the 
heavy-quark is concerned, based on LPHD one may expect that a purely perturbative treatment is dual to 
the sum over all possible hadronic excitations.  Therefore, without involving any phenomenological 
fragmentation function at the hadronization stage, one could attempt to describe the energy fraction 
distribution averaged over heavy-flavored hadron states, the mixture that often appears experimentally.  
Such purely perturbative approach is at least free from the 
problem of \lq\lq double counting\rq\rq\ which one may face when trying to combine the effects 
of perturbative and hadronization stages.

\begin{itemize}
\item[(iii)] {\it Average multiplicity of events containing heavy quarks}
\end{itemize}
As we discussed above, a forward suppression of soft-gluon radiation off an energetic massive quark 
$Q$ induces essential differences in the structure of the accompanying radiation in light- and 
heavy-quark-initiated jets.  According to the LPHD concept, this, in particular, should lead to corresponding 
difference in companion multiplicity of radiated light hadrons.$\,$\cite{dkt1,bas,dkt7}

It is a direct consequence of the perturbative approach that the difference of companion average 
multiplicities of hadrons, $\Delta N_{Qq}$ from equal energy (hardness) heavy- and light-quark 
jets should be energy-independent (up to power correction terms ${\cal O} (M^2/W^2)$.  This constant is 
different for $c$ and $b$ quarks and depends on the type of light hadron under study (e.g.\ all charged, 
$\pi^0$, etc.).  This is in marked contrast with the prediction of the models based on the idea of reduction 
of the energy scale,$\,$\cite{pcr} $N_{Q\bar{Q}} (W) = N_{q\bar{q}} ((1 - \langle x_Q \rangle) W)$, so that 
the difference of $q$- and $Q$-induced multiplicities grows with $W$ proportional to $N(W)$.  We shall 
elucidate this bright prediction of the perturbative scenario$\,$\footnote{The discussion of companion 
multiplicity presented below was borrowed from the paper by Yu.L. Dokshitzer, S.I. Troyan and one of the 
authors (VAK) which has never been published.}.

Recall, that companion multiplicity is an infrared-sensitive quantity dominated by the emission of 
relatively soft gluons.  In this domain the only difference between heavy- and massless-quark cases comes  
from the suppression of soft bremsstrahlung in the forward direction $\Theta_{1+} \lapproxeq \Theta_0 \equiv 
M/E_Q$, where $\Theta_{1+}$ is the angle between a primary gluon $(g_1)$ and the quark $(+)$.  In the 
region of large gluon radiation angles $\Theta_{1+} \gg \Theta_0$, the finite-mass effects are power-suppressed 
and do not affect the picture of strictly angular-ordered evolution of $g_1$ as a secondary jet.  It is 
straightforward to verify that within the MLLA accuracy the finite mass induces only a small integral 
correction to the angular-ordered prescription for the next-generation gluon $g_2$.  As a result, in the 
large-angle region the internal structure of secondary-gluon jets is identical to that for the light-$q$ case, 
and the emission angle $\Theta_{1+}$ should be taken as an evolution parameter to restrict the subsequent 
cascading.  As usual, the smaller the radiation angle $\Theta_{1+}$, the less populated with offspring 
partons the gluon subjet $g_1$ is.

The situation changes, however, when $\Theta_{1+}$ becomes smaller than $\Theta_0$.  The dead cone region 
gives a sizeable (though nonleading) contribution, of order $\sqrt{\alpha_S}$, and should be taken into 
account within the MLLA.  Here the opening angle of the jet $g_1$ \lq\lq freezes\rq\rq\ at the value 
$\Theta_0$ and no longer decreases with $\Theta_{1+} \rightarrow 0$.  The reason for this is rather simple.  
Normally, in the \lq\lq disordered\rq\rq\ angular kinematics $\Theta_{21} \geq \Theta_{1+}$ the 
destructive interference between emission of a soft gluon $g_2$ by quark and $g_1$ cancels the independent 
radiation $g_1 \rightarrow g_2$.

Meantime, in the massive-quark case the interference contribution enters the game only when $\Theta_{2+} > 
\Theta_0$, so that the cancellation of the independent $1 \rightarrow 2$ term inside the $\Theta_0$-cone 
does not occur.  In physical terms what happens is the loss of coherence between $+$ and $1$ as emitters 
of the soft gluon 2 due to accumulated longitudinal separation $\Delta z_{1+} > \lambda_{||}^{(2)} \approx 
\omega_2^{-1}$ between massless and massive charges $(v_1 = 1, v_+ \approx 1 - \Theta_0^2/2 < 1)$.  Indeed, 
during the formation time of $g_2, t_f^{(2)} \sim 1/\omega_2 \Theta_{21}^2$ the quark and the gluon 1 separate 
in the longitudinal direction by
\be
\label{eq:a8}
\Delta z_{1+} \sim t_f^{(2)} \: |v_+ - \cos \Theta_{1+}| \: \sim \: \lambda_{||}^{(2)} \cdot (\Theta_{1+}^2 
+ \Theta_0^2) \biggr / \Theta_{21}^2.
\ee
It is the last factor which determines whether interference is essential or not.  When this ratio is 
larger than unity, $Q$ and $g_1$ are separated enough for $g_2$ to be able to resolve them as two 
individual classical charges.  In these circumstances gluon $g_1$ acts as an independent source of the 
next-generation bremsstrahlung quanta.  Otherwise, no additional particles triggered by $g_1$ emerge on 
top of the yield determined by the quark charge (which equals total color charge of the $Qg_1$ system).  
In the massless limit $(\Theta_0 = 0)$ this reproduces the standard angular-ordered picture, $\Theta_{21} < 
\Theta_{1+}$.  In the massive case one concludes from (\ref{eq:a8}) that the upper limit of the relative 
gluon angle remains finite when $\Theta_{1+}$ falls inside the dead cone:  $\Theta_{21} < \Theta_0$ for 
arbitrarily small $\Theta_{1+} \ll \Theta_0$.

Thus, the proper evolution parameter for subsequent parton cascading of the primary gluon $g_1$ 
(generalization of an \lq\lq opening angle\rq\rq\ of jet 1) may be chosen as
\be
\label{eq:a9}
\tilde{\Theta}_{1+}^2 \; \equiv \; \Theta_{1+}^2 \: + \: \Theta_0^2.
\ee
Another comment is in order concerning generalization of the argument of the running coupling to the 
massive-quark case.  Here again the substitution similar to (\ref{eq:a9}) is applicable.  Namely, for the 
effective coupling that determines the probability of the emission $Q \rightarrow Q + g_1$ one has 
to use as an argument$\,$\cite{yia2}
\be
\label{eq:a10}
k_\perp^2 \; = \; \omega_1^2 \left [ \left (2 \sin \frac{\Theta_{1+}}{2} \right )^2 \: + \: \Theta_0^2 \right ] 
\; \approx \; \omega_1^2 \: \tilde{\Theta}_{1+}^2.
\ee

Within MLLA the expression for the multiplicity of light particles accompanying the production of a heavy-quark 
pair, $N_{Q\bar{Q}}$, can be obtained by convoluting the probability $dw$ of a single gluon bremsstrahlung 
off a heavy-quark $Q$ with the parton multiplicity initiated by the gluon subjet with the hardness parameter 
$k_\perp$.

The differential first-order cross section for $e^+ e^- \rightarrow Q\bar{Q}g$ integrated over the angles 
of the $Q\bar{Q}g$ system was first presented by Ioffe$\,$\cite{bli} (QED first-order results can 
be found elsewhere$\,$\cite{vnb}).  For the case of vector $Q\bar{Q}$-production current the differential over the 
gluon energy fraction $z$
\be
\label{eq:a11}
z \; = \; \frac{\omega_1}{E}, \quad\quad 2E \; = \; W \; \equiv \; \sqrt{s}
\ee
gluon emission probability can be written as$\,$\cite{dkt3,dks}
\be
\label{eq:a12}
dw_V \; = \; \frac{C_F \alpha_S}{\pi v} \: \frac{dz}{z} \: \frac{d\eta}{\sqrt{1 - \eta}} \left \{ 
2 (1 - z) \: \frac{\eta - \eta_0}{\eta^2} \: + \: z^2 \left [ \frac{1}{\eta} \: - \: \frac{1}{2} \right ] 
\: \zeta_V^{-1} \right \}.
\ee
Here
\bea
\label{eq:a13}
1 \: \geq \: \eta & = & 1 - \beta^2 \cos^2 \Theta_C \: \geq \: \eta_0 \: = \: \frac{4m^2}{1 - z}; \quad\quad 
m \; \equiv \; M/W \; \ll \; 1, \nonumber \\
& & \nonumber \\
\zeta_V & = & (3 - v^2)/2 \; = \; 1 \: + \: 2 m^2 
\eea
with $\beta$ the quark velocity and $\Theta_C$ the polar gluon angle in the $Q\bar{Q}$ rest frame,
\be
\label{eq:a14}
\beta^2 \; = \; \beta^2 (z) \; = \; 1 \: - \: \frac{4m^2}{1 - z} \; \leq \; v^2 \; = \; 1 \: - \: 4m^2 \; 
\geq \; z.
\ee
The first term in curly brackets contains the main (double logarithmic) contribution and corresponds to 
the universal soft gluon bremsstrahlung.  As it follows from the celebrated Low soft bremsstrahlung 
theorem,$\,$\cite{fel}
 extended to the fermion case by Burnett and Kroll,$\,$\cite{thb} both $z^{-1}$ and $z^0$ parts of the 
radiation density have classical origin and are therefore process 
(and quark spin) independent.$\,$\cite{dks} 
 This first term explicitly exhibits the dead cone behaviour:  forward \lq\lq soft\rq\rq\ 
radiation vanishes, $\eta \rightarrow \eta_0$ when $\sin \Theta_C \rightarrow 0$.  The second term which is 
proportional to $z^1$ (hard gluons) depends, in principle, on the $Q\bar{Q}$ production mechanism.  Namely, 
both $- \frac{1}{2}$ subtraction term and the $\zeta$-factor may be different for the different production 
currents.$\,$\cite{dks,dkt3}  This indicates that already at the level of $\alpha_S$ corrections 
the average companion multiplicity acquires a process-dependent contribution from the three-jet ensembles and, 
thus, cannot be treated any more as an intrinsic characteristic of the $Q\bar{Q}$ system.

Neglecting now the corrections of relative order $\alpha_S$ and $m^2$ the average companion multiplicity 
$N_{Q\bar{Q}} (W)$ can be given by
\bea
\label{eq:a15}
N_{Q\bar{Q}} (W) & = & \frac{C_F}{\pi} \: \int_{0}^{v^2} \: \frac{dz}{z} \: \int_{\eta_0}^1 \left \{ 
\frac{d \eta}{\eta} \: \frac{1 + (1 - z)^2}{\sqrt{1 - \eta}} \right . \nonumber \\
& & \nonumber \\
& & - \; \left . \frac{\eta_0 d \eta}{\eta^2} \: 2 (1 - z) 
\right \} \: \left ( \alpha_S (k_\perp) \: N_G (k_\perp) \right )
\eea
with $k_\perp^2 = (zW/2)^2 \eta$.  Note that the second term in the curly brackets is non-logarithmic 
in $\eta$ and generates next-to-leading correction $\delta_1 N \sim \sqrt{\alpha_S (M)} N(M)$, see below.

The main contribution to $N_{Q\bar{Q}}$ comes from the DL phase-space region.  Meanwhile, the expression 
(\ref{eq:a15}) keeps track of significant SL effects as well, provided that the multiplicity factor $N_G$ 
is calculated with the MLLA accuracy.

Introducing a convenient variable
\be
\label{eq:a16}
\kappa^2 \; = \; \frac{E^2}{\omega_1^2} \: k_\perp^2 \; = \; E^2 \left [ \left ( 2 \sin \: \frac{\Theta_{1+}}{2} 
\right )^2 \: + \: \Theta_0^2 \right ] ,
\ee
one can rewrite Eq.~(\ref{eq:a15}) as
\be
\label{eq:a17}
N_{Q\bar{Q}} (W) \; = \; 2 \int_{M^2}^{W^2} \frac{d \kappa^2}{\kappa^2} 
\left [ 1 - \frac{M^2}{\kappa^2} \right ] 
\: \int_{Q_0}^\kappa \; \frac{dk_\perp}{\kappa} \;\:P_q^g(z) % \Phi_F^G (z) 
\; \frac{\alpha_S (k_\perp)}{2 \pi} \: N_G (k_\perp) , 
\ee
with $k_\perp = z \kappa$.  Here $ P_q^g$   %$\Phi_F^G (z)$ 
stands for the standard DGLAP kernel and $Q_0$ denotes a transverse 
momentum cut-off as used in the previous sections.  In the massless limit $M \lapproxeq Q_0$ the contribution of 
the $M^2/\kappa^2$ term vanishes and (\ref{eq:a17}) reproduces the known result for the $q$-jet multiplicity.  
The integration can be performed with the use of the relation corresponding to the evolution equation for 
light-quark jet multiplicity:
\be
\label{eq:a18}
N_q^\prime (Q) \; \equiv \; \frac{\partial}{\partial \ln Q^2} \: N_q (Q/2) \; = \; \int_{Q_0}^Q \: 
\frac{dk_\perp}{Q} \: P_q^g(z)   %\Phi_F^G (z) 
\: \frac{\alpha_S (k_\perp)}{2 \pi} \; N_G (k_\perp),
\ee
with
\be
\label{eq:a19}
N_{q\bar{q}} (Q) \; = \; 2N_Q \: \left ( \frac{Q}{2} \right ),
\ee
see Section~3.  Then Eq.~(\ref{eq:a17}) takes the form
\be
\label{eq:a20}
N_{Q\bar{Q}} (W) \; = \; N_{q\bar{q}} (W) \: - \: 2N_q \left ( \frac{M}{2} \right ) \: - \: 2N_q^\prime 
\left ( \frac{M}{2} \right ) \: + \: {\cal O} (\alpha_S N_q; \Theta_0^2 N_q^\prime).
\ee
Notice that the factor 2 in the argument of $N_q$ generates $\sqrt{\alpha_S} N_q (M)$ correction and is 
under control in the present analysis whereas it could be omitted in the $N_q^\prime$ terms as producing 
$\alpha_S N_q (M)$ terms which we neglected systematically in (\ref{eq:a17}).  Within this accuracy the term 
$N^\prime \sim \sqrt{\alpha_S} N_q (M)$ can be embodied into the multiplicity factor by shifting its argument, 
namely
\be
\label{eq:a21}
N_q \left ( \frac{M}{2} \right ) \: + \: N_q^\prime \left ( \frac{M}{2} \right ) \: \approx \: N_q 
\left ( M \frac{\sqrt{e}}{2} \right ).
\ee
Finally we arrive at the formula expressing the companion multiplicity in $e^+ e^- \rightarrow Q\bar{Q}$ 
in terms of that of $e^+ e^- \rightarrow q\bar{q}$ (assuming $M \gg \Lambda$)
\be
\label{eq:a22}
N_{Q\bar{Q}} (W) \; = \; N_{q\bar{q}} (W) \: - \: N_{q\bar{q}} (\sqrt{e} M) \: [1 + {\cal O} (\alpha_S (M))].
\ee
The total particle multiplicity in $Q\bar{Q}$ events then reads as
\be
\label{eq:a23}
N^{e^+ e^- \rightarrow Q\bar{Q}} (W) \; = \; N_{Q\bar{Q}} (W) \: + \: n_Q^{dk},
\ee
where $n_Q^{dk}$ stands for the constant {\it decay} multiplicity of the heavy-quarks $(n_Q^{dk} = 11.0 \pm 0.2$ 
for $b$-quarks, $n_Q^{dk} = 5.2 \pm 0.3$ for $c$-quarks$\,$\cite{bas}).

Some comments are in order here.  The main consequence of (\ref{eq:a23}) is that the {\it difference} 
between particle yields from $q$- and $Q$-jets at fixed energy $W$ depends on the heavy-quark mass and remains 
$W$-independent.$\,$\cite{dfk1,dkt7,bas}  The question arises, to what extent this difference can be quantitatively 
predicted?  An important issue concerns non-leading corrections $\sim {\cal O} (\alpha_S (W))$ to the main 
term.  From the first sight one may expect that instead of (\ref{eq:a22}) the correct formula would look like
\bea
\label{eq:a24}
N_{Q\bar{Q}} (W) & = & \{ N_{q\bar{q}} (W) \}^{\rm MLLA} \: [1 + {\cal O} (\alpha_S (W))] \nonumber \\
& & \nonumber \\
& & \quad\quad\quad - \: \{ N_{q\bar{q}} (\sqrt{e} M) \}^{\rm MLLA} \: [1 + {\cal O} (\alpha_S (M))],
\eea
where $\{ N \}^{\rm MLLA}$ denotes the MLLA multiplicities accounting for the $\sqrt{\alpha_S} + \alpha_S$ 
effects in the anomalous dimension and $1 + \sqrt{\alpha_S}$ terms in the normalization (coefficient function).  
Meantime, because of the steep growth of the multiplicity (faster than any power of $\ln W$) the neglected, 
order $\sqrt{\alpha_S (W)} N(W)$ terms would dominate over the finite subtraction term in (\ref{eq:a24}) with 
increasing energy.  Thus, the very possibility to discriminate between the $Q$- and $q$-quark jets within the 
present theoretical accuracy may be endangered.  However, a close inspection of the subleading corrections 
to Eq.~(\ref{eq:a22}) (proportional to $N_{q\bar{q}} (W)$) shows that all of them are independent of the 
heavy-quark mass $M$.  Thus, the first corrections of order $\alpha_S (W) N_{q\bar{q}} (W)$ arise either from 
further improvement of the description of anomalous dimension $\Delta \gamma (\alpha_S) \sim \alpha_S^2$ 
determining intra-jet cascades, or from ${\cal O} (\alpha_S (W))$ terms in the coefficient function due to
\begin{itemize}
\item 3-jet configuration {\it quark} + {\it antiquark} + {\it hard gluon at large angles},
\item the so-called \lq\lq dipole correction\rq\rq\ to the angular-ordering scheme {\it quark} + 
{\it antiquark} + {\it two soft gluons with large emission angles},$\,$\cite{dkmt2}
\end{itemize}
which are insensitive to the $\Theta_0$ value with the {\it power} accuracy $\sim \Theta_0^2 \ll 1$.  Therefore, 
replacing the approximate MLLA multiplicity factors in (\ref{eq:a24}) by the real observable multiplicities, one 
arrives at (\ref{eq:a22}) which makes it possible to establish a phenomenological relation between light and 
heavy-quark jets with relative accuracy $\sqrt{\alpha_S (M)} \: M^2/W^2$.

Thus, the {\it difference} between particle yields from $q$- and $Q$-jets at fixed annihilation energy 
$W$ proves to be $W$-independent 
$\,$\cite{dfk1,dkt7,bas}
\bea
\label{eq:a25}
\delta_{Qq} & = & N^{e^+ e^- \rightarrow Q\bar{Q}} (W) \: - \: N^{e^+ e^- \rightarrow q\bar{q}} (W) \; = \; 
{\rm const}\:(W), \nonumber \\
& & \nonumber \\
\delta_{bc} & = & N^{e^+ e^- \rightarrow b\bar{b}} (W) \: - \: N^{e^+ e^- \rightarrow c\bar{c}} (W) \; = \; 
{\rm const}\:(W).
\eea
Let us emphasize that it is the QCD coherence which plays a fundamental role in the derivation of this result.  
Due to this the gluon bremsstrahlung off {\it massive} and {\it massless} quarks differ only at 
parametrically small angles $\Theta \lapproxeq \Theta_0 \equiv M/E$ where, due to the angular-ordering, 
cascading effects are majorated by the $N^\prime (M)$ factor.

The results of experiments on multiplicities in $b\bar{b}$ and $c\bar{c}$ events of $e^+ e^-$ 
annihilation$\,$\cite{ada1,delphi7} show that to the available accuracy the differences $\delta_{bq}$ and $\delta_{cq}$ are 
fairly independent of $W$.  This is in marked contrast to the steeply rising total multiplicity and confirms 
the MLLA expectations.  As illustrated in Fig.~\ref{figval1} the data on $\delta_{bq}$ are clearly inconsistent with 
a model$\,$\cite{pcr} based on the reduction of the energy scale.$\,$\cite{delphi7}

\begin{figure}[t]
\begin{center}
\mbox{\epsfig{file=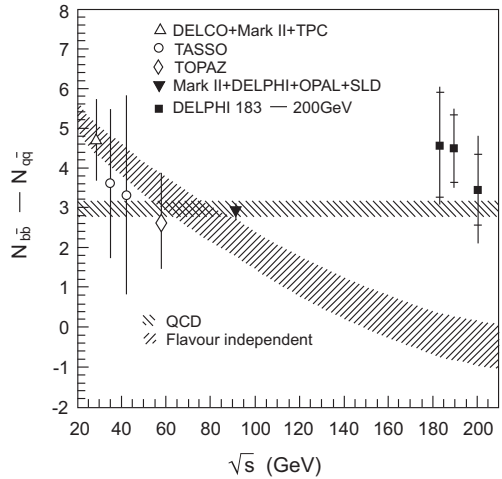,width=6.5cm%
%,bbllx=64bp,bblly=280bp,bburx=490bp,bbury=693bp%
}}
          \end{center}
\caption{The difference $\protect\delta_{bq}^{ch}$ 
between the average charged multiplicities of $b$- and 
light-quark events.$\,$\protect\cite{delphi7}  
The perturbative expectation is chosen as the average 
of the experimental values of $\protect\delta_{bq}^{ch}$ up to the $Z^0$. 
 Also shown is the prediction of the 
model based on the reduction of the energy scale 
(independently of the quark flavor).}
\label{figval1}
\end{figure}

We turn now to the absolute values of the charged multiplicity difference $\delta_{cq}^{ch}$.  Neglecting 
in Eq.~(\ref{eq:a22}) the subleading ${\cal O} (\alpha_S)$ corrections, the following MLLA expectation was 
found$\,$\cite{bas} by inserting the appropriate experimental numbers into the r.h.s.\ of 
Eqs.~(\ref{eq:a22})--(\ref{eq:a23}):
\be
\label{eq:a26}
\delta_{bq}^{ch} \; = \; 5.5 \: \pm \: 0.8.
\ee
This exceeds the experimental values in Fig.~\ref{figval1}, showing the essential role of the next-to-MLLA (order 
$\alpha_S (M) \cdot N(M)$) terms. % In Ref.~$\,$\cite{vap} 
An attempt was made$\,$\cite{vap} to improve Eq.~(\ref{eq:a22}).  
However, we have to mention here that the very picture of accompanying multiplicity being induced by a single 
cascading gluon$\,$\cite{vap} is not applicable at the level of subleading effects.  Therefore, 
a self-consistent reliable theoretical improvement of the MLLA predictions for the absolute values of $N_{Qq}$ 
remains to be achieved.

Further detailed experimental results could provide stringent tests of the perturbative predictions.  It will 
be very interesting to check whether the differences $\delta_{bq}^{h}$ of mean multiplicities for the identified 
particles remain energy independent.  Results from the $Z^0$ are available as reference values.  In particular, 
the DELPHI data$\,$\cite{delphi0} suggest $\delta_{bq}^{\pi^0} \simeq 0$ and therefore, using again 
Eqs.~(\ref{eq:a22})--(\ref{eq:a23}), one may expect the relation
\be
\label{eq:a27}
N_{\pi^0}^{e^+ e^- \rightarrow b\bar{b}} / N_{\pi^0}^{e^+ e^- \rightarrow q\bar{q}} \; \simeq \; 1.
\ee
The energy behaviour of $\delta_{Qq}^{h}$ should be watched closely at a future linear collider.

\begin{itemize}
\item[(iv)] {\it Spectra of light particles from the heavy-quark jets}
\end{itemize}
The bremsstrahlung-suppression effect on the energy spectra of light particles accompanying $Q\bar{Q}$ 
production can be examined in a very similar way to the average multiplicity considered above.  The 
resulting companion particle distribution $D_Q^h (x, \ln (E/\Lambda))$ is predicted to be depopulated in 
the hard momentum region compared with the case of the light-quark 
$D_q^h (x, \ln
E/\Lambda)$.$\,$\cite{dkt1,dkt7}  It is instructive to first gain insight by considering the double-logarithmic result.  
Due to the dead-cone phenomenon the difference between $D_Q^h$ and $D_q^h$ is connected with the gluon 
radiation at angles $\Theta < \Theta_0$.  Radiation in the restricted angular cone is similar to the 
$q$-jet production process with the characteristic hardness $M \gg \Lambda$.  This process induces a 
gluon radiation pattern with the corresponding cascading which leads to a certain final system of 
light hadrons.  Now, boosting this jet with the Lorentz-factor $\gamma = E/M$ along the direction of 
its momentum one arrives at an exact image of the dead cone:  an ensemble of energetic particles 
(with energies up to $E$) concentrated inside the cone $\Theta_0$.  Thus, with the double logarithmic 
accuracy a simple formula for the $Q$-jet particle distribution 
can be written$\,$\cite{dkt1,dkt7}
\be
\label{eq:a28}
D_Q^h \left ( x, \ln \: \frac{E}{\Lambda} \right ) \; = \; D_q^h \left ( x, \ln \: \frac{E}{\Lambda} 
\right ) \: - \: D_q^h \left ( x, \ln \: \frac{M}{\Lambda} \right ).
\ee
Here the structure of the heavy-quark jet is expressed in terms of the particle distribution 
generated by the light-quark at different energy scales:  $Q^2$ and $M^2$.  In Eq.~(\ref{eq:a28}) 
$x$ is the light-hadron energy fraction, $x = E_h/E \gapproxeq \Lambda/M$.  Therefore the depopulation 
occurs for quite energetic particles only, whereas in the soft domain
\be
\label{eq:a29}
\Lambda \; \lapproxeq \; E_h \; \lapproxeq \; E \: \frac{\Lambda}{M}
\ee
the light-hadron spectra in the $Q$- and light-quark jets should be identical.

In order to obtain the MLLA result one has to take into account, first of all, the energy loss by 
heavy-quark at the first steps of the evolution.  Using the evolution equations the expression for 
the $Q$-jet companion particle spectrum can be approximately given by
\be
\label{eq:a30}
D_Q^h \left ( x, \ln \: \frac{W}{\Lambda} \right ) \; = \; D_q^h \left ( x, \ln \: \frac{W}{\Lambda} 
\right ) \: - \: D_q^h \left ( \frac{x}{\langle y \rangle}; \; \ln \: \frac{M \sqrt{e}}{\Lambda} 
\right )
\ee
where $D_q^h (x, \ln W/\Lambda)$ is the standard spectrum of particles in a $q$-jet and $\langle y 
\rangle$ being the averaged scaled energy fraction of the $Q$.  This problem needs further detailed studies.

Measurements of the inclusive charged particle distributions in $b$- and $uds$-quark events at the 
$Z^0$$\,$\cite{opal0} clearly show that for $b$-quarks the momentum spectrum is significantly softer 
than for $uds$ quarks.  Unfortunately so far no comparison with the analytical QCD expectations has 
been performed.

\section{Color-Related Phenomena in Multi-Jet Events}

\subsection{Radiophysics of Hadronic Flows}

Coherence phenomenon is an intrinsic property of QCD (and in fact of any gauge theory).  Its observation 
is important in the study of strong interactions and in the search for deviations from the Standard Model.  
As we have already discussed, within the LPHD scenario, an intimate connection is expected between the 
multi-jet event structure and the underlying color dynamics at small distances.  The detailed features of 
the parton-shower system, such as the flow of color, governs the distribution of color-singlet hadrons in the 
final state (see e.g. Refs. 15,146).    %$\,$\cite{dkmt2,dkmt1}).  

Color coherence effects in hadron multiplicity flows in the inter-jet regions have been very well established 
from the early 1980's in $e^+ e^-$ annihilation, see 
Refs. 147,148,106,149,150,  %\cite{jade,tpcc,opal1,L30,delphi4} 
in what has been termed 
the \lq\lq string\rq\rq$\,$\cite{ags} or \lq\lq drag\rq\rq$\,$\cite{adkt2} effect.  As discussed below the 
existing data are clearly in favour of QCD coherence, in particular, particle production in the region 
between the quark and antiquark jets in $e^+ e^- \rightarrow q\bar{q}g$ events is suppressed.  In pQCD such 
effects arise from interference between the soft gluons radiated from the 
$q, \bar{q}$ and
$g$.$\,$\cite{adkt2,dkmt2}  
The PETRA/PEP data first convincingly demonstrated that the wide-angle particles do not belong to any particular 
jet but have emission properties dependent on the overall jet ensemble.  The inter-jet-coherence 
phenomena were then successfully studied at LEP, TRISTAN and TEVATRON 
(also discussed in recent 
reviews$\,$\cite{ko,nikos,kh1}).  The experiments have nicely demonstrated the connection 
between color and hadronic flows.  

Surely, it is entirely unremarkable that the quantum mechanical interference 
effects are observed in QCD.  Of real importance is that the experiment proves that these effects survive the 
hadronization phase.

The inter-jet coherence deals with the angular structure of particle flow when 
three or more hard partons are 
involved.  The hadron distribution proves to depend upon the geometry and color topology of the hard-parton 
skeleton.  The clear observation of inter-jet-interference effects gives another strong evidence in favor of the 
LPHD concept.  The collective nature of multiparticle production reveals itself here via the QCD wave properties 
of the particle flows.

The detailed experimental studies of the color-related effects are of particular interest for better 
understanding of the dynamics of hadroproduction in the multi-jet events.  For instance, under special 
conditions some subtle interference effects, breaking the probabilistic picture, may even become 
dominant.$\,$\cite{adkt2,dkt2}  We recall that QCD radiophysics predicts both attractive and repulsive 
forces between the active partons in the event.  Normally the repulsion effects are small, but in the case 
of color-suppressed ${\cal O} (1/N_C^2)$ phenomena they may play a leading role.  It should be noted that 
usually the inter-jet drag effects are viewed only on a completely inclusive basis, when all the constituents 
of the multi-element color antenna are simultaneously active.$\,$\cite{adkt2,dkt2}

A challenging possibility to operate within the perturbative scenario with the complete collective picture 
of an individual event (at least at very high energies) was first discussed
in Ref. 155. % $\,$\cite{dkt8}  
The 
topologometry on the event-by-event basis could turn out to be more informative than the results of 
measurements averaged over the events.$\,$\cite{orava,vpn,kh1}  Note, that there is an essential 
difference between the perturbative radiophysics and the parton-shower Monte Carlo models.  The latter not 
only allow but even require a completely exclusive probabilistic description.  Normally (such as in the case 
of $e^+ e^- \rightarrow q\bar{q}g$) the two pictures work in a quite peaceful coexistence; the difference 
only becomes drastic when one deals with the small color-suppressed effects.

Note that in spite of theoretical uncertainties, the 
phenomenological success of the  Monte Carlo models which include
interfering gluons
indicates that the color-suppressed interference terms do not induce a very large effect.

Let us emphasize that the relative smallness of the non-classical effects by no means diminishes their 
importance.  This consequence of QCD radiophysics is a serious warning against the traditional ideas of 
independently evolving partonic subsystems.  So far (despite the persistent pressure from the theorists) 
no clear evidence has been found experimentally in favour of the non-classical color-suppressed effects in jets, 
and the peaceful coexistence between the perturbative inter-jet coherence and color-topology-based fragmentation 
models remains unbroken.  However, these days the color-suppressed interference effects attract increased 
attention.  This is partly boosted by the findings that the QCD interference (interconnection) between the 
$W^+$ and $W^-$ hadronic decays could affect the precise $W$ mass reconstruction at LEP-2, 
for example,$\,$\cite{sk1,brw0}  QCD interconnection may affect also the top quark studies, in particular, its mass 
reconstruction.$\,$\cite{sk0}

It is worthwhile to mention that the relative smallness of the color-suppressed interference effects 
is supported by recent searches by OPAL of the so-called reconnection effects in hadronic $Z^0$ 
decays.$\,$\cite{opal02}
Finally, we recall that color-related collective effects could become a phenomenon of large potential value 
as a new tool helping to distinguish the new physics signals from the conventional QCD
backgrounds.$\,$\cite{dkt1,dkmt1,dkt8,eks,ks,hks}

In the next subsections we briefly survey the basic ideas related to QCD collective effects in multi-jet 
events and then we describe (wherever possible) the latest data on QCD radiophysics.  

\subsection{String/Drag Effect in $q\bar{q}g$ Events}

The first (and still best) example of the inter-jet color-related phenomena is the string/drag effect in 
$e^+ e^- \rightarrow q\bar{q}g$.  Since the new results based on the refined analysis of $q\bar{q}g$ events 
continue to pour out from the LEP groups it seems useful to recall the main ideas$\,$\cite{adkt2} of the pQCD 
explanation of this bright coherence phenomenon.  We consider first the angular distribution of particle 
flows at large angles to the jets in $e^+ e^- \rightarrow q\bar{q}g$.  The more general case including the 
inside jet particle flow will be discussed below.

Let all the angles $\Theta_{ij}$ between jets and the jet energies $E_i$ be large $(i = \{ + - 1\} \equiv 
\{ q\bar{q}g \}):  \Theta_{+-} \sim \Theta_{+1} \sim \Theta_{-1} \sim 1, E_1 \sim E_+ \sim E_- \sim E \sim 
W/3$.  As was discussed above, within the perturbative picture the angular distribution of soft inter-jet 
hadrons carries information about the coherent gluon radiation off the color antenna formed by three 
emitters ($q, \bar{q}$ and $g$).  The wide-angle distribution of a secondary soft gluon $g_2$ displayed 
in Fig.~\ref{figval2}
 can be written as
\be
\label{eq:a31}
\frac{8 \pi d N_{q\bar{q}g}}{d\Omega_{\vec{n}_2}} \; = \; \frac{1}{N_C} \: W_{\pm 1} (\vec{n}_2) \: 
N_g^\prime (Y_m) \; = \; \left [ (\widehat{1+}) \: + \: (\widehat{1-}) \: - \: \frac{1}{N_C^2} \: 
(\widehat{+-}) \right ] \: N_g^\prime (Y_m).
\ee

\begin{figure}[t]
\begin{center}
\mbox{\epsfig{file=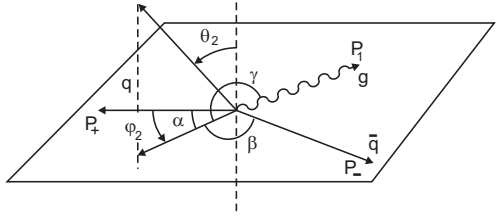,width=7.0cm%
%,bbllx=64bp,bblly=280bp,bburx=490bp,bbury=693bp%
}}
          \end{center}
\caption{Kinematics of inter-jet radiation in three-jet events.} 
\label{figval2}
\end{figure}

Here the \lq\lq antenna\rq\rq\ $(\widehat{ij})$ is represented as
\be
\label{eq:a32}
(\widehat{ij}) \; = \; \frac{a_{ij}}{a_i a_j}, \quad\quad a_{ij} \; = \; (1 - \vec{n}_i \vec{n}_j), \quad\quad 
a_i \; = \; (1 - \vec{n}_2 \vec{n}_i)
\ee
and $N_g^\prime (Y_m)$ is the so-called cascading factor taking into account that a final soft particle is 
a part of cascade (see Refs.  64,15  %$\,$\cite{adkt2,dkmt2} 
and Section 3), $N_g^\prime (Y) \equiv dN_g/dY$.  Furthermore, 
$Y_m = \ln E \Theta_m/\Lambda$, where one defines the angle $\Theta_m = {\rm min} \{\Theta_+, \Theta_-, 
\Theta_1 \}$ with $\cos \Theta_i = \vec{n}_2 \vec{n}_i$ for $i = \{+, -, 1\}$.

The radiation pattern corresponding to the case when a photon $\gamma$ is emitted instead of a gluon reads
\be
\label{eq:a33}
\frac{8 \pi d N_{q\bar{q}\gamma}}{d\Omega_{\vec{n}_2}} \; = \; \frac{1}{N_C} \: W_{+-} (\vec{n}_2) \: 
N_g^\prime (Y_m) \; = \; \frac{2C_F}{N_C} \: (\widehat{+-}) \: N_g^\prime (Y_m).
\ee
The dashed line in Fig.~\ref{figval3} displays the corresponding \lq\lq directivity diagram\rq\rq, which represents 
the particle density (\ref{eq:a33}) projected onto the $q\bar{q}\gamma$ plane:
%\newpage
\bea
\label{eq:a34}
W_{+-} (\varphi_2) & = & 2C_F \: \int \: \frac{d \cos \Theta_2}{2} \: (\widehat{+-}) \; = \; 2C_F \: a_{+-} \: 
V (\alpha, \beta), \nonumber \\
& & \nonumber \\
V (\alpha, \beta) & = & \frac{2}{\cos \alpha - \cos \beta} \: \left ( \frac{\pi - \alpha}{\sin \alpha} \: - \: 
\frac{\pi - \beta}{\sin \beta} \right ) ; \alpha \: = \: \varphi_2, \; \beta \: = \: \Theta_{+-} - 
\varphi_2. \nonumber \\
\eea
The expression $W_{+-} (\vec{n}_2)$ is simply related to the particle distribution in two-jet events $e^+ e^- 
\rightarrow q (p_+) + \bar{q} (p_-)$, Lorentz boosted from the quark {\it cms} to the lab system.

\begin{figure}[htb]
\begin{center}
\mbox{\epsfig{file=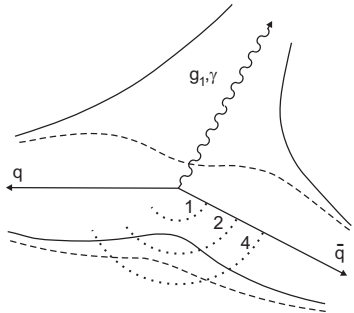,width=5.0cm%
%,bbllx=64bp,bblly=280bp,bburx=490bp,bbury=693bp%
}}
          \end{center}
\caption{Directivity diagram of soft gluon radiation, projected onto the $q\bar{q}\gamma$ (dashed line) 
and $q\bar{q}g$ (solid) event planes.  Particle flows of Eqs.~(\ref{eq:a34}) and (\ref{eq:a35}) are drawn 
in polar coordinates:  $\Theta = \varphi_2, r = \ln 2W (\varphi_2)$.  Dotted circles show the constant levels 
of particle flow:  $W (\varphi_2) = 1, 2, 4$.}
\label{figval3}
\end{figure}

Replacing $\gamma$ by $g_1$ changes the directivity diagram essentially because the antenna element $g_1$ now 
participates in the emission as well.  However, this change leads not only to an appearance of an additional 
particle flow in the $g_1$ direction.  Integrating (\ref{eq:a31}) over $\Theta_2$ one obtains $(\gamma = 
\Theta_{1+} + \varphi_2)$:
\be
\label{eq:a35}
W_{\pm 1} (\varphi_2) \; = \; N_C \: \left [a_{+1} V (\alpha, \gamma) \: + \: a_{-1} V (\beta, \gamma) \: - \: 
\frac{1}{N_C^2} \: a_{+-} V (\alpha, \beta) \right ].
\ee
Fig.~\ref{figval3} illustrates that the particle flow in the direction opposite to $\vec{n}_1$ appears to be 
{\it considerably lower} than in the photon case.  So, the destructive interference diminishes radiation in the 
region between the quark jets giving a surplus of radiation in the $q - g$ and $\bar{q} - g$ valleys.  One easily 
sees that the color-coherence phenomena strongly affect the total three-dimensional shape of particle flows in 
three-jet events, practically excluding the very possibility of representing it as a sum of three parton 
contributions.

Let us recall here that, owing to coherence, the radiation of a soft gluon $g_2$ at angles larger than the 
characteristic angular size of each parton jet proves to be insensitive to the jet internal structure:  $g_2$ 
is emitted by a color current which is conserved when the jet splits.  This is the reason why one may replace 
each jet by its parent parton with $p_i^2 \approx 0$.

For illustration of the particle drag phenomenon it is instructive$\,$\cite{adkt2} to examine the fully 
symmetric $q\bar{q}g$ events (Mercedes-type topology) where $\vec{n}_+ \vec{n}_- = \vec{n}_+ \vec{n}_1 = 
\vec{n}_- \vec{n}_1$.  Let us take $\vec{n}_2$ pointing in the direction opposite to $\vec{n}_1$, i.e.\ 
midway between quarks.  Thus, neglecting the weak dependence $N_g^\prime$ on $\Theta$, one arrives at the 
ratio (numbers for $N_C = 3$)
\be
\label{eq:a36}
\frac{dN_{q\bar{q}g}/d\vec{n}_2}{dN_{q\bar{q}\gamma}/d\vec{n}_2} \; = \; \frac{N_C^2 - 2}{2 (N_C^2 - 1)} \; 
\approx \; 0.44.
\ee
The corresponding ratio for the projected particle flows obtained from (\ref{eq:a34}) and (\ref{eq:a35}) reads
\be
\label{eq:a37}
\frac{dN_{q\bar{q}g}/d\varphi_2}{dN_{q\bar{q} \gamma}/d\varphi_2} \; \approx \; \frac{9 N_C^2 - 14}{14 
(N_C^2 - 1)} \; \approx \; 0.60.
\ee
We emphasize that Eq.~(\ref{eq:a31}) provides not only the planar picture, but the global 
three-dimensional wide-angle pattern of particle flows.

% xxx   other ratios

 For hadron states associated with the $\bar{q}qg$ and $\bar{q}q\gamma$ events the
above ratios of particle flows 
%given by Eqs.~(\ref{eq:a36})--(\ref{eq:a41}) 
should remain asymptotically correct, since non-perturbative 
hadronization effects are expected to cancel at high energies.  It is an interesting question, to what 
extent the non-perturbative effects are important at present energies for a quantitative description of the 
data.

The analysis of the bremsstrahlung pattern clearly demonstrates particle \lq\lq drag\rq\rq\ by the gluon 
jet $g_1$.  If one drops the color suppressed contribution, the two remaining terms in Eq.~(\ref{eq:a31}) 
may be interpreted as the sum of two independent $(\widehat{1+})$ and $(\widehat{1-})$ antenna patterns, 
boosted from their respective rest frames into the overall $q\bar{q}g$ {\it cms}.  The point is, that by 
neglecting the $1/N_C^2$ terms, the hard gluon can be treated as a quark-antiquark pair.  In this 
approximation each external quark line is uniquely connected to an external antiquark line of the same 
color, forming colorless $q\bar{q}$ antennae.  In the general case, when calculating the resultant 
soft-radiation pattern, one can only deal with a set of such color-connected $q\bar{q}$ pairs because the 
interference between gluons emitted from {\it non-color-connected} lines proves to be suppressed 
by powers of $1/N_C^2$.$\,$\cite{adkt2,dkt8,gg4,gg3}

Note that in the first order in $\alpha_S$ the $q\bar{q}$-antenna pattern (\ref{eq:a33}) can be 
presented in the form$\,$\cite{adkt2}
\be
\label{eq:a42}
d w_{q\bar{q}\gamma} \; = \; C_F \: \frac{\alpha_S (k_\perp)}{\pi} \: \frac{dk_\perp^2}{k_\perp^2} 
\: dy_{||}
\ee
where
\be
\label{eq:a43}
y_{||} \; = \; \frac{1}{2} \: \ln \frac{(p_+ \cdot k_2)}{(p_- \cdot k_2)} , \quad k_\perp^2 \; = \; 
\frac{2 (p_+ \cdot k_2)(p_- \cdot k_2)}{(p_+ \cdot p_-)}
\ee
are the Lorentz-invariant generalizations of the rapidity and the transverse momentum of $g_2$ in the 
$q\bar{q}$ {\it cms} correspondingly.  Eq.~(\ref{eq:a42}) describes the structure of the gluon 
radiation in $e^+ e^- \rightarrow q\bar{q}$ events.  Due to color coherence the particle multiplicity in 
a secondary gluon jet $g_2$ depends on the transverse momentum $k_\perp$ of the gluon and not on its 
energy.$\,$\cite{adkt2,dkt11}  The radiation antenna formalism$\,$\cite{adkt2,dkt8} is closely related to the 
Lund color dipole approach.$\,$\cite{gg1}  In both schemes the natural choice for the cut-off parameter in the 
cascades is the gluon transverse momentum in the {\it cms} of the emitting antenna-dipole.  This implies 
that the soft gluons connect the active hard partons in exactly the same way as the string in the Lund 
string fragmentation model,$\,$\cite{ags} which illustrates the connection between pQCD and the string 
model.$\,$\cite{adkt2}  Notice, however, that within the latter model there is no string piece spanned 
directly between the quark and antiquark, no particles are produced in between them, except by some 
minimum \lq\lq leakage\rq\rq\ from the other two regions.

When a gluon is emitted in, let us say, an $e^+ e^-$ annihilation event, then the radiation of a second 
softer gluon is described by two antennas-dipoles, one spanned between the $q$ and $g$ and another 
between the $g$ and $\bar{q}$.  The emission of a third, still softer, gluon is described in terms of 
three antennas-dipoles etc.  The gluons appear to be color-ordered in such a way that the dipole is 
stretched between the color charge of one gluon and the corresponding anticolor charge of the next.  
Thus, neglecting the $1/N_C^2$ terms, the QCD cascade is formulated as a branching process in which 
antennas-dipoles are successfully split into the smaller and smaller ones.

\subsection{Multiplicity Flows in Three-Jet Events}

We begin here by comparing particle flows in $q\bar{q}g$ and $q\bar{q}\gamma$ events and in the next 
subsection we consider the respective total event multiplicities.

In what follows we first do not refer to a particular three-jet-event selection method but focus on 
the soft radiation by the $q\bar{q}g$ system for three given angular
directions.$\,$\cite{dkmt1,dkt11}

In the leading order in $\alpha_S$ the massless parton kinematics is unambiguously fixed as follows:
\bea
\label{eq:a45}
x_+ \; = \; 2 \: \frac{\sin \Theta_{1-}}{\Sigma \sin \Theta_{ij}}, \quad x_- & = & 2 \: 
\frac{\sin \Theta_{1+}}{\Sigma \sin \Theta_{ij}}, \quad x_1 \; = \; 2 \: 
\frac{\sin \Theta_{+-}}{\Sigma \sin \Theta_{ij}}, \nonumber \\
& & \\
x & + & x_- \: + x_1 \; = \; 2, \nonumber
\eea
with $x_i = 2E_i/W$ being the normalized parton energies and $\Theta_{ij}$ the angles between partons 
$i$ and $j (+, - \equiv q, \bar{q}; 1 \equiv g_1)$.  We emphasize here, that, owing to color-coherence, 
the radiation of a secondary gluon $g_2 (k_2 \ll E_i)$ at angles larger than the aperture of each 
parton jet is insensitive to the jet internal structure.

Let us turn to the radiation pattern for $q\bar{q}\gamma$ events.  In the previous subsection we 
considered the radiation at large angles to the jet directions, here we include the small-angle 
radiation in MLLA.  The angular distribution of particle flow can be written as
\be
\label{eq:a46}
\frac{8 \pi dN_{q\bar{q}\gamma}}{d\Omega_{\vec{n}_2}} \; = \; \frac{2}{a_+} \: N_g^\prime 
(Y_{q+}, Y_q) \: + \: \frac{2}{a_-} \: N_q^\prime (Y_{\bar{q}-}, Y_{\bar{q}}) \: + \: 2 I_{+-} 
\: N_q^\prime (Y),
\ee
where
\be
\label{eq:a47}
Y_{q(\bar{q})} \; = \; \ln \frac{E_{q (\bar{q})}}{\Lambda}, \; Y_{q+} \: = \: \ln \left ( \frac{E_q 
\sqrt{a_+/2}}{\Lambda} \right ), \; Y_{\bar{q}_-} \: = \: \ln \left ( \frac{E_q \sqrt{a_-/2}}{\Lambda} 
\right ), \; Y \: \equiv \: \ln \frac{E}{\Lambda}
\ee
and
\be
\label{eq:a48}
I_{+-} \; = \; (\widehat{+-}) \: - \: \frac{1}{a_-} \: - \: \frac{1}{a_+} \; = \; 
\frac{a_{+-} - a_+ - a_-}{a_+ a_-}.
\ee

The factor $N_A^\prime (Y_i, Y) \equiv (d/dY_i) N_A (Y_i, Y)$ takes into account that 
the final registered hadron is a part of cascade.  $N_A (Y_i, Y)$ stands for the multiplicity in a 
jet $A (A = q, g)$ with the hardness scale $Y$ of particles concentrated in the cone with an angular 
aperture $\Theta_i$ around the jet direction $\vec{n}_i$.  To understand the meaning of the quantity 
$N_A (Y_i, Y)$ it is helpful to represent it as
\bea
\label{eq:a49}
N_A (Y_i, Y) & = & \sum_{B = q,g} \: \int_0^1 \: dz \: z \: \overline{D}_A^B (z, \Delta \xi) \: N_B 
(\overline{Y}_i), \nonumber \\
& & \nonumber \\
\Delta \xi & = & \frac{1}{b} \: \ln (Y/\overline{Y}_i), \quad \overline{Y}_i \; = \; Y_i \: + \: 
\ln \: z \; = \; \ln \left ( \frac{zE}{\Lambda} \: \sqrt{\frac{a_i}{2}} \right ). 
\eea
Here $N_B (\overline{Y}_i)$ is the multiplicity in a jet with the hardness scale $\overline{Y}_i$, 
initiated by a parton $B$ within the cone $\Theta_i$, and $\overline{D}_A^B$ denotes the structure 
function for parton fragmentation $A \rightarrow B$.

Eq.~(\ref{eq:a49}) accounts for the fact that due to the inside-jet coherence the radiation at small 
angles $\Theta_i \ll 1$ is governed not by the overall color of a jet $A$, but by that of a sub-jet $B$, 
developing inside a much narrower cone $\Theta_i$.

For the emission at large angles $(a_+ \sim a_- \sim 1)$ when all the factors $N^\prime$ are approximately 
the same, Eq.~(\ref{eq:a46}) reads, c.f.\ Eq.~(\ref{eq:a33})
\be
\label{eq:a50}
8 \pi \: \frac{dN_{q\bar{q}\gamma}}{d\Omega_{\vec{n}_2}} \; = \; 2 (\widehat{+-}) \: N_q^\prime (\ln E/\Lambda).
\ee
The cascading factor here can be presented as
\be
\label{eq:a51}
\frac{N_C}{C_F} \: N_q^\prime \left ( \ln \frac{E}{\Lambda} \right ) \; \approx \; N_g^\prime \left ( \ln 
\frac{E}{\Lambda} \right ) \; = \; \int^E \: \frac{dE_g}{E_g} \: 4 N_C \: \frac{\alpha_S (E_g)}{2 \pi} \: 
N_g \left ( \ln \frac{E_g}{\Lambda} \right ).
\ee
One can easily see that for the radiative two-jet events the emission pattern is given by the 
$q\bar{q}$ sample Lorentz-boosted from the quark {\it cms} to the {\it lab} system (i.e.\ the {\it cms} of 
$q\bar{q}\gamma$), and the corresponding particle multiplicity should be equal to that in $e^+ e^- 
\rightarrow q\bar{q}$ at $W_{q\bar{q}}^2 = (p_q + p_{\bar{q}})^2$.

Now we consider the three-jet event sample when a hard photon is replaced by a gluon $g_1$.  For a 
given $q\bar{q}g_1$ configuration the particle flow can be presented as
\bea
\label{eq:a52}
\frac{8 \pi d N_{q\bar{q}g}}{d \Omega_{\vec{n}_2}} & = & \frac{2}{a_+} \: N_q^\prime (Y_{q+}, Y_q) \: + \: 
\frac{2}{a_-} \: N_q^\prime (Y_{\bar{q}-}, Y_{\bar{q}}) \: + \: \frac{2}{a_1} \: N_g^\prime (Y_{g1}, Y_g) 
\nonumber \\
& & \\
& & + \; 2 \left [I_{1+} \: + \: I_{1-} \: - \: \left ( 1 \: - \: \frac{2C_F}{N_G} \right ) \: I_{+-} \right ] 
\: N_g^\prime (Y), \nonumber
\eea
where, in addition to the definitions in Eq.~(\ref{eq:a47}) one has
\be
\label{eq:a53}
Y_g \; = \; \ln \frac{E_g}{\Lambda}, \quad\quad Y_{g1} \; = \; \ln \left ( \frac{E_g \sqrt{a_1/2}}{\Lambda} 
\right ).
\ee
This formula accounts for both types of coherence:  the angular ordering inside each of the jets and the 
collective nature of the inter-jet flows.  The first three terms in Eq.~(\ref{eq:a52}) are collinear 
singular as $\Theta_i \rightarrow 0$ and contain the factors $N^\prime$, describing the evolution of each 
jet initiated by the hard emitters $q, \bar{q}$ and $g_1$.  The last term accounts for the interference 
between these jets.  It has no collinear singularities and contains the common factor $N_g^\prime (Y, Y)$ 
independent of the direction $\vec{n}_2$.  Eq.~(\ref{eq:a52}) predicts the energy evolution of particle 
flows in $q\bar{q}g$ events.

As follows from (\ref{eq:a46})--(\ref{eq:a52}) when one is replacing a hard photon by a gluon $g_1$ with 
otherwise identical kinematics, an additional particle flow arises:

%\vfill
\bea
\label{eq:a54}
\left ( \frac{8\pi dN}{d\Omega_{\vec{n}_2}} \right )_g & = & \frac{8\pi dN_{q\bar{q}g}}{d\Omega_{\vec{n}_2}} 
\: - \: \frac{8\pi dN_{q\bar{q}\gamma}}{d\Omega_{\vec{n}_2}} \nonumber \\
& & \\
& = & \frac{2}{a_1} \: N_g^\prime (Y_{g1}, Y_g) \: + \: [I_{1+} \: + \: I_{1-} \: - \: I_{+-}] \: 
N_g^\prime (Y). \nonumber
\eea
Note that for the case of large radiation angles the two cascading factors $N^\prime$ become approximately 
equal and one has, c.f.\ Eq.~(\ref{eq:a31})
\be
\label{eq:a55}
\left ( \frac{8 \pi dN}{d\Omega_{\vec{n}_2}} \right )_g \; = \; \left [ (\widehat{1+}) \: + \: (\widehat{1-}) 
\: - \: (\widehat{+-}) \right ] \: N_g^\prime (Y).
\ee
An instructive point is that this expression is not positively definite.  One clearly observes the net 
destructive interference in the region between the $q$ and $\bar{q}$ jets.  The soft radiation in this 
direction proves to be less than that in the absence of the gluon jet $g_1$.

\subsection{Topology Dependence of Three-Jet-Event Multiplicity}

Now we turn to the connection between the average particle multiplicities in the two-jet and three-jet 
samples of $e^+ e^-$ annihilation.  The particle multiplicity in an individual quark jet is formally defined 
from the process $e^+ e^- \rightarrow q\bar{q} \rightarrow$ hadrons by
\be
\label{eq:a56}
N_{e^+ e^-}^{ch} (W) \; = \; 2 \: N_q^{ch} (E) \: \left [1 + {\cal O} \left ( \frac{\alpha_S (W)}{\pi} 
\right ) \right ], \quad W \; = \; 2E.
\ee

As we have already discussed, when three or more partons are involved in a hard interaction the multiplicity 
cannot be represented simply as a sum of the independent parton pieces, but rather it becomes dependent 
on the geometry of the whole ensemble.

So, the problem arises of describing the multiplicity in three-jet events, $N_{q\bar{q}g}$, in terms of 
the characteristics of the individual $q$ and $g$ jets.  The quantity $N_{q\bar{q}g}$ should depend on the 
$q\bar{q}g$ geometry in a Lorentz-invariant way and should have a correct limit when the event is transformed 
to the two-jet configuration by decreasing either the energy of the gluon $g_1$ or its emission angle.

The angular integral of Eq.~(\ref{eq:a46}) can be easily checked to reproduce the total multiplicity.  One 
can write $N_{q\bar{q}\gamma}$ as
\be
\label{eq:a57}
N_{q\bar{q}\gamma} \; = \; \int \: \frac{d N_{q\bar{q}\gamma}}{d\Omega_{\vec{n}_2}} \; = \; N_q (Y_q) + 
N_q (Y_{\bar{q}}) + 2 \ln \sqrt{\frac{a_{+-}}{2}} \: N_q^\prime (Y).
\ee
Now let us transform this formula to the Lorentz-invariant expression.  To do this we rewrite
\be
\label{eq:a58}
Y_{q(\bar{q})} \; = \; Y + \ln x_{+ (-)}, \quad x_{+ (-)} \; \equiv \; E_{q(\bar{q})}/E
\ee
and use the expression
\be
\label{eq:a59}
N_q (Y_q) \; = \; N_q (Y) + \ln x_+ N_q^\prime (Y) + {\cal O} \left ( \frac{\alpha_S}{\pi} \: N_q \right ).
\ee
Then
\be
\label{eq:a60}
N_{q\bar{q}\gamma} \; = \; 2N_q (Y) + \ln \frac{x_+ x_- a_{+-}}{2} \: N_q^\prime (Y) + {\cal O} (\alpha_S N) 
\; = \; 2N_q (Y_{+-}^*) [1 + {\cal O} (\alpha_S)]
\ee
with
\be
\label{eq:a61}
Y_{+-}^* \; = \; Y + \ln \sqrt{\frac{x_+ x_- a_{+-}}{2}} \; = \; \ln \sqrt{\frac{(p_+ \cdot 
p_-)}{2\Lambda^2}} \; 
= \; \ln \frac{E^*}{\Lambda}.
\ee
Here $E^*$ is the quark energy in the {\it cms} of $q\bar{q}$, i.e.\ the Lorentz-invariant generalization 
of a parameter of hardness of the process.

The multiplicity $N_{q\bar{q}g}$ can be written by analogy as
\bea
\label{eq:a62}
N_{q\bar{q}g} & = & \int \: \frac{dN_{q\bar{q}g}}{d\Omega_{\vec{n}}} \; = \; N_q (Y_q) + N_q (Y_{\bar{q}}) + 
N_g (Y_g) \nonumber \\
& & \\
& & + \; \left [ \ln \sqrt{\frac{a_{1+} a_{1-}}{2a_{+-}}} \: + \: \frac{2 C_F}{N_C} \: \ln 
\sqrt{\frac{a_{+-}}{2}} \right ] \: N_g^\prime (Y) \nonumber
\eea
where $Y_g = Y + \ln x_1$.  Proceeding as before, one arrives finally at the Lorentz-invariant 
result$\,$\cite{dkt11}
\be
\label{eq:a63}
N_{q\bar{q}g} \; = \; \left [ 2N_q (Y_{+-}^*) + N_g (Y_g^*) \right ] \: \left ( 1 + {\cal O} \left ( 
\frac{\alpha_S}{\pi} \right ) \right )
\ee
with
\be
\label{eq:a64}
Y_g^* \; = \; \ln \sqrt{\frac{(p_+ \cdot p_1)(p_- \cdot p_1)}{2 (p_+ \cdot p_-) \Lambda^2}} \; = \; \ln 
\frac{p_{1\perp}}{2\Lambda},
\ee
where $p_{1\perp}$ the transverse momentum of $g_1$ in the $q\bar{q}$ {\it cms}, c.f.\ Eq.~(\ref{eq:a43}).

Comparing (\ref{eq:a57}) with (\ref{eq:a63}), we see that replacement of a photon $\gamma$ by a gluon $g_1$ 
leads to the additional multiplicity
\be
\label{eq:a65}
N_g (Y_g^*) \; = \; N_{q\bar{q}g} (W) \: - \: N_{q\bar{q}\gamma} (W),
\ee
which depends not on the gluon energy but on its {\it transverse momentum}, 
i.e.\ on the hardness of the 
primary process.$\,$\footnote{Within antenna pattern formalism the observation that a gluon adds multiplicity 
related to its $p_\perp$ was made.$\,$\cite{dkt11}  A similar result was derived in the string-length 
approach to the dipole picture,$\,$\cite{agu} 
 more details can be found elsewhere.$\,$\cite{egu,egk}}  Eq.~(\ref{eq:a63}) 
reflects the coherent nature of soft emission and has a correct limit when the event is transformed to 
the two-jet configuration.  
Eq.~(\ref{eq:a63}) can be rewritten in terms of the observed multiplicity in 
$e^+ e^-$ collisions$\,$\cite{ko} as
\be
\label{eq:a66}
N_{q\bar{q}g} \; = \; \left [ N_{e^+ e^-} (2E^*) \: + \: \frac{1}{2} \: r (p_{1\perp}) \: N_{e^+ e^-} 
(p_{1\perp}) \right ] \: (1 + {\cal O} (\alpha_S)).
\ee
In this formula $r (p_{1\perp})$ denotes the ratio of multiplicities in gluon and quark jets at {\it cms} 
energy $W = p_{1\perp}$.  The expression for $N_{q\bar{q}g}$ can be presented in another form$\,$\cite{dkt11}
\be
\label{eq:a67}
N_{q\bar{q}g} \; = \; \left [N_g (Y_{1+}) + N_g (Y_{1-}) + 2N_q (Y_{+-}) - N_g (Y_{+-}) \right ] \: 
[1 + {\cal O} (\alpha_S)],
\ee
where $Y_{ij} = \ln \left (\sqrt{(p_i \cdot p_j)/2\Lambda^2} \right ) = \ln (E_{ij}^*/\Lambda)$.  Such a representation 
deals with multiplicities of two-jet events at appropriate invariant pair energies $W_{ij} = 2E_{ij}^*$ 
in the ($ij$) {\it cms} frame.  Expression (\ref{eq:a67}) has a proper limit, $2N_g (W/2)$, when the 
${q\bar{q}g}$ configuration is forced to a quasi-two-jet one, $g(8) + q\bar{q}(8)$, with a sufficiently 
small angle between the quarks.  The experimental tests of Eq.~(\ref{eq:a67}) can be performed with the 
tagged heavy quarks, see also Ref. 89.  %$\,$\cite{jwg}  
Analogously to (\ref{eq:a66}), Eq.~(\ref{eq:a67}) could be 
presented as
\bea
\label{eq:a68}
N_{q\bar{q}g} & \simeq & \left [ N_{e^+ e^-} (2 E^*) \: \left (1 \: - \: \frac{r (E^*)}{2} \right ) 
\right . \nonumber \\
& & \nonumber \\
& & + \; \frac{r (E_{1+}^*)}{2} \: N_{e^+ e^-} \: (2 E_{1+}^*) \\
& & \nonumber \\
& & + \; \left . \frac{r (E_{1-}^*)}{2} \: N_{e^+ e^-} \: (2 E_{1-}^*) \right ]. \nonumber
\eea

Until now we have confined ourselves to the so-called \lq unbiased\rq\ case and have not 
addressed an issue of the three-jet-selection procedure.  As well known, different approaches to the 
three-jet selections employ different definitions of the $q\bar{q}g$-kinematics.  Moreover, most jet 
definitions give three-jets which are all biased.

To gain some insight on how the jet resolution works, let us first consider a sample of two-jet 
$q\bar{q}$-events selected in such a way that there are no sub-jets with $p_\perp > p_{\perp {\rm cut}}$ 
(within a $k_T$-clustering scheme with a resolution parameter $p_{\perp {\rm cut}}$ this means that 
these are only two primary $q$- and $\bar{q}$-jets).  If $N_{q\bar{q}} (L)$ is the \lq unbiased\rq\ 
$q\bar{q}$-multiplicity, ($N_{q\bar{q}} (L) \equiv N_{e^+ e^-} (L)$ with $L \equiv \ln (s/\Lambda^2)$), 
then in events with precisely two jets at resolution $y_{\rm cut} = p_{\perp {\rm cut}}^2/s$ there is a 
rapidity plateau of length $\ln (1/y_{\rm cut})$, see Fig.~\ref{figval4} and the multiplicity $N_{q\bar{q}} (L, 
\kappa_{\rm cut})$ is$\,$\cite{dkt8,agns,cwdf}
\bea
\label{eq:a69}
N_{q\bar{q}} (L, \kappa_{\rm cut}) & \approx & N_{q\bar{q}} (\kappa_{\rm cut}) \: + \: 
(L - \kappa_{\rm cut}) \: N_{q\bar{q}}^\prime (\kappa_{\rm cut}); \nonumber \\
& & \\
\kappa & = & \ln \left (\frac{p_\perp^2}{\Lambda^2} \right ). \nonumber
\eea
The first term corresponds to two cones around the $q$ and $\bar{q}$ jet directions.  This also 
corresponds exactly to an unbiased $q\bar{q}$ system with {\it cms} energy $p_{\perp {\rm cut}}$.  The 
second term describes a central rapidity plateau of width $(L - \kappa_{\rm cut}) = \ln 1/y_{\rm cut}$, 
in which the limit for gluon emission is given by the constraint $\kappa_{\rm cut}$.  This expression for 
a two-jet event can be generalized for a biased multi-jet configuration.$\,$\cite{agns,cwdf,egu,egk}
\begin{figure}[htb]
\begin{center}
\mbox{\epsfig{file=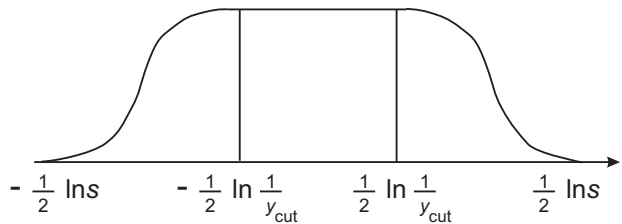,width=8.5cm%
%,bbllx=64bp,bblly=280bp,bburx=490bp,bbury=693bp%
}}
          \end{center}
\caption{Rapidity plateau in two-jet events.}
\label{figval4}
\end{figure}

The average particle multiplicity in the selected two-jet sample is smaller than in an unbiased sample.  
The modification due to the bias is similar to the suppression from a Sudakov form factor.  It is 
formally ${\cal O} (\alpha_S)$, but it also contains a factor $\ln^2 (s/p_\perp^2)$.  Thus, it is small 
for large $p_\perp$-values but it becomes significant for smaller $p_\perp$.  This clearly demonstrates 
that the multiplicity in this restricted case depends on {\it two} scales, $\sqrt{s}$ and $p_{\perp {\rm cut}}$.  
The $p_\perp$ of an emitted gluon is related to the virtual mass of the radiating parent quark.  Therefore, 
the two scales $\sqrt{s}/2$ and $p_{\perp {\rm cut}}$ represent the energy and virtuality of the quark and 
antiquark initiating the jets.

Though the leading-logarithmic result in Eq.~(\ref{eq:a69}) describes qualitatively an impact of the bias, 
for a quantitative analysis an account of the subleading effects may be required.  These subleading 
corrections were calculated$\,$\cite{egu} where the MLLA formula for the biased two-jet-event multiplicity 
was presented as
\be
\label{eq:a70}
N_{q\bar{q}} (L, \kappa_{\rm cut}) \; \approx \; N_{q\bar{q}} (\kappa_{\rm cut} + c_q) \: + \: (L - 
\kappa_{\rm cut} - c_q) \: N_{q\bar{q}}^\prime (\kappa_{\rm cut} + c_q); \quad c_q \; = \; \frac{3}{2}.
\ee
The results from the ARIADNE Monte Carlo are demonstrated$\,$\cite{egk} to be
in good agreement with the 
prediction of Eq.~(\ref{eq:a70}).

Let us return now to the three-jet-event case.  As was already discussed, in the large-$N_C$ limit the 
emission of softer gluons from a $q\bar{q}g$ system corresponds to two antennas-dipoles which emit gluons 
independently.  If a gluon jet is resolved with the transverse momentum $p_\perp$, this imposes a constraint 
on the emission of sub-jets from the two antennas.  Thus, the contribution from each antenna is determined 
by an expression like Eqs.~(\ref{eq:a69},\ref{eq:a70}).

An essential issue concerns the definition of the jet resolution scale $p_\perp$ for three-jet events.  
In the Lund dipole approach$\,$\cite{gg1,agns,gg3} $p_\perp$ is defined as 
\be
\label{eq:a71}
p_{\perp {\rm Lu}}^2 \; = \; \frac{4 (p_+ \cdot p_1) (p_- \cdot p_1)}{s},
\ee
c.f.\ Eqs.~(\ref{eq:a43}, \ref{eq:a64}).

These two $p_\perp$-definitions agree for soft gluons but deviate for harder gluons.  While $p_{\perp {\rm Lu}}$ 
is always bounded by $\sqrt{s}/2$ the scale parameter $p_{1 \perp}$ in Eq.~(\ref{eq:a64}) has no kinematic upper 
limit in the massless case.  For hard gluons $p_{\perp {\rm Lu}}$ is of the same order as the parent quark 
virtuality.  It was shown$\,$\cite{agus} that the ${\cal O} (\alpha_S^2)$ matrix elements are well described if 
$p_{\perp {\rm Lu}}$ is chosen as an ordering parameter in the perturbative branching.  Therefore, it looks 
natural to assume$\,$\cite{egu,egk} that the constraint on further emissions is described by the identification 
$p_{\perp {\rm cut}} = p_{\perp {\rm Lu}}$.  The multiplicity in a $qg$-dipole with restricted $p_\perp$ can 
be described, analogously to the $q\bar{q}$-case, as two forward regions and a central plateau.

We conclude this subsection by focusing on the three-jet configurations obtained by iterative clustering until 
exactly three jets remain, {\it without} a specified resolution scale, where, hence, the constraint on sub-jet 
$p_\perp$ is described by $p_{\perp {\rm cut}} = p_{\perp {\rm Lu}}$.  
This allows 
us to avoid \lq biasing\rq\ the gluon jet sample, which makes this selection procedure convenient for 
extracting the unbiased $gg$-multiplicity.$\,$\cite{egk}  This strategy is chosen in the recent DELPHI analysis 
$\,$\cite{hkls1,ms1} while each event is clustered to precisely three jets.

By clustering each event to three jets one gets$\,$\cite{egk} for the multiplicity in three-jet 
events (c.f.\ Eqs.~(\ref{eq:a63}--\ref{eq:a66}))
\be
\label{eq:a72}
N_{q\bar{q}g} \; = \; N_{q\bar{q}} (L_{q\bar{q}}, \kappa_{\rm Lu}) \: + \: \frac{1}{2} \: N_{gg} (\kappa),
\ee
with 
\bea
\label{eq:a73}
L_{q\bar{q}} & = & \ln \frac{2 (p_+ \cdot p_-)}{\Lambda^2}, \quad \kappa_{\rm Lu} \; = \; \ln 
\frac{4 (p_+ \cdot p_1) (p_- \cdot p_1)}{s \Lambda^2}, \nonumber \\
& & \\
\kappa & = & \ln \frac{2 (p_+ \cdot p_1) (p_- \cdot p_1)}{(p_+ \cdot p_-) \Lambda^2} \; = \; \ln 
\frac{p_{1_{\perp}}^2}{\Lambda^2}. \nonumber
\eea
An alternative expression for $N_{q\bar{q}g}$ is given as$\,$\cite{egk}
\be
\label{eq:b72}
N_{q\bar{q}g} \; = \; N_{q\bar{q}} (L, \kappa_{\rm Lu}) \: + \: \frac{1}{2} \: N_{gg} (\kappa_{\rm Lu}).
\ee
Note that the consistency between Eqs.~(\ref{eq:a72}) and (\ref{eq:b72}) follows from the fact that the 
total rapidity range in the $qg$ and $\bar{q}g$-dipoles,
\be
\label{eq:a74}
L_{qg} \: + \: L_{g\bar{q}} \; = \; \ln \frac{2 (p_+ \cdot p_1)}{\Lambda^2} \: + \: \ln 
\frac{2 (p_- \cdot p_1)}{\Lambda^2},
\ee
can be described in two different ways by the equalities
\be
\label{eq:a75}
L_{qg} + L_{g\bar{q}} \; = \; L_{q\bar{q}} + \kappa \; = \; L \: + \: \kappa_{\rm Lu}.
\ee
Let us emphasize that due to QCD coherence the argument in $N_{gg}$ is $p_{1\perp}^2$ in (\ref{eq:a72}) 
and $p_{1 \perp {\rm Lu}}^2$ in (\ref{eq:b72}) and not $(2 p_\perp)^2$ as one would naively expect.  The 
difference between the results of Eqs.~(\ref{eq:a72}) and (\ref{eq:b72}) arises from the subleading 
corrections (for example, recoil effects) which are not controllable within
this framework.$\,$\cite{egk}

Formally the bias is related to order $\alpha_S$ effects, and while it practically does not affect the 
gluon piece it can be important for the $q\bar{q}$-contribution.  Selecting events with comparatively 
large values of $p_\perp$, where the bias is not essential, we arrive at the 
old result$\,$\cite{dkt11} 
(c.f. Eq.~(\ref{eq:a63}))
\be
\label{eq:a76}
N_{q\bar{q}g} \; = \; \left [N_{q\bar{q}} (L_{q\bar{q}}) \: + \: \frac{1}{2} \: N_{gg} (\kappa) \right ] \: 
\left [1 + {\cal O} (\alpha_S) \right ].
\ee
The effect of the bias becomes more and more important for smaller values of $p_{\perp {\rm cut}}$.  
On the 
other hand,$\,$\cite{egk} the bias is negligible if one selects gluon system recoiling against 
two quark jets in the same hemisphere.  
%This is good news for OPAL analysis.$\,$\cite{jwg,opal02}

\subsection{Experimental Studies of Three-Jet Events}

In this subsection we focus mainly on the new experimental evidence of the inter-jet color coherence effects 
from the analyses of the LEP-1 data, 
also discussed in recent reviews.$\,$\cite{ko,vpn,kh1,alephr,ms}  The main lesson 
from the recent impressive studies is that now we have (quite successfully) entered the stage of 
quantitative tests of the details of color drag phenomena.  Owing to the massive LEP-1 statistics, one 
can perform a very detailed study allowing to test even some subtle features of the theoretical 
expectations.

Let us mention a few new facts concerning comparison with the analytical QCD predictions.  DELPHI$\,$\cite{delphi4} 
have performed the first quantitative verification of the perturbative prediction$\,$\cite{adkt2} for the ratio 
$R_\gamma$
\be
\label{eq:a77}
R_\gamma \; = \; \frac{N_{q\bar{q}} (q\bar{q}g)}{N_{q\bar{q}} (q\bar{q}\gamma)}
\ee
of the particle population densities in the interquark valley in the $e^+ e^- \rightarrow q\bar{q}g$ and 
$e^+ e^- \rightarrow q\bar{q}\gamma$ events.  For a clearer quantitative analysis a comparison was 
performed for the $Y$-shaped symmetric events using the double-vertex method for the $q$-jet tagging.  The 
ratio $R_\gamma$ of the charged particle flows in the $q\bar{q}$ angular interval [$35^\circ, 115^\circ$] was 
found to be
\be
\label{eq:a78}
R_\gamma^{\exp} \; = \; 0.58 \: \pm \: 0.06.
\ee
This value is in fairly good agreement with the expectation following from Eqs.~(\ref{eq:a34}) and 
(\ref{eq:a35}) at $N_C = 3$, for the same angular interval
\be
\label{eq:a79}
R_\gamma^{th} \; \approx \; \frac{0.65 N_C^2 - 1}{N_C^2 - 1} \; \approx \; 0.61.
\ee
(Note that this ratio is slightly larger than the one predicted for Mercedes-type events in (\ref{eq:a36})).  
The string/drag effect is quantitatively explained by the perturbative prediction and the above ratio does 
not appear to be affected by hadronization effects in an essential way.

\begin{figure}[tb]
\begin{center}
\mbox{\epsfig{file=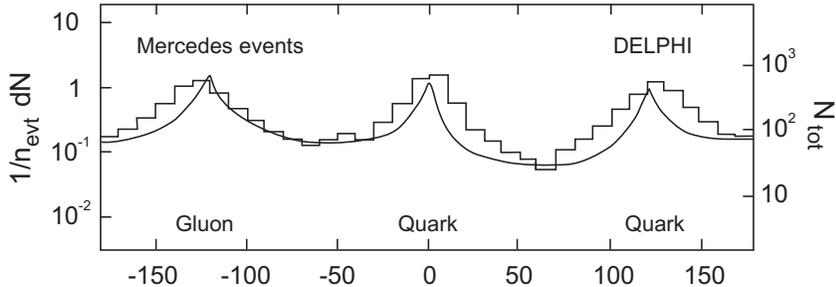,width=11.0cm%
%,bbllx=64bp,bblly=280bp,bburx=490bp,bbury=693bp%
}}
          \end{center}
\caption{Charged particle flow in $q\protect\bar{q}g$-events% 
$\,$\protect\cite{delphi4} in comparison with analytical %\linebreak 
prediction.$\,$\protect\cite{ko}}
\label{figval5}
\end{figure}

Another DELPHI result$\,$\cite{delphi4} concerns the analysis of the threefold-symmetric $e^+ e^- \rightarrow 
q\bar{q}g$ events using the double-vertex tagging method.  It is shown that the string/drag effect is 
clearly present in these fully symmetric events and it cannot be an artefact due to kinematic selections.  
The azimuthal-angle dependence of particle density in the event plane is shown in
Fig.~\ref{figval5}.  Quantitatively, 
comparing the minima located at $\pm [50^\circ, 70^\circ]$, the particle population ratio $R_g = 
N_{qg}/N_{q\bar{q}}$ in the $q - g$ and $q - \bar{q}$ valleys is found to be
\be
\label{eq:a80}
R_g^{\exp} \; = \; 2.23 \: \pm \: 0.37.
\ee
This is to be compared with the asymptotic prediction $R_g = 2.46$ for projected rates at central angles, 
whereas for the above angular interval one finds$\,$\cite{ko,adkt2}
\be
\label{eq:a81}
R_g^{th} \; \approx \; 2.4
\ee
in good agreement with the experimental value.  The number (\ref{eq:a81}) was obtained from the prediction 
for the full angular distribution$\,$\cite{dkt11} (see subsection 2.3) which is also displayed in
Fig.~\ref{figval5} with the 
normalization adjusted.  The calculation takes into account both the inside-jet and inter-jet coherence.  
As can be seen, the relative depth of the two valleys between the jets is well reproduced.  Also reasonable 
is the distribution around the gluon jet direction.  On the other hand, more particles than expected are 
found within the quark jets.  In part this is a consequence of using in the calculation the asymptotic 
value 4/9 for $1/r$, which, as we have already discussed, is approached rather slowly from above.  Despite 
these shortcomings, the main effect, the ratio $R_g$ of particle densities in the valleys between the two 
types of jets, has been correctly predicted by the analytical calculations.

Another prediction concerns the beam-energy dependence.  At a higher energy the angular density in
Fig.~\ref{figval5} 
gets essentially multiplied by an overall factor:$\,$\cite{dkt11}  the particle density in between the 
jets rises at a rate comparable to the density within the jets.

%It was realized long ago$\,$\cite{dkt8} that 
An instructive test of the inter-jet coherence can be
performed when studying the particle flow in the direction 
transverse to the 3-jet event plane.$\,$\cite{klo}
Consider the radiation of a soft gluon perpendicular to the plane of a 3-jet
$q\overline q g$ event. This radiation can be obtained from (\ref{eq:a31}) 
where the contribution from one antenna is simply 
$(\widehat{ij}) = 1-\cos \Theta_{ij}$. The ratio 
$R_\perp=dN^{q\overline q g}/dN^{q\overline q}$ of soft perpendicular radiation in three
and two-jet events is obtained as$\,$\cite{klo}
\begin{equation}
R_\perp=\frac{N_C}{4C_F}[2-\cos\Theta_{1+}-\cos\Theta_{1-} -\frac{1}{N_C^2}
(1-\cos \Theta_{+-})].
\label{rperp}
\end{equation}
In the large $N_C$ limit we have just the superposition of two dipoles with
contribution $\sim 1-\cos\Theta_{1q}$.  
Note the limiting cases $R_\perp=1$ for collinear primary gluon
emission and the proper $gg$ limit $R_\perp=N_C/C_F$ for the configuration
with parallel $q\overline q$ ($\Theta_{+-}=0)$ recoiling against the gluon.
   
DELPHI$\,$\cite{delphi03} have presented the first results on 
the particle yield in the transverse direction for the $Y$-shaped symmetric $q\bar{q}g$-events. 
Fig.~\ref{figval6} 
shows the multiplicity within the cone with the fixed opening angle of $30^\circ$ perpendicular to the 
event plane.  The abscissa is the angle $\Theta_1$ between the low-energy jets
 (the non-leading $q$-jet and the gluon).  
The plotted curve corresponds to the perturbative
prediction from Eq.~(\ref{rperp}) using
$\Theta_{1+}=\Theta_{+-}=\pi-\Theta_1/2$ and $\Theta_{1-}=\Theta_1$
with the normalization left free.  
The rise of particle flow                                                   
observed for increasing opening angle $\Theta_1$ signals
an increase of the ``effective color charge'' corresponding to a 
transition from a $q\overline q$ to a $gg$ type antenna. 
The experimental data prove to be in a
%\be
%\label{eq:a82}
%N_\perp^{q\bar{q}g} \: \sim \: 2 \: + \: \cos \frac{\Theta_1}{2} \: - \: \cos \Theta_1 \: - \: 
%\frac{1}{N_C^2} \: \left ( 1 + \cos \frac{\Theta_1}{2} \right )
%\ee
quite good agreement with the 
perturbative predictions, providing 
a new test of the analytical results independent of the hadronization 
models.

\begin{figure}[t]
\begin{center}
\mbox{\epsfig{file=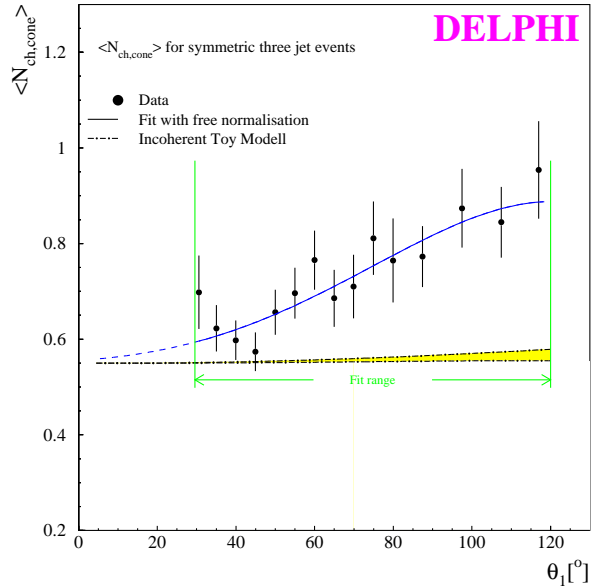,width=8.0cm%
%,bbllx=64bp,bblly=280bp,bburx=490bp,bbury=693bp%
}}
          \end{center}
\caption{Multiplicity within a $30^\circ$-cone perpendicular to three-jet-event
plane as a function 
of the event topology.$\,$\protect\cite{delphi03} The curve represents the
perturbative prediction (\ref{rperp}).}
\label{figval6}
\end{figure}

Of special interest is also the measurement of the yield of particles 
perpendicular to the production plane in the low momentum range.$\,$\cite{klo}  
Within the perturbative scenario the 
very soft particle production should be sensitive again 
only to the color charge topology of the primary emitters.  
%the density of perpendicular radiation roughly doubles when going from the primary $q\bar{q}$- to the 
%$gg$-antenna.  
It looks quite challenging to test whether the angular dependence of the particle yield given 
by Eq.~(\ref{rperp}) holds downwards up to small momenta $p \sim Q_0$.

If one allows for arbitrary three-jet kinematic configurations, new information can be obtained about 
the evolution of the event portrait with the variation of toplogy, see previous subsection. 
ALEPH,$\,$\cite{aleph01} DELPHI$\,$\cite{delphiglu,kh1,hkls1,ms1} and OPAL$\,$\cite{opal02} have convincingly 
demonstrated that, in agreement with the QCD radiophysics, the event multiplicity in three-jet events 
depends both on the jet energies and on the angles between the jets.  These results clearly show the 
predicted toplogical dependence of jet properties.

Recently DELPHI$\,$\cite{hkls1} have reported the new results on the total charged particle multiplicity 
in $Y$-shaped three-jet events.  A crucial point in the DELPHI analysis 
$\,$\cite{delphiglu,hkls1} is that 
each event is clustered to precisely three jets. 
 This allows to determine $N_{q\bar{q}g}$ as a 
function of the opening angle, $\Theta_1$, 
between the low-energy jets and to extract the \lq unbiased\rq\ 
$gg$-multiplicity, $N_{gg}$.  Recall that in 
symmetric $Y$-events there are only two scale parameters, 
$\sqrt{s} = M_Z$ and $\Theta_1$,
 so the event-topology dependence can be expressed as a function of $\Theta_1$ 
only, assuming that the leading jet is not the gluon jet,
 which is true for most cases.  %In$\,$\cite{hkls1} 
A comparison is performed$\,$\cite{hkls1} with analytical results based on 
Eqs.~(\ref{eq:a72}, \ref{eq:b72}).  The values 
of $N_{q\bar{q}}$ entering the expressions (\ref{eq:a72}, \ref{eq:b72})
 via Eq.~(\ref{eq:a70}) are 
extracted from the existing data on the energy dependence of $N_{e^+ e^-}$, 
using the standard 
parametrizations.$\,$\cite{delphiglu}

\begin{figure}[tbh]
\begin{center}
\mbox{\epsfig{file=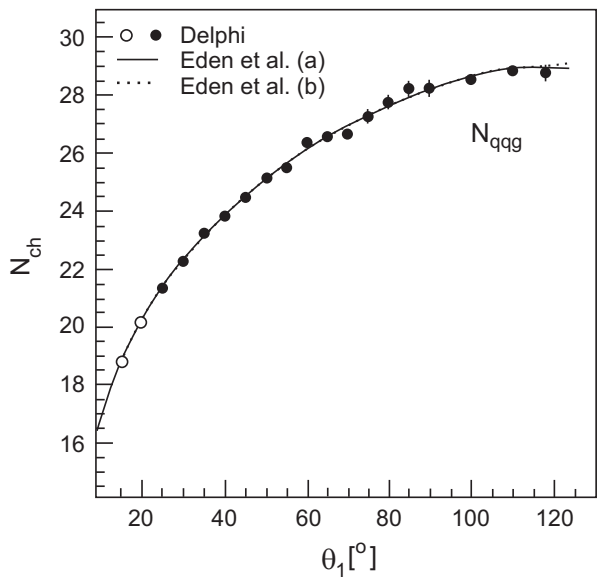,width=8.0cm%
%,bbllx=64bp,bblly=280bp,bburx=490bp,bbury=693bp%
}}
          \end{center}
\caption{Fits of Eqs.~(\ref{eq:a72})(a) and Eq.~(\ref{eq:a73})(b) to the data.  The full points 
depict the range of the fit.$\,$\protect\cite{hkls1}}
\label{figval7}
\end{figure}

Fig~\ref{figval7}$\,$\cite{hkls1} shows the results of comparison with the predictions of Eqs.~(\ref{eq:a72}) 
and (\ref{eq:b72}).  In both cases one finds a good agreement with the data.  DELPHI analysis clearly 
demonstrates that Eqs.~(\ref{eq:a72}, \ref{eq:b72}) fit the data in the wide $\Theta_1$-range better than 
the \lq unbiased\rq\ Eq.~(\ref{eq:a76}).  

\subsection{Collective Phenomena in High-$p_\perp$ Hadronic Reactions}

Color coherence leads to a rich diversity of interference drag-effects in high-$p_\perp$ multi-jet events 
in hadronic collisions (see, for example, Refs. 128,146,155,179, 161,162)
                  %$\,$\cite{dkt1,dkmt1,dkt8,emw,eks,ks}) 
and in $\gamma p$ and DIS 
processes (see, for example, Refs. 59,180,181). %$\,$\cite{klo,bko,akl}).  
The analysis of collective drag phenomena in these 
processes is considerably more subtle than in $e^+ e^-$ annihilation due to the presence of colored 
constituents in both initial and final states.

Recall that during a hard interaction, color is transferred from one parton to another and the 
color-connected partons act as color antennas, with interference effects taking place in the initial 
or final states, or between the initial and final states.  Radiation from the incoming and 
outgoing partons forms jets of hadrons around the direction of these colored emitters.  The soft 
gluon radiation pattern accompanying a hard partonic system can be represented, to leading order in 
$1/N_C$, as a sum of contributions corresponding to the color-connected
partons.$\,$\cite{dkt8,emw}  In 
hadronic high-$p_\perp$ reactions by varying the experimental conditions (triggers) one may separate 
the dominant partonic subprocesses and switch from one subprocess to another.  Recall also that the 
length and height of the hadronic \lq\lq plateau\rq\rq\ depend here on different parameters:  the length 
is determined by the total {\it energy} of the collision, and the height and the plateau structure depend 
on the {\it hardness} of the process governed, as a rule, by the {\it transverse} energy-momentum 
transfer.  Thus, information becomes available that is inaccessible in $e^+ e^-$ annihilation where both 
energy and hardness were given by the value of $\sqrt{s}$.  Therefore, the high-$p_\perp$ hadronic 
collisions provide us with a very prospective laboratory for the detailed studies of the color-related 
phenomena.

As a simple illustrative example let us consider the topology of events, resulting from the high-$p_\perp$ 
$qq$-scattering$\,$\cite{dkt8,emw} which has been studied recently in detail.$\,$\cite{eks}
                 %(for recent detailed studies see$\,$\cite{eks})
\be
\label{eq:a83}
- \hat{t} \: \sim \: \hat{s} \; = \; x_1 \cdot x_2 \: s \quad\quad (x_1, x_2 \sim 1).
\ee
In this case the two crossing processes shown in Fig.~\ref{figval8}(a) and
\ref{figval8}(b) have approximately equal probabilities.  
However, each of them has its own color topology, and therefore, specific particle flows, as schematically 
shown in Fig.~\ref{figval8}(c) and \ref{figval8}(d).

\begin{figure}[htb]
\begin{center}
\mbox{\epsfig{file=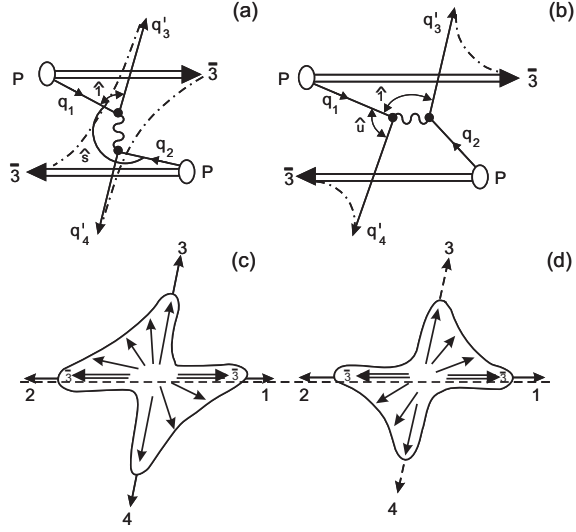,width=8.5cm%
%,bbllx=64bp,bblly=280bp,bburx=490bp,bbury=693bp%
}}
          \end{center}
\caption{Color antennas for two crossing subprocesses of $qq$-scattering (a,b) and the drawings of the 
corresponding hadronic flows (c,d).}
\label{figval8}
\end{figure}

\ For the subprocess of Fig.~\ref{figval8}(a) the soft particle radiation pattern is
\bea
\label{eq:a84}
& & \quad\quad\quad\quad\quad\quad\quad \frac{4 \pi dN^{q_1 q_2}}{d \Omega_{\vec{n}}} \; = \; \frac{C_F}{N_C} \: 
N_g^\prime \left ( \ln \frac{E}{\Lambda} \right ) \nonumber \\
& & \\
& & \times \left \{ (\widehat{14}) + (\widehat{23}) \: + \: \frac{1}{2N_C C_F} \: \left (2 [(\widehat{12}) + 
(\widehat{34})]  -  (\widehat{14}) - (\widehat{23}) - (\widehat{13}) - (\widehat{24}) \right ) \right \}, 
\nonumber 
\eea
where antenna $\widehat{ij}$ is given by Eq.~(\ref{eq:a32}).  In full analogy with the discussions of the drag 
effect in subsection~2.2 one may say that the leading contribution (the first term in (\ref{eq:a84})) has the 
structure of the sum of two independent $q\bar{q}$-antennas $(\widehat{14})$ and $(\widehat{23})$.  This fact 
also can be mimicked by means of the topological picture of the Lund string model.  Let us emphasize that 
in our case to each contribution (single antenna) a dynamical distribution corresponds which takes into 
account the effects of cascade multiplication.  Furthermore, the perturbative approach permits one to control 
not only the leading color contribution, but also the ${\cal O} (1/N_C)$ corrections.

The complete set of antenna patterns for various parton scattering subprocesses for arbitrary $N_C$ 
is listed in.$\,$\cite{eks}  

It has been known for a long time$\,$\cite{dkmt1,dkt8} that an especially bright color interference effect 
arises in the case of large-$E_T$ production of color singlet objects, for instance, in $V +$ jet events 
(with $V = \gamma, W^\pm$ or $Z$).  The hadronic antenna patterns for such processes are entirely 
analogous to that in the celebrated string-drag effect in $e^+ e^- \rightarrow q\bar{q}g$ events.

Recently the first (very impressive) data on $W +$ jet production from D0$\,$\cite{d00,d001} have become 
available.  The color-coherence effects are clearly seen and these studies have a very promising future.  
They may play the same role for hadron colliders as the important series of results on inter-jet studies 
at $e^+ e^-$ colliders. % In$\,$\cite{ks} 
The quantitative predictions were presented$\,$\cite{ks} for color interference 
phenomena in the distribution of soft particles and jets in $V +$ jet production at hadron colliders, 
in particular the Tevatron $\sqrt{s} = 1.8~{\rm TeV}~p\bar{p}$ collider.  There are two leading-order 
processes, $q\bar{q} \rightarrow Vg$ and $qg \rightarrow Vq$.  Each has its own distinctive antenna 
pattern.  In principle, the antenna pattern could be used as a \lq partonometer\rq\ to identify the 
dominant scattering process.

Omitting the CKM factors for clarity, the matrix elements squared for the lowest-order processes 
(for the case $V = W$) can be written as
\bea
\label{eq:a85}
\overline{\sum} |{\cal M}|^2 \left (q (p_1) \bar{q} (p_2) \rightarrow W (p_3) g (p_4) \right ) & = & 
\frac{g_s^2 g_W^2}{4} \left ( 1 - \frac{1}{N_C^2} \right ) \frac{t^2 + u^2 + 2s M_W^2}{tu}, 
\nonumber \\
& & \\
\overline{\sum} |{\cal M}|^2 \left (q (p_1) g (p_2) \rightarrow W (p_3) q (p_4) \right ) & = & 
\frac{g_s^2 g_W^2}{4} \: \frac{1}{N_C} \: \frac{t^2 + s^2 + 2u M_W^2}{-ts}, \nonumber
\eea
where $s = (p_1 + p_2)^2, t = (p_1 - p_3)^2$ and $u = (p_1 - p_4)^2$.  In the soft-gluon approximation, 
the corresponding $2 \rightarrow 3$ matrix elements are
\bea
\label{eq:a86}
\overline{\sum} |{\cal M}|^2 (q\bar{q} \rightarrow Wg (g)) & = & g_s^2 N_C \left ( [14] + [24] - 
\frac{1}{N_C^2} \: [12] \right ) \: \overline{\sum} |{\cal M}|^2 (q\bar{q} \rightarrow Wg) \nonumber \\
& & \\
\overline{\sum} |{\cal M}|^2 (qg \rightarrow Wq (g)) & = & g_s^2 N_C \left ( [12] + [24] - 
\frac{1}{N_C^2} \: [14] \right ) \: \overline{\sum} |{\cal M}|^2 (qg \rightarrow Wq) \nonumber
\eea
with 
\be
\label{eq:a87}
[ij] \; \equiv \; \frac{p_i \cdot p_j}{p_i \cdot k \: p_j \cdot k} \; = \; \frac{1}{E_k^2} \: 
(\widehat{ij}).
\ee
Note that for these processes, the effect of the soft gluon emission is simply to multiply the 
lowest-order matrix elements squared by an overall factor consisting of three different antennas, defined 
according to (\ref{eq:a86}), one of which is suppressed in the large $N_C$ limit.  This structure 
is universal for any electroweak boson + jet production, i.e.\ $V = W^\pm, Z^0, \gamma, \gamma^*$.

It is worthwhile to mention that the study of soft radiation pattern accompanying dijet photoproduction 
can provide a new tool to distinguish the so-called direct and resolved
mechanisms.$\,$\cite{klo,bko}  Of 
special interest here is the measurement of the soft particle yield perpendicular to 
the reaction plane.$\,$\cite{klo,bko}

Inter-jet color coherence effects were successfully studied by both the CDF and D0 Collaborations in 
$p\bar{p}$ collisions at the Fermilab Tevatron Collider.$\,$\cite{cdf01,d002,nikos}  The 
experimental results look rather promising.  Thus, the measurements$\,$\cite{cdf01,d002} of the spatial 
correlations between the softer third jet and the second leading-$E_T$ jet in $p\bar{p} \rightarrow$ 3 jet 
$+ X$ events has clearly demonstrated the presence of initial-to-final interference effects in $p\bar{p}$ 
interactions.

In the most direct way the drag-effect signal is extracted in the recent D0 analysis$\,$\cite{d00} of $W +$ jet 
events by comparing the soft particle angular distributions around the colorless $W$ boson and opposing 
leading-$E_T$ jet in the same event.  In this study the $W$ boson provides a template against which the 
soft particle pattern around the jet can be compared.  Such comparison reduces the sensitivity to various 
biases that may be present in the vicinity of both the $W$ boson and the jet.

\begin{figure}[tb]
\begin{center}
\mbox{\epsfig{file=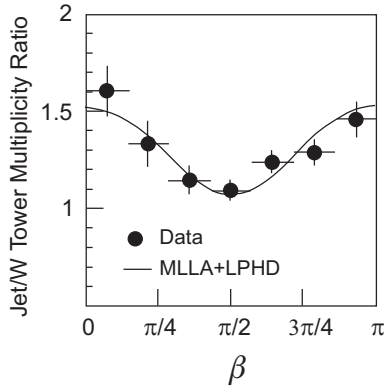,width=5.0cm%
%,bbllx=64bp,bblly=280bp,bburx=490bp,bbury=693bp%
}}
          \end{center}
\caption{Comparison of the jet/$W$ multiplicity ratio from data to 
the MLLA-LPHD %\linebreak 
prediction$\,$\protect\cite{ks}}
\label{figval9}
\end{figure}

Inter-jet coherence effects clearly manifest themselves as an enhancement of soft particle radiation 
around the tagged jet in the event plane (the plane defined by the direction of the $W$ boson and the beam 
axis) relatively to the transverse plane when compared with the particle production around the $W$ boson.  
This is illustrated in Fig.~\ref{figval9} which shows D0 results for the jet/$W$ particle flow ratio as a function 
of the azimuthal angle $\beta$ together with the perturbative expectation based on MLLA and
LPHD.$\,$\cite{ks}  
The analytical prediction proves to be in good quantitative agreement with the data, thus providing 
new evidence supporting the LPHD picture.

\section{Conclusions}

Perturbative QCD proves to be very successful in its applications
to multiparticle production in hard processes. Still, the problem of the soft
limit of the theory and of color confinement is not solved.
Therefore, at present
stage hadroproduction phenomena cannot be derived solely from the
perturbation theory without additional model-dependent assumptions.
The complexity
of the existing popular models requires 
Monte-Carlo methods to derive their predictions.

In this review we have concentrated 
on the question of the extent to which the semisoft
phenomena in multiparticle production in hard processes reflect the properties of the
perturbative QCD cascades.

During the last  years the experiments have collected exceedingly rich new 
information on the dynamics of hadronic jets --- the footprints of QCD partons.
New QCD physics results from LEP2, TEVATRON and HERA continue to pour out 
shedding light on various aspects  of hadroproduction in multi-jet events.  
The existing data convincingly show that the analytical  perturbative approach 
to QCD jet physics is in a remarkably healthy shape.

The key concept of this approach is the hypothesis of local parton- hadron duality,
according to which the parton cascade can be evolved down to a  low virtuality scale
of the order of the hadronic masses, where 
the conversion of partons into hadrons
occurs. Therefore it is the physics of QCD branching  which
governs the gross features of the jet development. 

  In particular, the perturbative 
universality of jets appearing in different hard processes is nicely confirmed.
Moreover, the data demonstrate that the transition between the perturbative
and non-perturbative stages of jet evolution is quite smooth, and that the 
bright color coherence phenomena successfully survive the hadronization 
stage.

LEP1 proves to be a priceless source of information on the QCD dynamics.  It 
has benefited from the record statistics and the substantial lack of background.  
We have learned much interesting physics, but 
there is still a need for further refined
analyses of the data recorded at LEP1. 

The programs of QCD studies at the LHC and at a  future linear $e^+ e^-$ collider
look quite promising.  The semisoft QCD physics steadily becomes one of the important
topics for investigation in the TEVATRON and HERA experiments.

Concluding this review let us emphasize once more that, of course, there is no
mystery within the perturbative QCD framework.  One is only supposed to perform
the calculational routine properly.  So it is entirely unremarkable that the
quantum mechanical interference effects should be observed in the perturbative
results.  Of real importance is that the experiment demonstrates that the
transformer between the perturbative and non-perturbative phases acts very
smoothly.  This message could (some day) shed light on the mechanism of color
confinement.

\section*{Acknowledgments}
We are grateful to
N. Brook, Yu. Dokshitzer, K. Hamacher, A. Korytov, 
S. Lupia, R. Orava, A. Safonov,
W.J. Stirling, N. Varelas, B. Webber and L. Zawiejski 
for useful discussions.
VAK and JW thank the theory group of the Max-Planck-Institute,
Munich, for their
hospitality.    
We are thankful to Sharon Fairless and Pauline Russell for their kind
assistance.
VAK thanks the Leverhulme Trust for a Fellowship.
This work was supported in part
by the EU Framework TMR programme, contract FMRX-CT98-0194 (DG 12-MIHT)
and by the Polish Committee for
Scientific Research under the grant no. PB 2 PO3B 01917.

\end{document}